\documentclass[twocolumn,traditabstract]{aa}

\usepackage{txfonts}
\usepackage{graphicx}
\usepackage{natbib}
\usepackage{hyperref}

\renewcommand{\deg}{\mbox{$^{\circ}$}}

\begin{document}

   \title{The thermal emission of Centaurs and Trans-Neptunian objects at millimeter wavelengths from ALMA observations}

\author{E. Lellouch\inst{1}
\and R.~Moreno\inst{1}
\and T.~M\"uller\inst{2}
\and S.~Fornasier\inst{1}
\and P.~Santos-Sanz\inst{3}
\and A.~Moullet\inst{4}
\and M.~Gurwell\inst{5}
\and J.~Stansberry\inst{6}
\and R.~Leiva\inst{1,7}
\and B.~Sicardy\inst{1}
\and B.~Butler\inst{8}
\and J. Boissier\inst{9}
}

 \institute{LESIA, Observatoire de Paris, PSL Research University, CNRS, Sorbonne Universit\'es, UPMC Univ. Paris 06, Univ. Paris Diderot, Sorbonne Paris Cit\'e, 5 place Jules Janssen, 92195 Meudon, France\\ 
\email{emmanuel.lellouch@obspm.fr}
\and Max-Planck-Institut f\"ur Extraterrestrische Physik, 
Giessenbachstra\ss e, 85748 Garching, Germany   
\and Instituto de Astrof\'isica de Andaluc\'ia-CSIC, Glorieta de la Astronom\'ia s/n, 18008-Granada, Spain.
\and National Radio Astronomy Observatory
520 Edgemont Road 22903 Charlottesville, VA, USA
\and Harvard-Smithsonian Center for Astrophysics, Cambridge, MA 02138, 
USA
\and Space Telescope Science Institute, 3700 San Martin Drive, Baltimore, MD 21218 USA
\and Instituto de Astrof\'isica, Facultad de F\'isica, Pontificia Universidad Cat\'olica de Chile, Av. Vicu\~na Mackenna 4860, Santiago, Chile
\and National Radio Astronomy Observatory, Socorro, NM 87801, USA
\and IRAM, Domaine Universitaire, 300 Rue de la Piscine, 38400 Saint-Martin-d'H\`eres                                      
             }

   \titlerunning{Thermal emission of TNOs/Centaurs from ALMA}
   \authorrunning{Lellouch et al.}

   \date{Received, \today; Revised, }


  \abstract
   {The sensitivity of ALMA makes it possible to detect thermal mm/submm emission from small and/or distant Solar System bodies
at the sub-mJy level. While the measured fluxes are primarily sensitive to the objects' diameters, deriving precise
sizes is somewhat hampered by the uncertain effective emissivity at these wavelengths. Following the
work of Brown and Butler (2017) who presented ALMA data for four TNOs with satellites, we report on
ALMA 233 GHz (1.29 mm) flux measurements of four Centaurs (2002 GZ$_{32}$, Bienor, Chiron, Chariklo) and two TNOs (Huya and Makemake), sampling a range of size, albedo and composition. 
These thermal fluxes are combined with previously published fluxes in the mid/far-infrared in order to derive their relative emissivity at radio (mm/submm) wavelengths, using NEATM and thermophysical models. We reassess earlier thermal measurements of these and other objects -- including Pluto/Charon and Varuna -- exploring in particular effects due to non-spherical shape and varying apparent pole orientation whenever information is available, and show that these effects can be key for reconciling previous diameter determinations and correctly estimating the spectral emissivities. We also evaluate the possible contribution to thermal fluxes of established (Chariklo) or claimed (Chiron) ring systems. For Chariklo, the rings do not impact the diameter 
determinations by more than $\sim$5 \%; for Chiron, invoking a ring system does not help in improving the consistency between
the numerous past size measurements.
As a general conclusion, all the objects, except Makemake, have radio emissivities significantly lower than
unity. Although the emissivity values show diversity, we do not find any significant trend  with physical parameters such as diameter, composition, beaming factor, albedo, or color, but we suggest that the emissivity could be correlated with grain size. The mean relative radio emissivity is found to be 0.70$\pm$0.13, a value that we recommend for the analysis of further mm/submm data. 
  
}

   \keywords{Kuiper belt objects: Centaurs: individual: 2002 GZ$_{32}$, Bienor, Chiron, Chariklo, Huya, Makemake, Pluto, Charon, Varuna, 1999TC$_{36}$. Planets and satellites: surfaces. Methods: observational. Techniques: photometric.
               }

\maketitle

\section{Introduction}
Size determinations of Kuiper Belt objects (KBOs) are now available for well over a hundred objects. 
Stellar occultations provide by far the most accurate method, whereby multiple chords with kilometric accuracy,
combined with lightcurve information, 
may yield three-dimensional shapes and even topography \citep[e.g.][]{dias17}.  
For objects with known mass (i.e. those having a much smaller satellite), this permits a precise determination
of the density, the most important geophysical parameter, bearing information on formation mechanisms \citep{brown13a}.
The occultation technique also has the unique potential to probe an object's environment \citep[rings, dust;][] {braga14,ruprecht15} with sensitivity
far superior to direct imaging. Although growing in number at a steady pace thanks to constantly improving occultation
track predictions, these studies have been however so far limited to a dozen objects or so, in particular due to the practical
involvement required to organize multi-site campaigns \citep[a list of successful occultations can be found in][]{santos16, leiva17}.

Thus, as of today, most size determinations
were obtained by thermal radiometry, whereby an optical measurement is combined with ore or more thermal
measurement(s) to derive an object's diameter and albedo. After 
pioneering attempts from the ground \citep[starting with][for Varuna]{jewitt01} and with the Infrared Space Observatory (ISO), most measurements  were
achieved from {\em Spitzer} \citep[$\sim$60 objects at 24 and 70 $\mu$m;][]{stansberry08,brucker09} and {\em Herschel} 
\citep[$\sim$120 objects at 70-160 $\mu$m, plus $\sim$10 of them at 250-500 $\mu$m;][and a few other papers dedicated to
specific objects]{lim10,muller10,santos12,vilenius12,vilenius14,mommert12,fornasier13,lellouch13,duffard14,lacerda14}. About
50 Centaurs/Scattered Disk objects were also detected by WISE  \citep[at 12 and/or 22 $\mu$m]{bauer13}.
The strength of these measurements is in their multi-wavelength character, which by sampling both sides
of the Planck peak, provides information on 
the object's thermal regime, significantly alleviating important
uncertainties on the size determination.

After the demise of these spaceborne facilities, and at least until JWST/MIRI observations become available, ALMA
(the Atacama Large Milimeter Array)
is the tool of choice to measure thermal emission from small and/or distant Solar System bodies. The primary strength of this
mm/submm facility is in its sensitivity. \citet{moullet11} estimated that $\sim$500 of the known KBOs at that time were detectable by ALMA, with a 5-$\sigma$ detection limit of D= 200 km at 40 AU and D = 400 km at 70 AU for 1 hour on source. Consistent with these numbers,  \citet{gerdes17} used ALMA (3 h on source) to detect the newly discovered 2014 UZ$_{224}$ at 92 AU with  S/N of 7 and inferred a $\sim$ 635 km diameter. Thus ALMA should be capable of considerably expanding the sample of TNOs with measured diameters.
Observations at mm/submm wavelengths can be in principle multi-band, but are not very sensitive to the temperature distribution. This is both a disadvantage -- information on the object thermal inertia is unlikely to be gained -- and an advantage -- 
the size determination is insensitive to details of the thermal modeling. The more serious source of uncertainty for the quantitative interpretation of mm/submm data is
the likely occurrence of `̀`emissivity effects" of various origins, depressing the emitted fluxes to lower values than expected, and possibly compromising the diameter determination (e.g. a 40 \% lower than unity emissivity will cause, if not accounted for,
a 20 \% underestimate of the diameter). Such effects are still poorly characterized in the case of transneptunian objects,
although already apparent in the far-IR \citep{fornasier13,lellouch16}. Recently \citet{brown17} observed four binary
KBOs at $\sim$230 and $\sim$350 GHz with ALMA, and determined spectral emissivities at these wavelengths systematically lower 
than the adopted bolometric emissivity, with ratios in the range 0.45-0.92.

Here we expand the \citet{brown17} study by reporting measurements of six additional Centaurs/TNOs with ALMA, as well as 
re-interpreting several ancient or recent observations of the same and other bodies (Pluto/Charon and Varuna in particular) from the ground. This study
thus contributes to the long-term goal of providing a benchmark description of the spectral emissivity of TNOs, its physical origin, and how it may vary with surface composition and other physical parameters, with the practical application of helping
the interpretation of future observations of TNOs at mm/submm wavelengths.


\section{ALMA observations}
We obtained thermal photometry at 1.29 mm of four Centaurs (2002 GZ$_{32}$, Chariklo, Bienor, Chiron) and two Trans-Neptunian objects (Huya and Makemake) with the ALMA 12-m array (proposal
2015.1.01084.S). While the number of objects in this study was necessarily limited due to telescope time constraints, 
the targets were chosen so as to sample different classes of albedo (from $\sim$ 4~\% to $\sim$80~\%), 
surface composition (water ice, volatile ices, featureless spectra), color (from neutral to red), and diameter (from $\sim$200 to $\sim$1500 km). 
Four out of the six objects (i.e. all but Bienor and 2002 GZ$_{32}$) had been investigated by Herschel/SPIRE, three
of them (Chiron, Chariklo and Huya) showing emissivity effects longwards of $\sim$200 $\mu$m \citep{fornasier13}. 
Complementing the study by \citet{brown17} on four mid-size (D = 700-1100 km) binary TNOs (Quaoar, Orcus, 2002 UX$_{25}$ and Salacia), we emphasized Centaurs (four objects, with diameter 180-250 km), three of which had (or now have) shape and/or pole orientation information, permitting more detailed modelling.

\begin{table*}
\caption{Observational details} 
\label{observations}      
\centering                          
\begin{center}
\begin{tabular}{lcccccccc}
Object & UT Date (start/end) & R$_h$ & $\Delta$ & Integration  & Beam ('') & Primary & Secondary  & 233 GHz flux \\
& & (AU) & (AU) & time & & calibrator  & calibrator & density (mJy) \\
\hline

\\

Huya & 25-Jan-2016/11:41 - 12:04 & 28.51 & 28.55 & 1119 sec & 1.30" x 0.97"  & Titan & J1550+0527& 0.614$\pm$0.072 \\
 &  &  &  & &  &  & J1549+0237& \\

\\

2002 GZ$_{32}$ & 29-Jan-2016/11:32 - 11:55 & 18.17 & 18.62 & 1119 sec & 1.26" x 0.91" & J1751+0939 & J1733-1304 &  0.540$\pm$0.026 \\

\\

Makemake & 02-Mar-2016/06:54 -  07:16 & 52.44  & 52.62 & 1119 sec & 1.09" x 0.86" & J1229+0203 & J1303+2433 &  1.185$\pm$0.085\\

\\

Chariklo & 05-Mar-2016/11:28 - 11:50 & 15.29 & 15.62 & 1119 sec & 0.84" x 0.75" & J1924-2914 & J1826-3650 &   1.286$\pm$0.090 \\

\\

Chiron & 26-Mar-2016/14:56 -   15:17 & 18.30 & 19.27 & 1119 sec & 0.81" x 0.75" & Pallas & J0006-0623&  0.311$\pm$0.027\\

\\

Bienor & 15-May-2016/15:06 -  15:28 & 15.57  & 16.51 & 1119 sec & 1.05" x 0.62" & J0237+2848 & J0238+1636 &  0.353$\pm$0.029\\

\\

\hline

\end{tabular}
\end{center}
\end{table*}

All observations were taken in the ALMA Band 6 (211-275 GHz), in the continuum (``TDM") mode. We use the standard frequency tuning
for that band, yielding four 1.875-GHz broad windows centered at 224, 226, 240 and 242 GHz.
The array was in a compact 
configuration (typically C36-2/3), yielding a synthetic beam of $\sim$0.8-1.2", much larger than the object themselves ($<$0.05"). 
All observations were obtained in dual polarization mode, 
with the two polarizations combined at data reduction stage to provide a measurement of the total flux for each object.

Interferometric observations require absolute flux calibrators  -- for which well-modelled Solar System planets or
satellites are optimum choices -- as well as point-like secondary calibrators
for calibration of the atmospheric and instrumental amplitude and phase gains as a function of time. Including calibration overheads,
each observation lasted $\sim$45 min, including $\sim$19 min on source. The requested sensitivity of 30 $\mu$Jy was 
met or exceeded for all sources.

Observational details for each object are given in Table~\ref{observations}. Except in one case, when Titan could be observed,
it turned out that the standard absolute flux calibrators (Titan, Uranus, Neptune, Ganymede...) were not accessible, therefore
radio-source (quasars) -- and in one occasion asteroid 2~Pallas -- were used instead. Some of these quasars are actually 
variable, but routinely monitored from various radio-telescopes, including ALMA, the Sub-Millimeter Array (SMA) and the NOEMA (a.k.a. Plateau de Bure) interferometer. As detailed in Appendix~\ref{sec:calibration}, we took great care in deriving the best flux estimates for these 
calibrators at the time of our TNO measurements, as well as realistic error bars on those. 

\begin{figure}[h]
\centering






\hspace{0cm}\includegraphics[width=4.cm,angle=90]{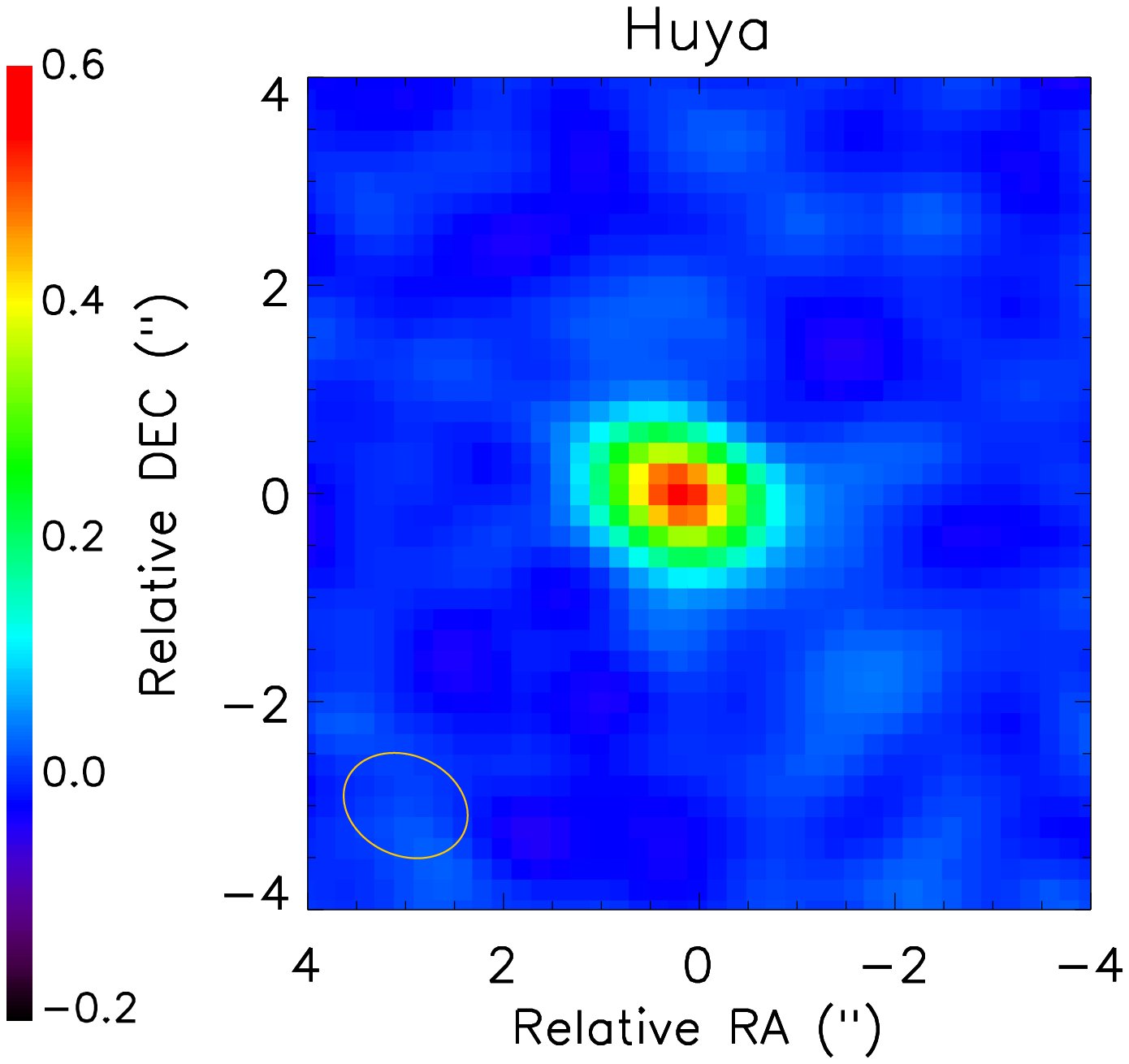}
\hspace{.5cm}\includegraphics[width=4.cm,angle=90]{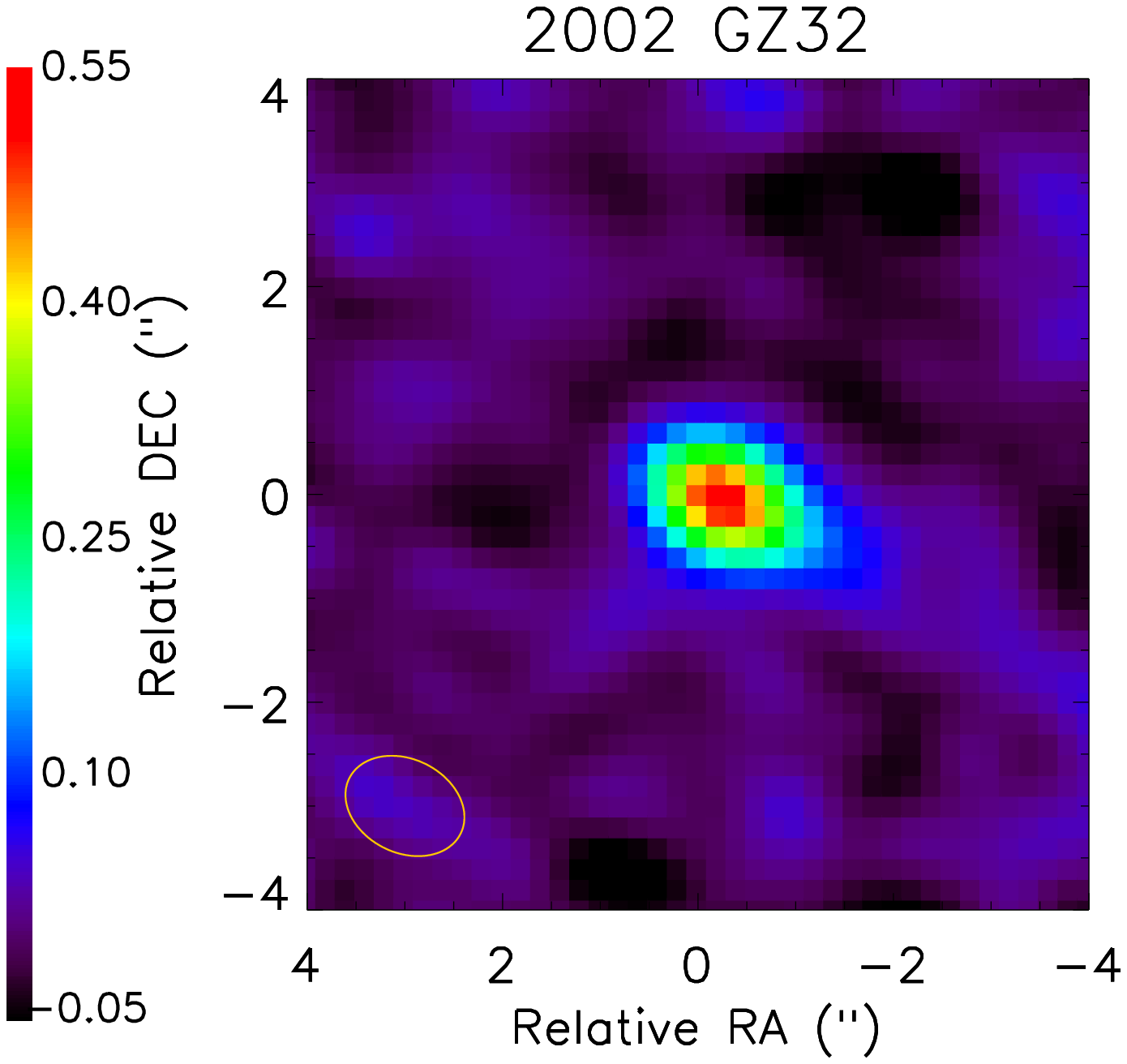} 

\hspace{-0.cm}\includegraphics[width=4.cm,angle=90]{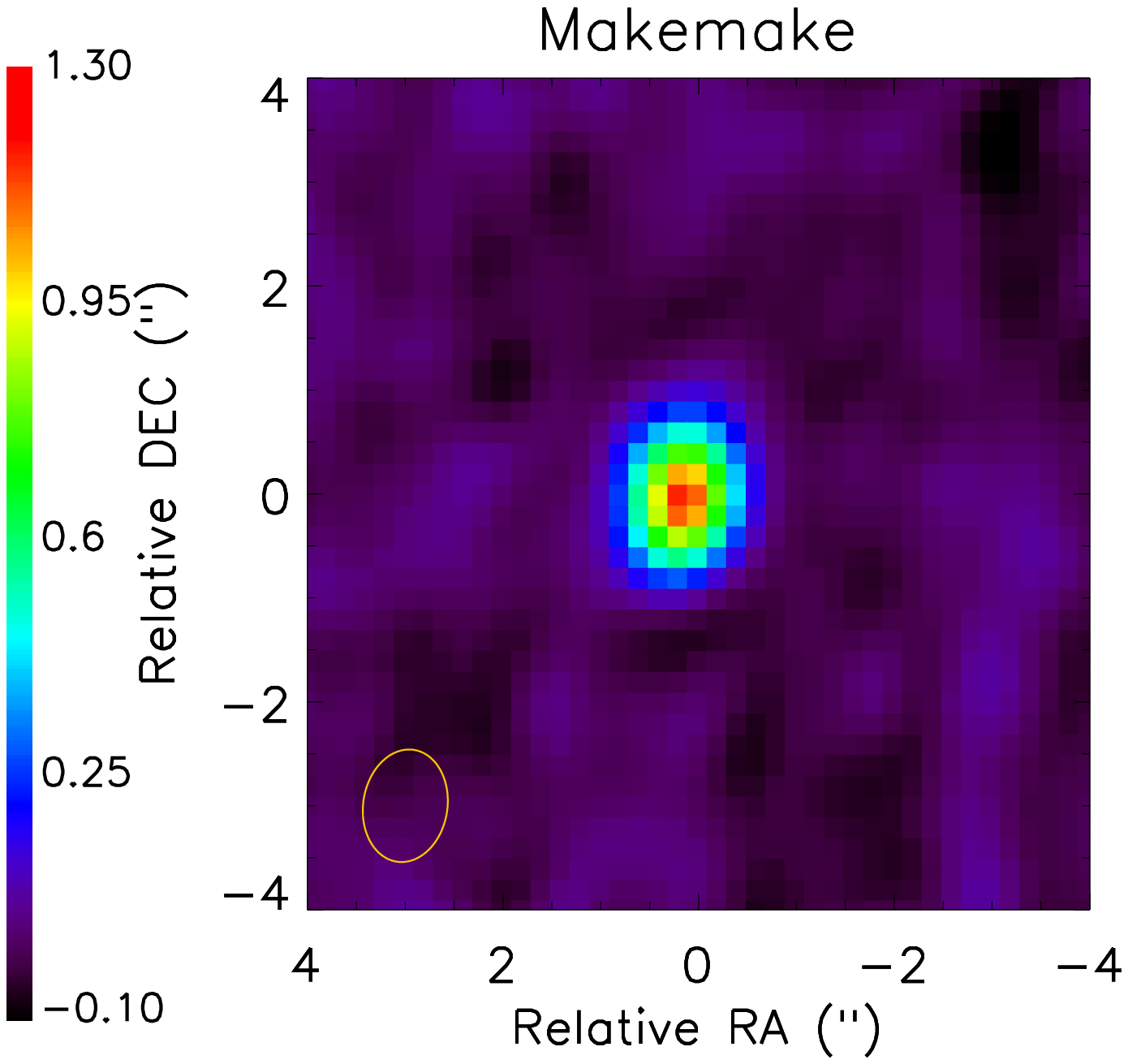}
\hspace{.5cm}\includegraphics[width=4.cm,angle=90]{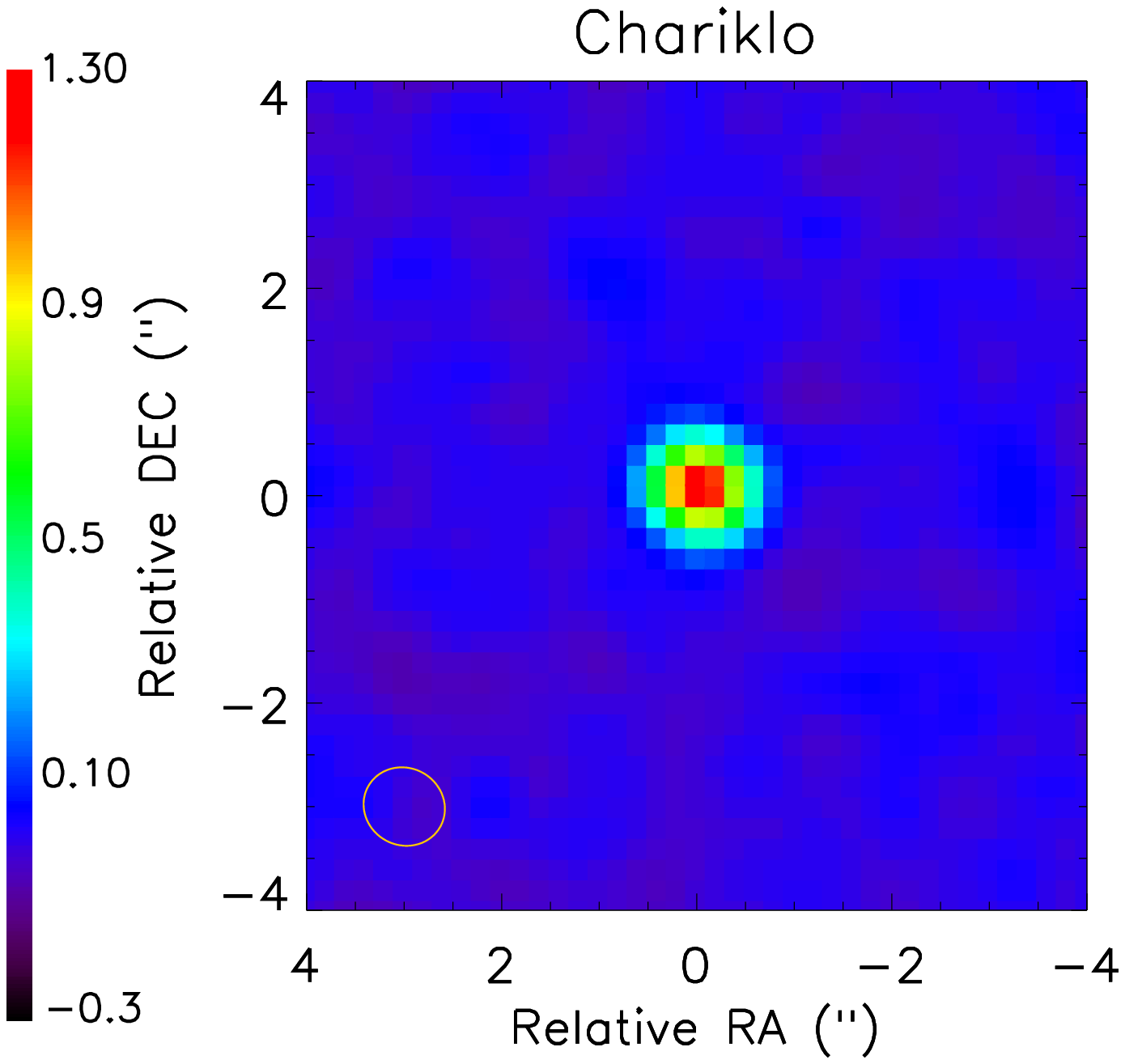}

\hspace{-0.cm}\includegraphics[width=4.cm,angle=90]{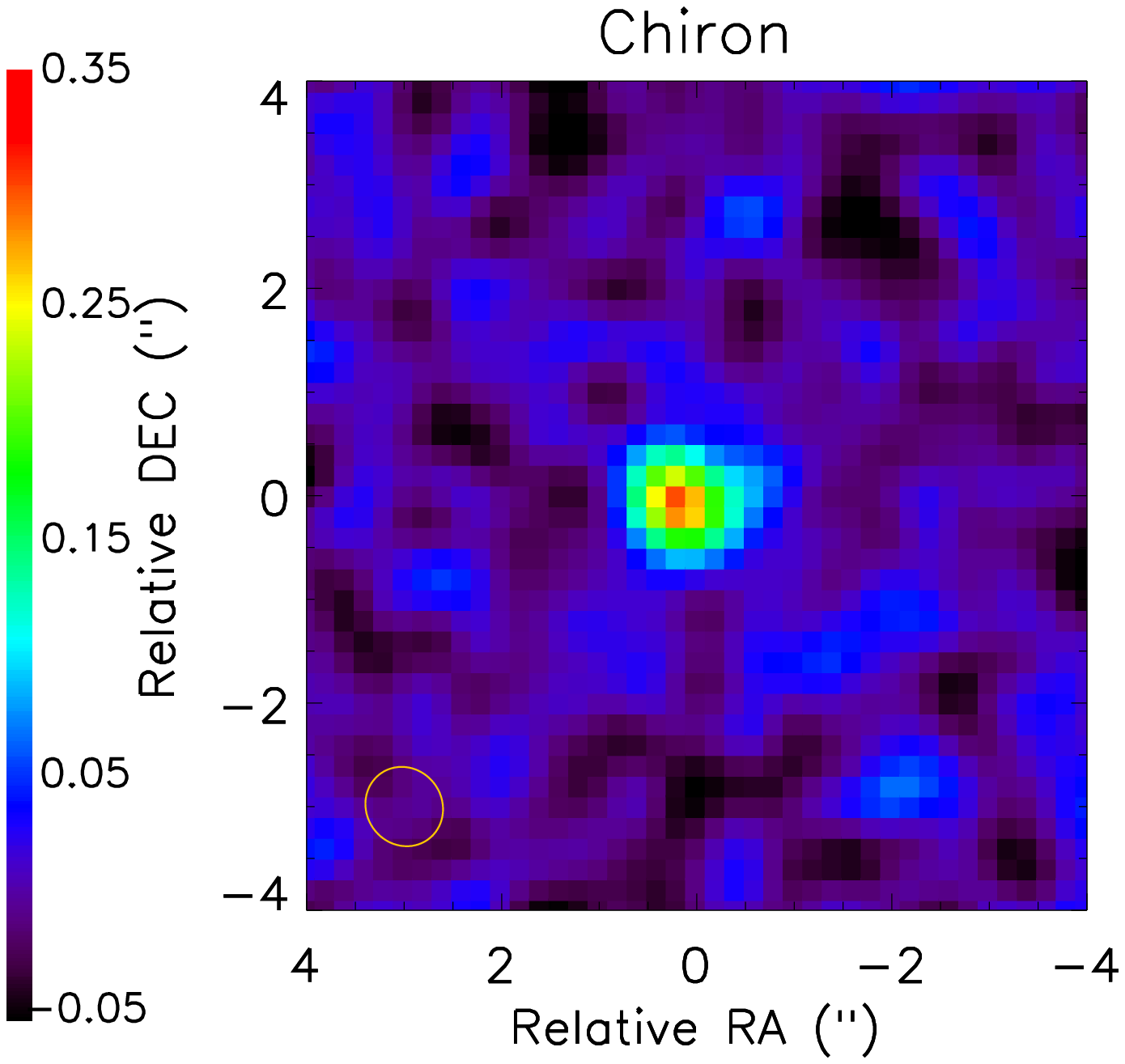}
\hspace{.5cm}\includegraphics[width=4.cm,angle=90]{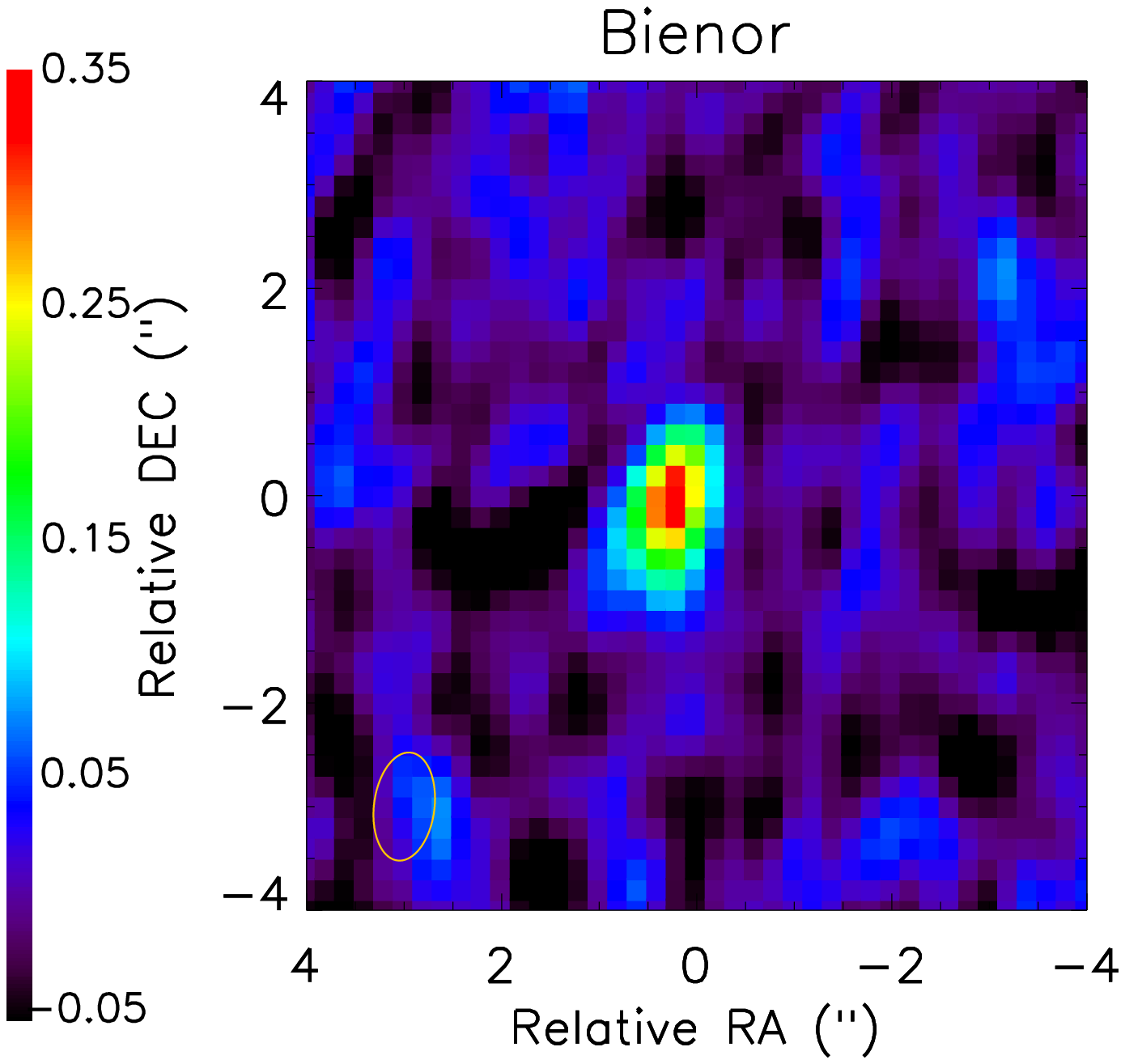}

\caption{ALMA 233 GHz images of the six targets, combining the four spectral windows. {\bf The color bar indicates the flux scale
(mJy/beam).}}
\label{fig:alma_images}
\end{figure}

Initial steps of the data reduction were performed in the CASA reduction package via the ALMA pipeline \citep{muders14}, 
providing a set of visibilities as a function of baselines between each antenna pair.
Results were then exported into the GILDAS package, under which the objects (targets and calibrators) were imaged.
Results of the imaging process for the six targets are shown in Fig~\ref{fig:alma_images}. For each target, visibility fitting (using a point-source model) provided the flux density in each of the four spectral windows, from
which the flux at a 233 GHz reference frequency (1287 $\mu$m) and its standard deviation were obtained. The latter was
quadratically combined with the uncertainty on the flux calibrator scale.  In half of the cases (Huya, Makemake, Chariklo), the precision on the calibrator flux was the limiting factor on the final TNO flux accuracy. Details are given in Appendix~\ref{sec:calibration}. Overall, the precision of the measured TNO flux ranges from 5 to 12 \%, with 8 \% in average.


\section{Modelling}
The essential approach to determine the radio emissivity of the observed TNOs / Centaurs is to combine the modelling 
of their mm/submm flux with previous measurements from far-IR facilities, namely {\em Spitzer}, {\em Herschel} (and WISE if
available). For most objects, in the absence of information on shape and/or rotational parameters (spin period and orientation),
we performed standard NEATM (Near Earth Asteroid Standard Model) fits, assuming sphericity. For several objects, for which such
information may be available, we explored more detailed thermophysical and non-spherical models. 
This is the case especially of Chariklo and Chiron, for which we also investigated the possible contribution
of rings to the observed fluxes.

\subsection{Spherical NEATM fits}
We started with standard NEATM fits of the thermal data. In this approach \citep{harris98}, 
the thermal flux is calculated as the one resulting from instantaneous equilibrium of the object (assumed spherical) with solar insolation, but correcting the temperatures
 by a semi-empirical beaming factor $\eta$, a wavelength-independent quantity which phenomenologically describes the combined effects of 
thermal inertia, spin state, and surface roughness. The NEATM approach is suitable when multiple
thermal wavelengths are available; in this case $\eta$ is, in essence, adjusted to match the object's SED. The local temperature
is then expressed as a function of solar zenith angle SZA:
\begin{equation}
T(SZA) = \left(\frac{S_{\odot}(1 - p_vq)}{\eta\epsilon_b~r_h^2}\right)^{0.25}~cos^{0.25}(SZA)
\end{equation}
where $S_{\odot}$ is the solar constant at 1 AU, $r_h$ is the heliocentric distance in AU, and $p_v$, $q$ and $\epsilon_b$ are the object's geometrical albedo, phase integral, and  bolometric emissivity, respectively. Within NEATM, multiple thermal measurements
are combined with the H$_v$ magnitude to determine the object's diameter, geometric albedo, and beaming factor.
This still involves assumptions on $q$ and $\epsilon_b$. For TNOs, most models have used the $q$ vs  $p_v$ dependence
shown in \cite{brucker09} and a fixed $\epsilon_b$ of 0.9 or 0.95.

\begin{table*}
\caption{Spherical NEATM fit results} 
\label{fitresults}      
\centering                          
\begin{center}
\begin{tabular}{lccccc}
Object    & Diameter  & Geometric           &  Beaming  & Ref. for & Relative mm/submm\\
 & (km) & albedo & factor  & submm/mm data & emissivity\\
\hline 
\\
2002~GZ$_{32}$  & 237$_{-11}^{+12}$  & 0.036$_{-0.005}^{+0.006}$ & 0.97$_{-0.08}^{+0.07}$  & This work & 0.80$_{-0.05}^{+0.05}$ (1.29 mm)\\
\\
Bienor  & 199$_{-12}^{+9}$  & 0.041$_{-0.012}^{+0.016}$ & 1.58$_{-0.11}^{+0.15}$  & This work & 0.62$_{-0.07}^{+0.07}$ 
(1.29 mm) \\
\\

Chariklo  & 241$_{-8}^{+9}$  & 0.037$_{-0.008}^{+0.009}$ & 1.20$_{-0.08}^{+0.10}$ & This work &  1.25$_{-0.10}^{+0.09}$ 
(1.29 mm)\\
\\

 &  &   &    & A01$^a$ &  1.08$_{-0.16}^{+0.16}$  (1.20 mm)\\
\\


Chiron &  210$_{-14}^{+11}$  & 0.172$_{-0.036}^{+0.043}$ & 0.93$_{-0.10}^{+0.13}$  & This work &  0.63$_{-0.07}^{+0.07}$ 
(1.29 mm) \\
\\

 &  &   &   & J92$^a$ &  0.81$_{-0.69}^{+0.56}$ (0.80 mm) \\

\\
 &  &  &   & A95$^a$ &  0.55$_{-0.13}^{+0.13}$ (1.20 mm) \\
\\

Huya & 458$_{-21}^{+22}$  & 0.081$_{-0.008}^{+0.008}$ & 0.93$_{-0.08}^{+0.09}$  & This work & 0.73$_{-0.10}^{+0.08}$ 
(1.29 mm)\\

\\
2002~UX$_{25}$ $^b$&   695$_{-29}^{+30}$  & 0.104$_{-0.010}^{+0.011}$ & 1.07$_{-0.07}^{+0.10}$  &  BB17$^a$ &  0.66$_{-0.06}^{+0.05}$  (0.87 mm)\\
\\
 &  &  &   & &  0.65$_{-0.05}^{+0.05}$ (1.30 mm) \\
\\

Orcus $^b$ &   960$_{-42}^{+45}$  & 0.231$_{-0.022}^{+0.017}$ & 0.97$_{-0.13}^{+0.08}$  & BB17$^a$ &  0.75$_{-0.07}^{+0.07}$ (0.87 mm)\\
\\
 &  &  &  &   &  0.92$_{-0.05}^{+0.07}$ (1.30 mm) \\
\\

Quaoar $^b$  & 1071$_{-57}^{+53}$  & 0.124$_{-0.012}^{+0.016}$ & 1.73$_{-0.19}^{+0.16}$  & BB17$^a$&  0.68$_{-0.04}^{+0.05}$   (0.87 mm)\\
\\
 &  &  &  &  &  0.46$_{-0.03}^{+0.03}$ (1.30 mm) \\
\\

Salacia $^b$ &   909$_{-38}^{+39}$  & 0.042$_{-0.004}^{+0.004}$ & 1.15$_{-0.08}^{+0.11}$  & BB17$^a$&  0.62$_{-0.04}^{+0.03}$  (0.87 mm) \\
\\
 &  &  &   & &  0.69$_{-0.04}^{+0.03}$ (1.30 mm) \\
\\

\hline
\multicolumn{6}{l}{$^a$ A01:\cite{altenhoff01}; BB17:\cite{brown17}; JL92:\cite{jewitt92}, } \\
\multicolumn{6}{l}{\hspace*{.1cm}     A95: \cite{altenhoff95}} \\
\multicolumn{6}{l}{$^b$ Objects from \cite{brown17}} 
\end{tabular}
\end{center}
\end{table*}

\begin{figure*}[ht]
\includegraphics[width=4.2cm,angle=-90]{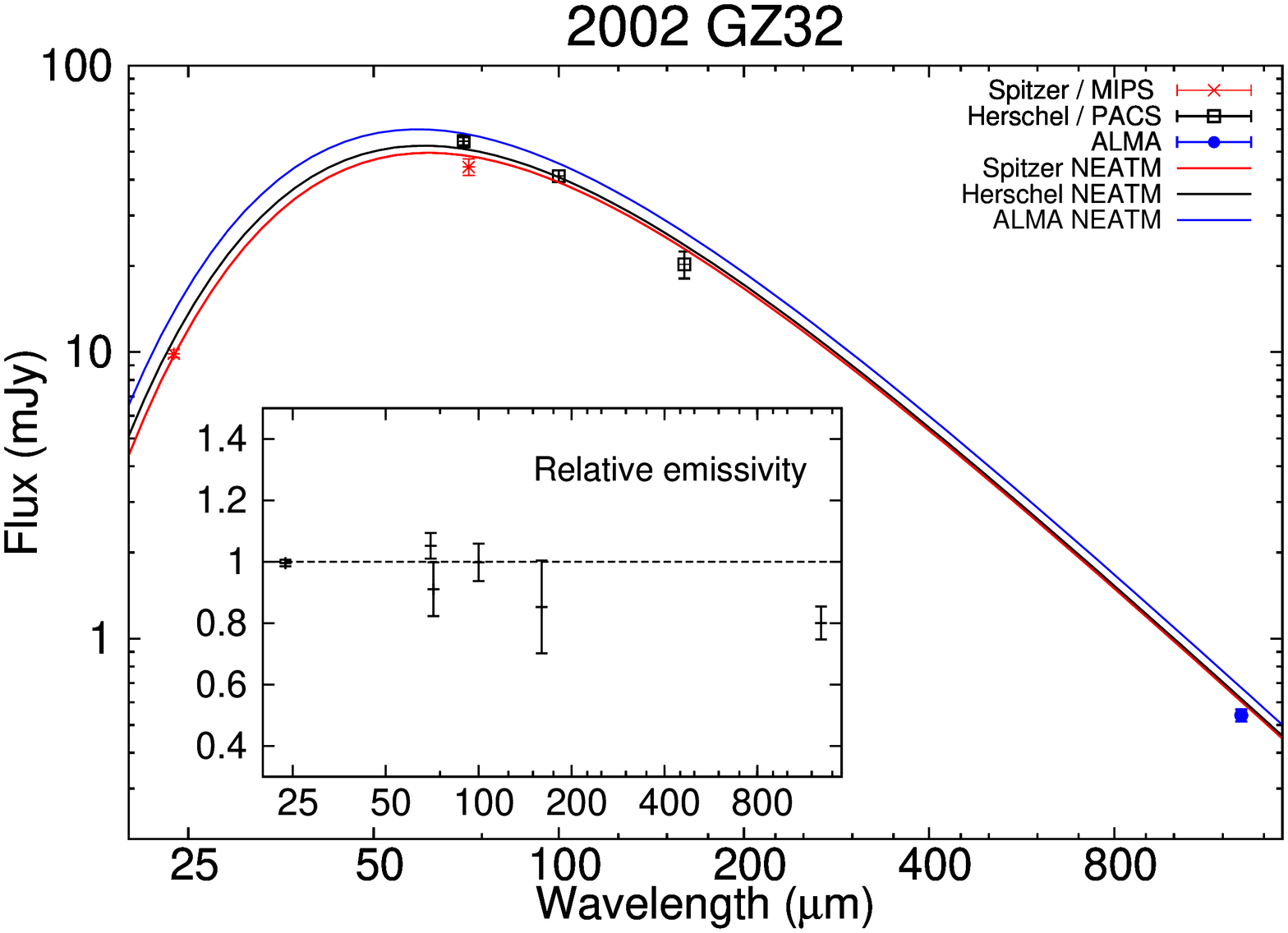}
\includegraphics[width=4.2cm,angle=-90]{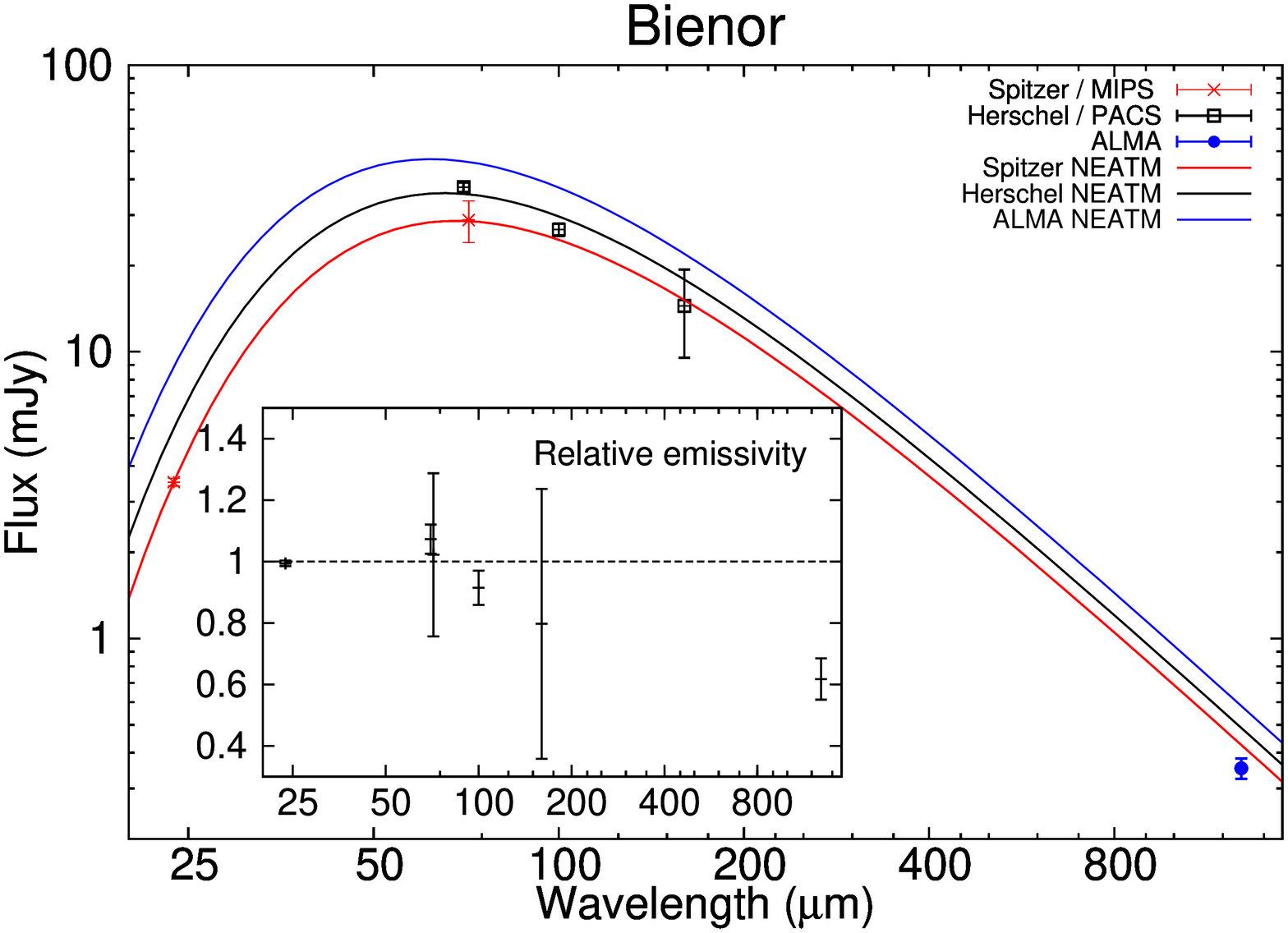}
\includegraphics[width=4.2cm,angle=-90]{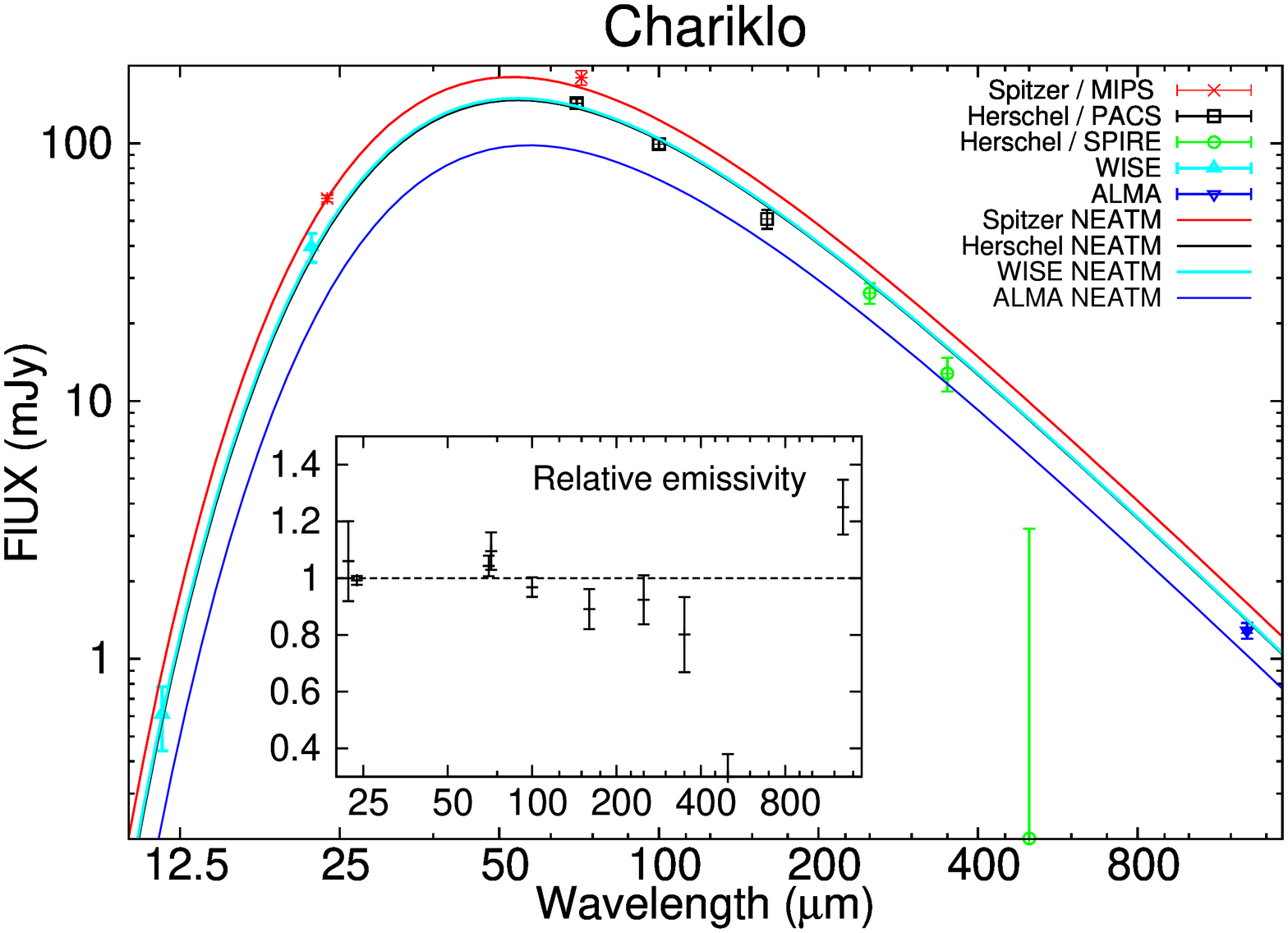}
\includegraphics[width=4.2cm,angle=-90]{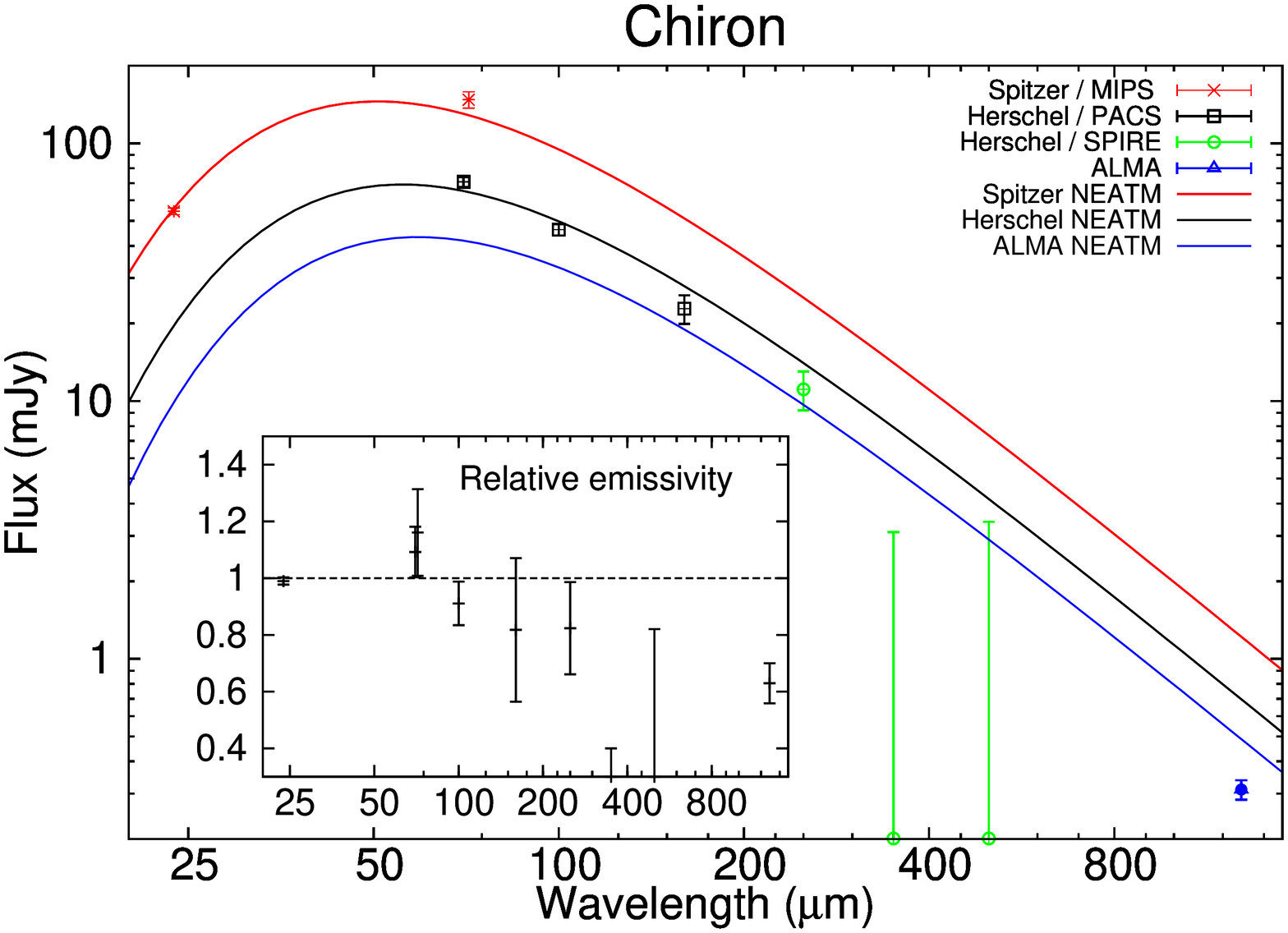}
\includegraphics[width=4.2cm,angle=-90]{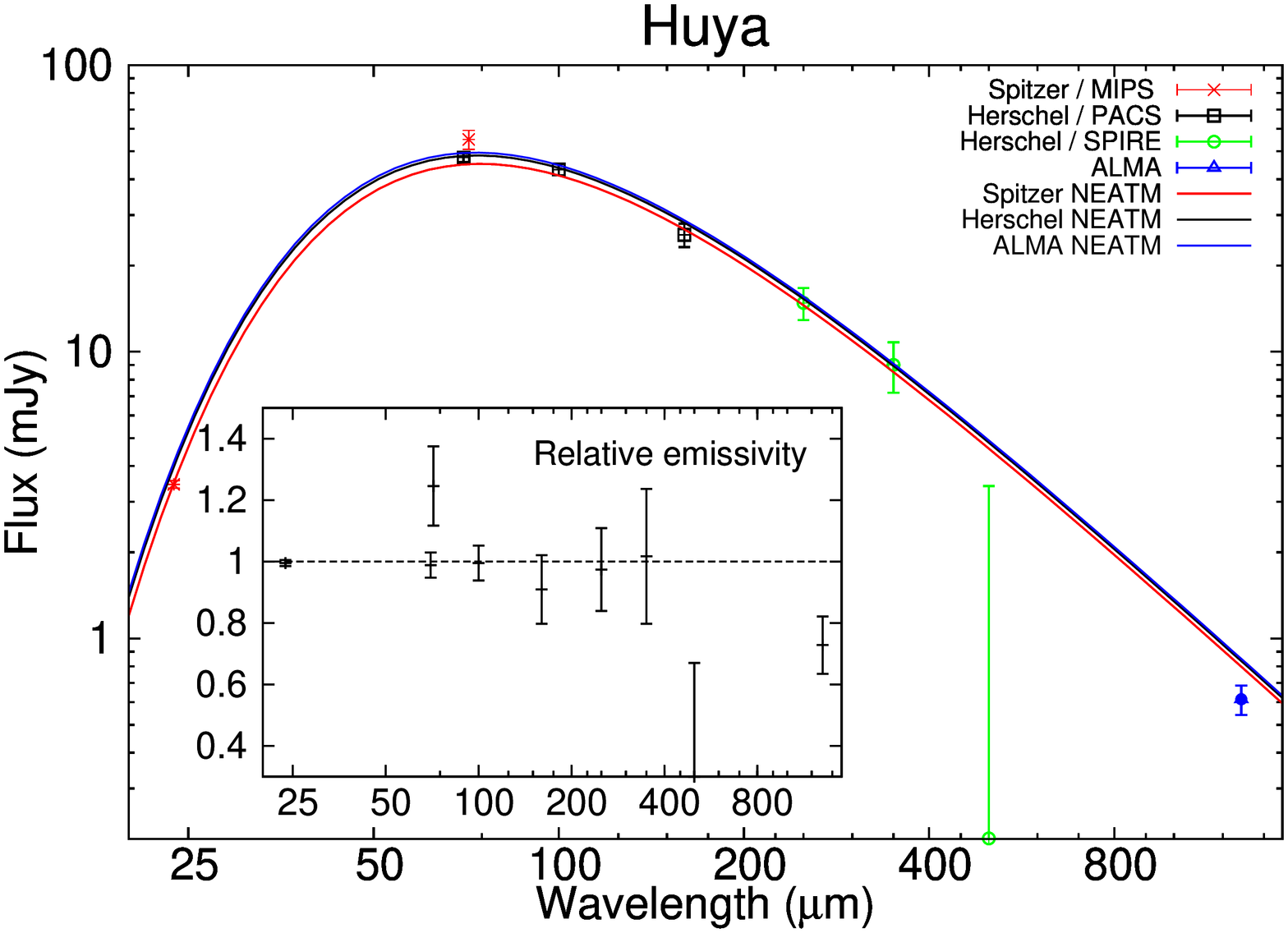}
\includegraphics[width=4.2cm,angle=-90]{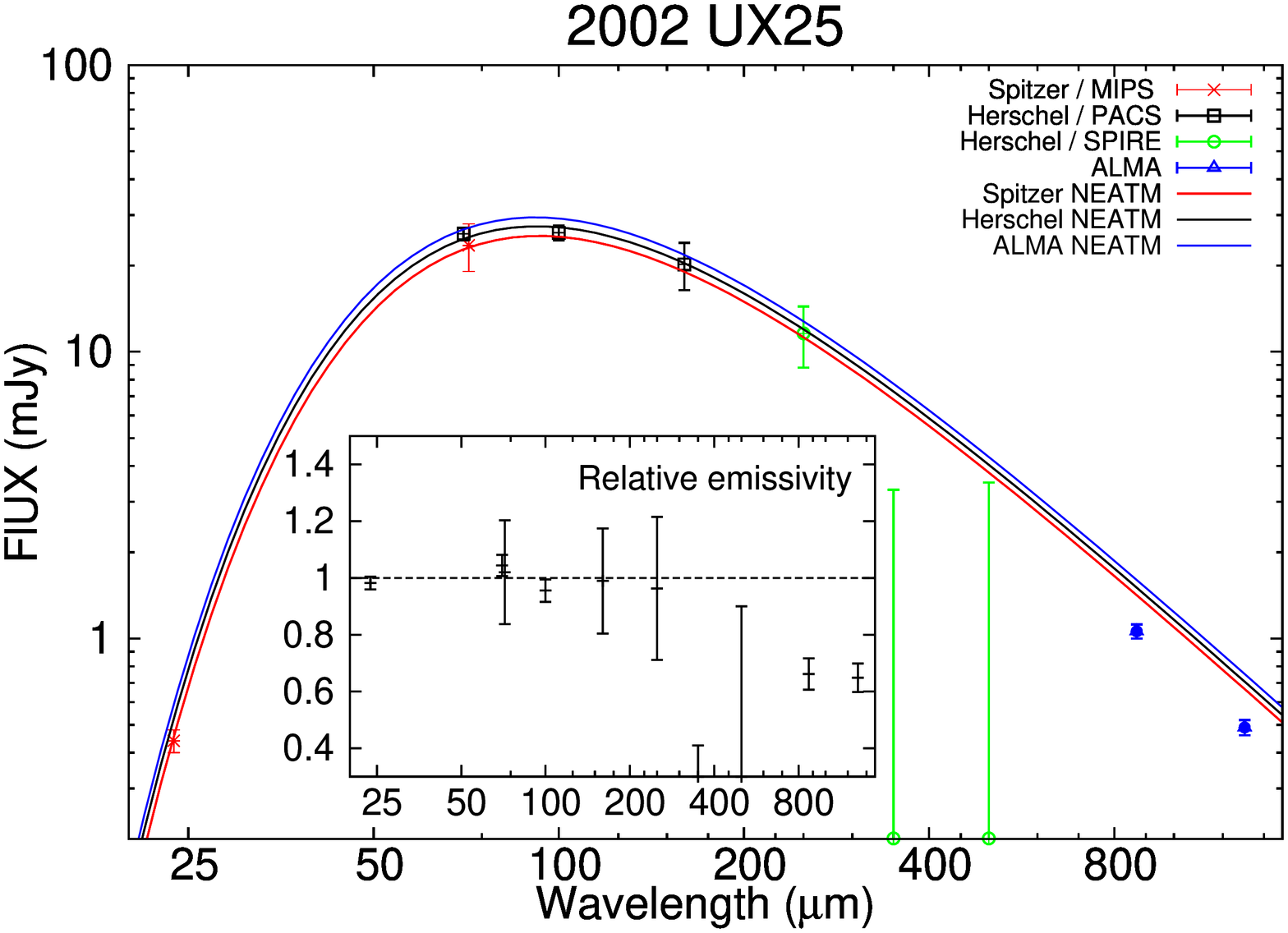}
\includegraphics[width=4.2cm,angle=-90]{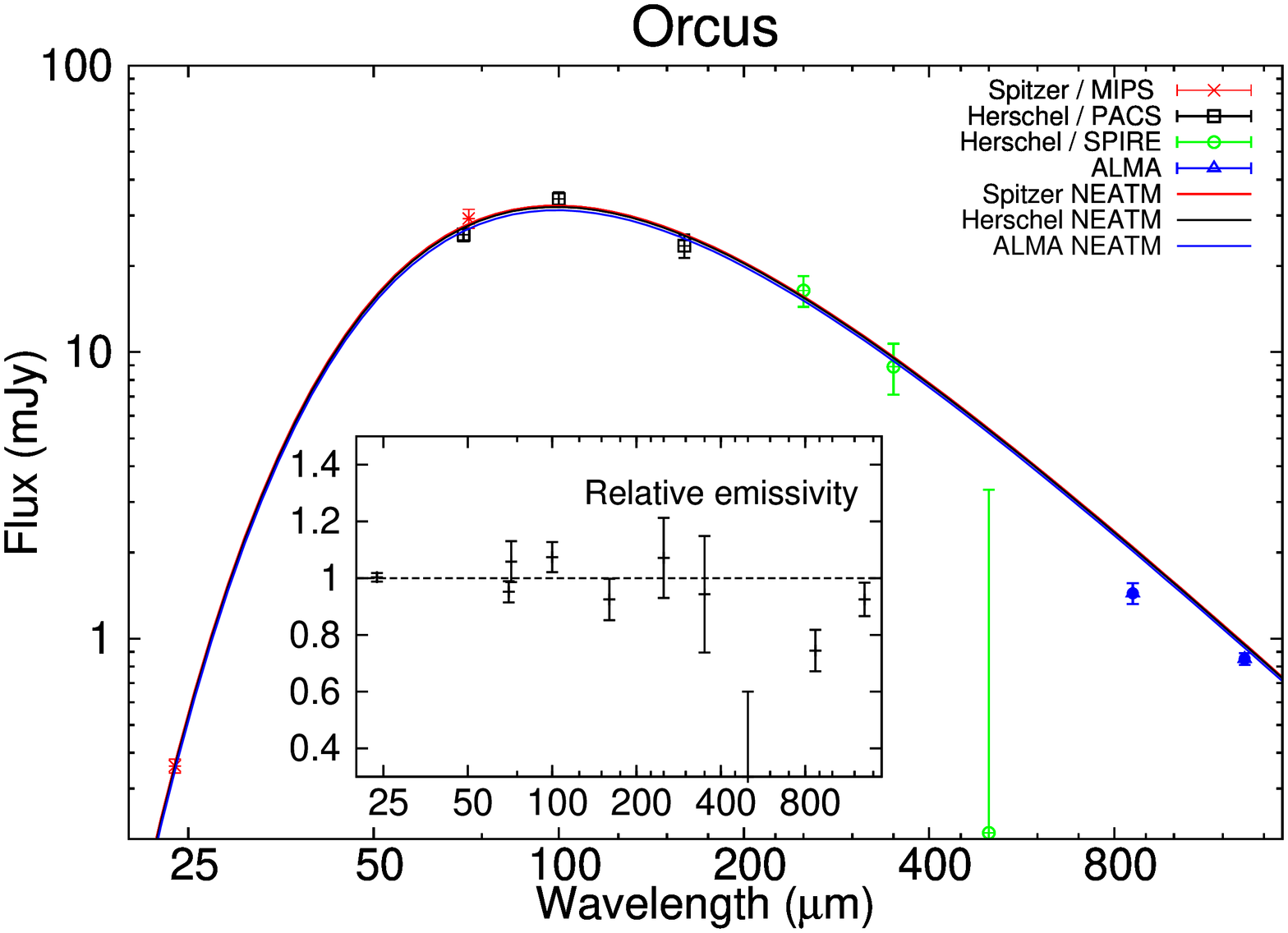}
\hspace*{1.9mm}\includegraphics[width=4.2cm,angle=-90]{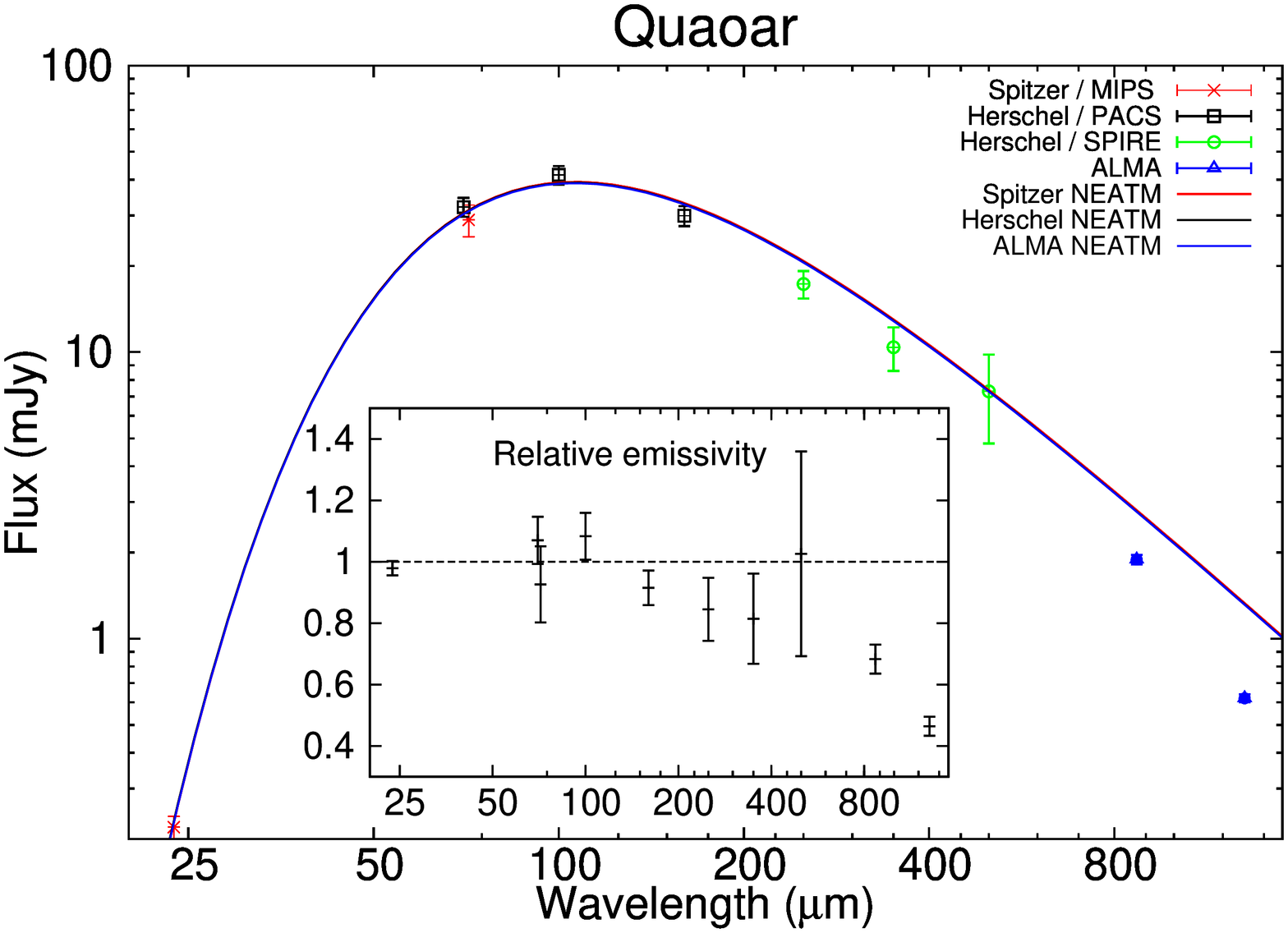}
\hspace*{1.1mm}\includegraphics[width=4.2cm,angle=-90]{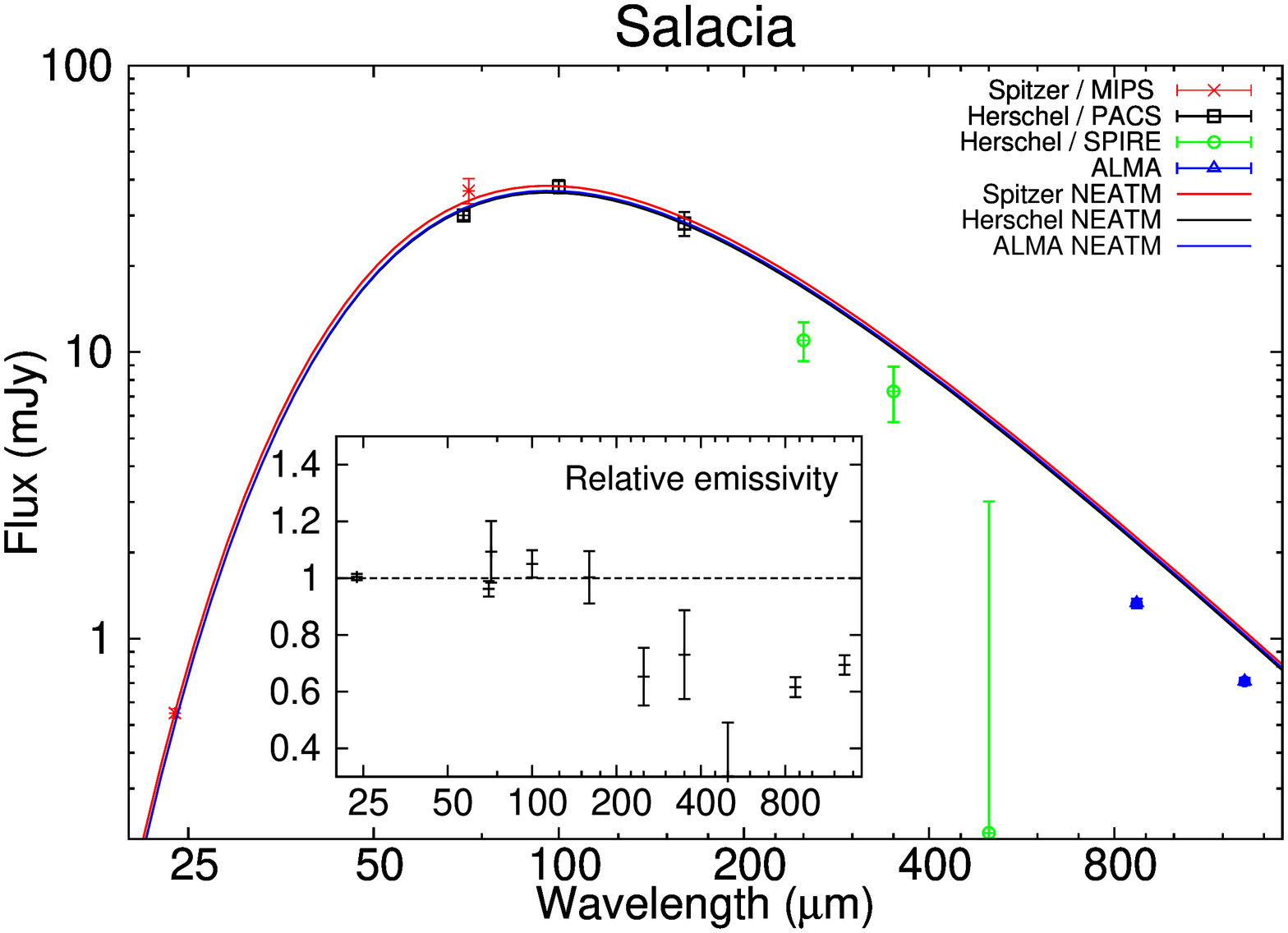}
\caption{Flux measurements, spectral energy distribution and relative emissivity spectra (in inset) for nine TNOs, including the four objects from \citet{brown17}. Different colors show NEATM fits applied to the conditions of the different
observations ({\em Spitzer}, {\em Herschel}, ALMA and WISE). 
}
\label{fig:overview}
\end{figure*}

{\em Herschel} data include PACS 3-band (70, 100, 160 $\mu$m) photometry for $\sim$120 objects and SPIRE
(250, 350, 500 $\mu$m) 3-band photometry for a much more restricted sample ($\sim$10 objects). Out of the six objects in this study and the four objects observed by \citet{brown17}, eight (i.e. all except 2002~GZ$_{32}$ and Bienor) are part of the SPIRE sample. Nonetheless, based on \cite{fornasier13} 
who found that significant emissivity effects occur longwards of 200 $\mu$m, we did not include the SPIRE
data when deriving the objects' diameter, geometric albedo, and beaming factor. For this purpose, we used only the {\em Spitzer}, {\em Herschel}/PACS (and WISE) data, which were fit simultaneously -- using the observing circumstances appropriate
for each individual measurement -- to derive the above three parameters.
The flux measurements from ALMA and SPIRE were then fit to derive the relative spectral emissivity (i.e.
$\epsilon_\lambda$~/~$\epsilon_b$) at the corresponding wavelengths (Fig.~\ref{fig:overview}).  
H$_v$ magnitudes were taken from literature, normally
sticking to the values used in the {\em Herschel} ``TNOs are Cool" papers. For objects whose H$_v$ magnitude
is known to vary on orbital timescales (Chariklo, Chiron, and Bienor), we at this point simply used the H$_v$ magnitude in the relevant time 
period (i.e. 2006-2010), noting that the inferred $p_v$ is of limited significance as being possibly affected by
ring and/or dust contamination; specifically we used H$_v$~=~7.30$\pm$0.2 for Chariklo, 5.92$\pm$0.2 for Chiron, and 
7.57$\pm$0.34 for Bienor.

An important issue stressed by \citet{brown17} is the need to evaluate realistic error bars on the fitted parameters.
With respect to previous modelling of TNOs, their model included (a) a Monte Carlo Markov Chain (MCMC) approach (b) the inclusion of uncertainties
on the input parameters $q$ (adopting a factor of 2 variability above and below the \cite{brucker09} relationship)
and $\epsilon_b$ (considering the 0.80-1.00 range instead of a fixed value). For their four objects, they found somewhat different  -- and larger in average by 25 \% -- uncertainties compared to the results of \citet{fornasier13}.  Here, we kept the technical approach to error bars of \citet{mueller11} adopted in the series of ``TNOs are Cool" papers, generating multiple datasets 
of synthetic fluxes to be fitted and of the input model parameters, based on the observational and model uncertainties. However,
in addition to the uncertainty on H$_v$, we included uncertainties
in $q$ and $\epsilon_b$, following \citet{brown17}. Consistent with these authors, we found that the latter, here implemented as a gaussian distribution of $\epsilon_b$ with mean and rms values of 0.90 and 0.06 respectively, has a significant impact on the range of solution parameters.
  
Spherical NEATM fits, including the retrieved emissivity curves, are gathered in Fig.~\ref{fig:overview}, with solution parameters  given in Table \ref{fitresults} for all our objects except Makemake -- which is discussed separately below. We also include our solution fits of the data presented in \cite{brown17}, mainly to verify the consistency of our approach to theirs. In particular, allowing for the uncertainty in the 
bolometric emissivity, we redetermine the equivalent diameters of 2002~UX$_{25}$, Orcus, Quaoar and Salacia as
695$_{-29}^{+30}$  km, 960$_{-42}^{+45}$  km, 1071$_{-57}^{+53}$  km and 909$_{-38}^{+39}$ km, respectively, to be compared
with 698$\pm$40  km, 965$\pm$40 km, 1083$\pm$50 km and 914$\pm$39 km in \cite{brown17}. Except for 2002~UX$_{25}$, for which
our error bar is 25 \% smaller than theirs\footnote{For this object \cite{brown17} mis-quoted a 40 km
uncertainty from \cite{fornasier13}, while these authors gave 24 km.}, all diameter central values and uncertainties are 
fully consistent. We are unsure of the reason for the slightly discrepant result on the diameter uncertainty
for 2002~UX$_{25}$, but feel that the difference should not be overstated, as the resulting error on the
density would be only 13.5 \% if our error on the diameter is adopted, vs 18.2 \% for the value of \cite{brown17}. In any case, our results on the mm/submm emissivity of the four objects, shown in the insets Fig.~\ref{fig:overview}, also show excellent consistency with \cite{brown17} (see their Fig. 6).

Prior to our ALMA measurements, two observations of Chiron \citep[][using the JCMT and IRAM-30m, respectively]{jewitt92,altenhoff95} and one of Chariklo \citep[][using IRAM-30m]{altenhoff01}
had been acquired at mm/submm wavelengths, respectively in 1991, 1994 and 1999-2000. Observing parameters (geocentric and heliocentric distances) and fluxes can be found in these papers as well as in \citet{groussin04}. We apply our NEATM fit solution parameters to these observations to obtain additional determinations of their mm/submm emissivity.  Results are included in Table \ref{fitresults}. In spite of vastly different observational distances (e.g. Chiron was at $\sim$ 10.5-8 AU in 1991-1994 vs $\sim$18.5 AU in 2016, and Chariklo was at $\sim$13 AU in 1999-2000 vs 15.5 AU in 2016), these earlier measurements are nicely consistent with the
emissivities estimated from ALMA.

Of the nine objects presented in Fig.~\ref{fig:overview} and Table \ref{fitresults}, eight clearly show relative emissivities lower than unity. With an emissivity of 1.25$_{-0.10}^{+0.09}$ derived in this model from the ALMA data, Chariklo seems to be an outlier. As we show
in the next section, however, this apparently anomalous emissivity can result from several factors that we will now study.

\begin{table*}
\caption{Additional thermal modelling fit results: Chariklo} 
\label{othermodelfitresultsChariklo}      
\centering                          
\begin{center}
\begin{tabular}{llcccccc}
Object  & Model  & Diameter or  & Geometric           &  Beaming & Thermal & Ref. for & Relative submm\\
 & & a,b,c (km) & albedo & factor & inertia (MKS) & submm/mm data & emissivity\\
\hline 
\\



Chariklo &  TPM$^a$, no rough. & 240$_{-8}^{+11}$  & 0.037$_{-0.008}^{+0.007}$ & N/A  & 2.6$_{-1.4}^{+1.7}$ & This work &  1.24$_{-0.09}^{+0.10}$  (1.29 mm) \\
\\
Chariklo &  TPM$^a$, small rough. & 240$_{-8}^{+11}$  & 0.037$_{-0.008}^{+0.007}$ & N/A  & 4.3$_{-1.4}^{+1.6}$ & This work &  1.23$_{-0.10}^{+0.10}$  (1.29 mm)\\
\\
Chariklo &  TPM$^a$, interm. rough. & 240$_{-8}^{+11}$  & 0.037$_{-0.008}^{+0.007}$ & N/A  & 5.9$_{-1.8}^{+1.5}$ & This work &  1.22$_{-0.10}^{+0.10}$ (1.29 mm)\\
\\
Chariklo &  TPM$^a$, large rough. &  240$_{-8}^{+11}$  & 0.037$_{-0.008}^{+0.007}$ & N/A  & 8.0$_{-1.7}^{+2.0}$ & This work &  1.21$_{-0.10}^{+0.10}$  (1.29 mm)\\
\\

Chariklo &  NEATM, Maclaurin & 143 x 143 x 96  & 0.038 & 1.09$_{-0.07}^{+0.07}$ & N/A & This work &  1.00$_{-0.07}^{+0.07}$ (1.29 mm) \\
\\

 &  &  &  &  &  & A01$^b$ &  0.81$_{-0.11}^{+0.14}$ (1.20 mm)\\
\\

Chariklo  &  NEATM, Jacobi & 160 x 142 x 88  & 0.041 & 1.04$_{-0.07}^{+0.08}$ & N/A & This work &  0.92$_{-0.06}^{+0.07}$ 
(1.29 mm) \\
\\

 &  &  &  &  &  & A01$^b$ &  0.76$_{-0.12}^{+0.11}$ (1.20 mm) \\
\\

Chariklo  &  NEATM, Triaxial & 146 x 131 x 101  & 0.038 & 1.11$_{-0.09}^{+0.08}$ & N/A & This work &  1.06$_{-0.08}^{+0.07}$ (1.29 mm) \\
\\
 &  &  &  &  &  & A01$^b$ &  0.87$_{-0.14}^{+0.14}$ (1.20 mm)\\
\\

Chariklo,  ring-corr. &  NEATM & 230$_{-9}^{+8}$  & 0.040$_{-0.008}^{+0.008}$ & 1.15$_{-0.10}^{+0.08}$ & N/A & This work &  1.21$_{-0.10}^{+0.10}$ (1.29 mm) \\
\\
 &  &  &  &  &  & A01$^b$ &  1.01$_{-0.17}^{+0.18}$  (1.20 mm)\\
\\
Chariklo,  ring-corr. &  NEATM, Jacobi & 154 x 136 x 84  & 0.041 & 1.00$_{-0.08}^{+0.07}$ & N/A & This work &  0.89$_{-0.08}^{+0.07}$ (1.29 mm) \\
\\
 &  &  &  &  &  & A01$^b$ & 0.71$_{-0.11}^{+0.13}$  (1.20 mm)\\
\\






\hline
\multicolumn{8}{l}{$^a$ TPM: thermophysical model}\\
\multicolumn{8}{l}{$^b$ A01:\cite{altenhoff01}}\\ 

\end{tabular}
\end{center}
\end{table*}

\subsection{Chariklo: more detailed modelling}
On June 3, 2013, observation of a stellar occultation by Chariklo resulted in the discovery of rings around
the body \citep{braga14} and in bringing entirely new information on its shape. In addition to their huge significance in itself, this observation adds valuable constraints here, making possible a more detailed modelling of the object's thermal flux. 
Indeed (i) the ring orientation provides information on Chariklo's spin direction (ii) information
on Chariklo's shape permits us to consider non-spherical models and (iii) based on the rings physical characteristics
deduced from the occultation, we can estimate their thermal emission.

\subsubsection{Thermophysical model}
\label{sec:TPM}
\cite{braga14} report the pole orientation of Chariklo's ring system to be given by J2000
$\alpha_P$ = 151.30$^{\circ}$$\pm$0.49$^{\circ}$, $\delta_P$ = 41.48$^{\circ}$$\pm$0.21$^{\circ}$.
A reasonable assumption is that the body's pole orientation coincides with the ring pole orientation. Under this
assumption, Chariklo's sub-solar and sub-earth latitudes are approximately 0$^{\circ}$ in early 1980 and late 2007, and
reach extremal values of --58$^{\circ}$ in mid-1996 and +58$^{\circ}$ in late 2021. The sub-solar latitude was
in the range --12$^{\circ}$ to +14$^{\circ}$ for the WISE, {\em Spitzer} and {\em Herschel} observations in 2006-2010, but
reached +45$^{\circ}$ in 2016 for the ALMA observations. Using this knowledge, along with
Chariklo's rotation period \citep[7.004$\pm$0.036h;][]{fornasier14}, we fit the WISE, {\em Spitzer} and {\em Herschel} data
with a spherical thermophysical model (TPM), accounting for the proper sub-solar/sub-earth latitude for each data set.
The model free parameters are the equivalent diameter, geometric albedo, and now the thermal inertia ($\Gamma$). As there is a known degeneracy between thermal inertia and surface roughness, we specify four levels of roughness (``none", ``small", ``intermediate", "large"), and incorporate the effect of roughness as a `̀`thermophysical model beaming factor"  $\eta_{TPM}(\Theta)$, which
for a given roughness scenario is a single-valued function of the thermal parameter\footnote {The thermal parameter, $\Theta$, related to thermal inertia $\Gamma$, is defined by \citet{spencer89} and represents the ratio of the radiation timescale of subsurface heat to the diurnal timescale.}
\citep[see discussion and Fig. 6 in][]{lellouch11}. Once the solution thermal inertia is found, the model
is applied to the ALMA data, again with its proper sub-solar/sub-earth latitude. Within this framework,
solutions for the diameter and geometric albedo (see Table~\ref{othermodelfitresultsChariklo}) are virtually identical to those obtained from NEATM, and thermal
inertias of 2.5--8 MKS are found. The apparent mm relative emissivity from the ALMA data is now nominally 1.23, instead of 1.25 from NEATM. A lower value is to be expected in relation to the more poleward sub-solar latitude in 2016 (causing the object to be in average warmer) than in 2006-2010, but the effect is very small, given the weak ($\sim$ linear) dependence of the mm flux to
temperatures. 
Applying similarly the thermophysical model solutions to the IRAM observations of \cite{altenhoff01} leads to emissivity values that are insignificantly different from that obtained with NEATM (1.07$\pm$0.16 instead of 1.08$\pm$0.16, not detailed in Table~\ref{othermodelfitresultsChariklo}).

\subsubsection{Shape effects}
Four additional multi-chord occultations by Chariklo were observed in April 29, 2014, June 28 2014, August 8, 2016 and October 1, 2016. Combining these data with those from June 2013, and using the pole orientation given by the rings, \citet{leiva17} estimated Chariklo's size, shape and density. For this, they considered a variety of models, including two hydrostatic equilibrium models
(Maclaurin spheroid and Jacobi ellipsoid), and a more generic triaxial ellipsoid shape in which the hydrostatic constraint was relaxed. Best fit solutions are: (i) Maclaurin spheroid: equatorial radius, a~= b~=~143$^{+3}_{-6}$ km; polar radius, c~=~96$^{+14}_{-4}$ km; density = 970$^{+300}_{-180}$ kg m$^{-3}$; (ii) Jacobi ellipsoid: semi-major axes, 
a = 157$\pm$4 km, b~= 139$\pm$4 km, c = 86$\pm$1 km; density = 796$^{+2}_{-4}$ kg m$^{-3}$ (these numbers override those given in \citet{leiva16} which did not include the occultations from 2016) (iii) triaxial ellipsoid: a = 148$^{+6}_{-4}$ km, b~= 132$^{+6}_{-5}$ km, c = 102$^{+10}_{-8}$ km. These non-spherical shapes lead to the object presenting a higher cross-section in 2016 vs 2006-2010, presumably contributing to the high apparent emissivity derived from ALMA. Below, we study this effect quantitatively.

\citet{braga14,duffard14} and \citet{fornasier14} noted that the discovery of Chariklo's rings provides a natural explanation of 
the brightness variability of the Chariklo+rings system, i.e. its fading from 1998 to 2009 and re-increase in brightness after
that date. Fitting the H$_v$ magnitude data with the sum of a spherical \citep[or quasi-spherical in the
case of][]{duffard14}\footnote{\citet{duffard14} indicate they used a = 122 km, b = 122 km, and c = 117 km but this
must be in error as they also state that such a figure ``can reproduce the
observed rotational amplitude in 2013 and the non-detection in
1997-2010, if all the variability is due to shape".} body with a ring of known dimensions but with changing aspect, they determined a ring
albedo (I/F reflectivity) of $\sim$9 \%, noting that the variations of the projected area of the object with season would also
contribute to the secular variation of Chariklo's H$_v$ magnitude, thereby reducing the required ring albedo.

\begin{figure}[ht]
\centering
\includegraphics[width=8cm,angle=0]{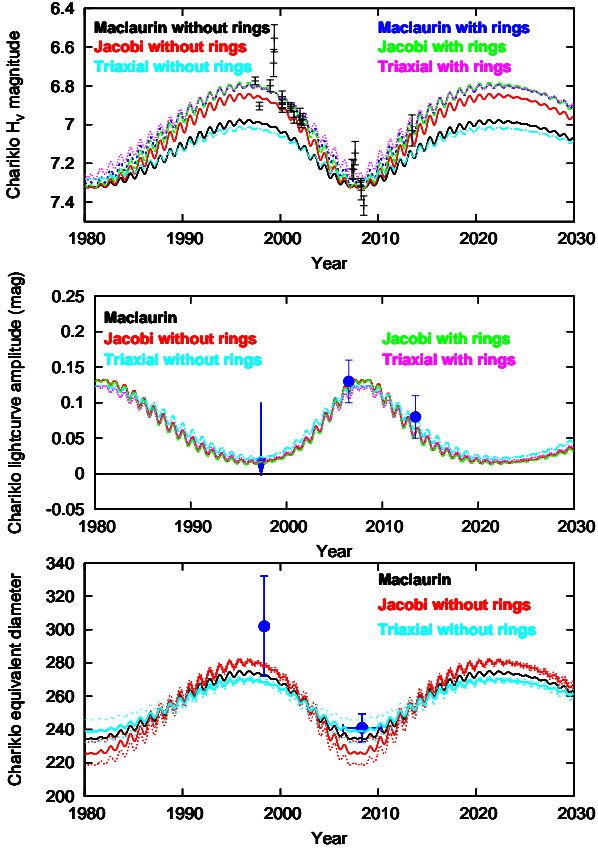}
\caption{Models of: (top) H$_v$ magnitude; (middle) optical lightcurve amplitude; and (iii) apparent diameter
of Chariklo for the three shape models (Mclaurin spheroid, Jacobi ellipsoid, and generic triaxial ellipsoid) of \citet{leiva17}. In the top panel, photometric data 
are from \citet{fornasier14}. For the two ellipsoid cases, the model H$_v$ is the mean value over the lightcurve. In the middle panel, lightcurves amplitudes $\Delta m$ = 0.13$\pm$0.03 and 0.08$\pm$0.03
mag in 2006 and 2013 \citep{galiazzo16,fornasier14}  and an upper limit ($\Delta m$ $\lessapprox$0.1 mag) in 1997
\citep{davies98} are shown; the spheroid model does not produce any lightcurve. In the bottom panel, the diameter reported by \citet{jewitt98} for 1998 and in this work 
for 2006-2010 ({\em Spitzer}, WISE, {\em Herschel}) are shown \citep[an additional measurement from WISE in 2010--][--giving D = 226.1$\pm$29.3
km, is not reported for plotting clarity]{bauer13}.  For the two ellipsoid cases, the apparent diameter is calculated at lightcurve
maximum and minimum (dotted lines) and at the intermediate phase (solid lines). No contribution of Chariklo's rings to the thermal emission is assumed here.
}
\label{fig:chariklo_optical_new.ps}
\end{figure}

We briefly reconsidered these models by including the above three 
 shape models from \citet{leiva17}. As in the above papers,
the adjustable parameters are Chariklo's geometric albedo $p_v$ and the ring I/F reflectivity. As in \citet{braga14} and \citet{leiva17}, but unlike in \citet{duffard14} and \citet{fornasier14}, we only included the contribution from Chariklo's inner
ring (C1R, radius 391 km, optical depth $\tau$=0.4), as the outer ring (C2R, radius 405 km, optical depth $\tau$=0.06) is $\sim$7 times optically thinner and likely has a negligible contribution to the system H$_v$ magnitude. 
For the three shape models,  Fig.~\ref{fig:chariklo_optical_new.ps} compares the calculated (i)  H$_v$ magnitude; (ii) lightcurve amplitude; and (iii) surface-equivalent diameter
of the solid body, as a function of time, with relevant data. As Chariklo's rotational period is not known
with sufficient accuracy to rephase optical or thermal measurements, we calculate a mean H$_v$ magnitude over the lightcurve in the case of the two ellipsoid models, while the equivalent diameter is calculated at maximum, minimum and intermediate phases. For the Maclaurin (resp. Jacobi, generic triaxial) model,  the solution parameters are $p_v$= 0.038 (resp. 0.041, 0.038) and (I/F)$_{ring}$ = 0.030 (resp. 0.010, resp. 0.035). Thus, as mentioned by \citet{leiva17} who derived very similar numbers, the ring contribution is vastly subdued when the object's non-sphericity is taken into account. The Jacobi and generic triaxial shapes 
additionally predict a large variability of the optical lightcurve amplitude; this agrees with
the detection of the optical lightcurve in 2006 and 2013  \citep[]
{galiazzo16,fornasier14}\footnote{Although \citet{fornasier14} mention a $\sim$0.11 mag lightcurve, inspection of their
Fig.~1 indicates that the amplitude is rather $\sim$0.08$\pm$0.03 mag. For \citet{galiazzo16}, we estimate $\sim$0.08$\pm$0.03
from their Fig. 2. These values are reported in Fig.~\ref{fig:chariklo_optical_new.ps}.  } but not in 1997 \citep{davies98}.
 Finally, for all three models, the variation of the projected area with time
is qualitatively consistent with the report of a significantly larger apparent diameter (302$\pm$30 km) in 1998 \citep{jewitt98} 
than in 2006-2010 (241$\pm9$ km; see Table \ref{fitresults}); this is explored more quantitatively in the next paragraph.

With these results in mind, we re-fitted the Chariklo thermal data with non-spherical models. Since the thermophysical models
lead to very similar emissivity results as the NEATM, we mostly explored the effect of non-sphericity in NEATM, the difference
between an ``elliptic NEATM" and the standard NEATM being in the expression of the solar zenith angle at the surface
of the ellipsoid, as well of course as the variability of the projected area in the former. The effect of ellipsoidal geometry in standard asteroid radiometric models has been first studied by \citet{brown85}. Specifying as input the three above shape models
from \citet{leiva17}, the WISE, {\em Spitzer} and {\em Herschel} data were refit in terms of (i) the beaming factor $\eta$ (ii) a scaling
factor ($f_{scale}$) to the overall object dimensions. Chariklo's geometric albedo was held fixed at the value determined
previously for each shape model. The model was run for an intermediate phase
between lightcurve maximum and minimum. Applying the model to the mm fluxes (ALMA, IRAM) then provided the radio emissivity. Uncertainty in the observed fluxes and other input parameters (H$_v$, $\epsilon_b$, $q$) were handled as before. Results are summarized in Table \ref{othermodelfitresultsChariklo}. For all three shape models, the scaling factor is very close to unity ($f_{scale}$ = 1.00$\pm$0.04 for the Maclaurin model, 1.02$\pm$0.04 for the Jacobi ellipsoid model, and 0.99$\pm$0.04 for the generic
triaxial model), confirming the consistency of the occultation-determined sizes with thermal radiometry. The beaming factors are slightly smaller (nominally 1.04--1.11) than in the spherical NEATM, in accordance with the study of \citet{brown85}. The most important result for our purpose is that for these non-spherical models, the inferred relative radio emissivities have decreased in most cases to values less than 1 ($\sim$0.92-1.05 from ALMA and $\sim$0.76-0.87 from IRAM), demonstrating that for both the ALMA (March 2016) and IRAM (1999-2000) data, the ``anomalously large" flux likely results from the large apparent cross section of Chariklo at those times (sub-solar/sub-earth latitude $\beta$$\sim$46$^{\circ}$ in 2016 and -50$^{\circ}$ in 1999-2000 vs $\beta$ = -12$^{\circ}$ to 14$^{\circ}$ over 2006-2010).
Applying also the models to the conditions of the \citet{jewitt98} 20.3 $\mu$m measurements, we calculate  fluxes of 
58$\pm$2 mJy, 70$\pm$2 mJy, and 52$\pm$2 mJy for the  Maclaurin, Jacobi, and generic triaxial models, respectively. 
All of these values agree reasonably well  with the measured flux (66$\pm$12 mJy), albeit notably less so
for the triaxial shape than for the Jacobi.
In contrast, with the spherical NEATM solution (Table \ref{fitresults}),      
the calculated 20.3 $\mu$m flux is only 34$\pm$1.5 mJy, at odds with the observations.
All this clearly favors non-spherical over spherical models. Although 
\citet{leiva17} favored the generic triaxial shape as permitting to avoid an extremely low ($\lesssim$ 1 \%) ring I/F,
we prefer here the Jacobi ellipsoid as 
providing the best agreement to the \citet{jewitt98} flux; in what follows, this shape model is adopted.

\subsubsection{Ring emission}

We constructed a simple model to estimate the thermal emission from Chariklo's rings.
The model is based on a simplified version of Saturn's rings models developed by \citet{froidevaux81,ferrari05,flandes10}.
Direct absorbed radiation from the Sun $F_{abs}$ is the only considered source of energy for ring particles, i.e. we neglect heat sources
due to thermal emission and reflected solar light from Chariklo itself, as well as mutual heating between ring particles. 
Under this assumption, the ring particle temperature $T_P$ will satisfy the following energy balance equation: 
\begin{equation}
F_{abs} = \epsilon_{r,b}~\sigma~f~T_P^4~(1 - \frac{\Omega_P}{4\pi})
\end{equation}
where
\begin{equation}
F_{abs} = (1 - A_r)~C(B',\tau)~\frac{S_{\odot}}{r_h^2}
\end{equation}
Here 
$A_r$ is the Bond albedo of the ring particles
and $\epsilon_{r,b}$ is their bolometric emissivity. $C$ is the ring shadowing function (i.e. the non-shadowed fractional area of a ring particle), which depends on the ring optical depth $\tau$ and the solar elevation above ring plane, $B'$.
${\Omega_P}$ represents the angle subtended by neighboring particles, and can be expressed as ${\Omega_P}$ = 6 (1 - e$^{-\tau}$) \citep{ferrari05}. Finally $f$ is the rotation rate factor of the particles; following \citet{flandes10}, we considered slow
rotators, i.e. $f$ = 2. For the ring shadowing function, we adopted the approximate expression from \cite{altobelli08}:
\begin{equation}
C(B',\tau) = \frac{sin~B'}{1 - e^{-\tau}}~(1 - e^{-\tau/B'}) 
\end{equation}

Once the ring particle temperature $T_P$ is calculated in this manner, the ring radiance B$_\lambda$ at wavelength $\lambda$ as seen from Earth and the associated brightness temperature $T_B(\lambda)$ are obtained as:
\begin{equation}
\textrm{B}_\lambda(T_B(\lambda)) ~ = ~\epsilon_{r,\lambda} ~ (1 - e^{-\tau}) ~C({B},\tau) ~\frac{\textrm B_\lambda(T_P)}{sin~{B}}
\label{eqn:tbrings}
\end{equation}
where B$_\lambda$ is the Planck function, $\epsilon_{r,\lambda}$ is the ring particle emissivity at wavelength $\lambda$, $B$ is the elevation of the observer above ring plane and $C(B,\tau)$ is the fractional emitting area of
a ring particle as seen from elevation $B$ \citep{ferrari05}. Equation~\ref{eqn:tbrings} reduces to 
B$_\lambda(T_B(\lambda))$ ~ = $\epsilon_{r,\lambda}$ ~ (1 - e$^{-\tau/B}$) B$_\lambda(T_P)$. 

In the above model, we nominally assumed $\epsilon_{r,b}$ = $\epsilon_{r,\lambda}$ = 1. This likely provides an upper limit of
the ring emission at the longest wavelengths. Specifically, Cassini/CIRS measurements of Saturn's rings
indicate that $\epsilon_{r,\lambda}$ declines at $\lambda$ longwards of $\sim$200 $\mu$m, in relation
to the progressive increase in single scattering albedo of water ice particles \citep{spilker05}. Similarly, 1.3-3 mm observations
of Saturn's B-ring at high solar elevation indicate brightness temperatures of 15-30 K \citep{dunn05}, vs $\sim$90 K 
as measured by Cassini/CIRS shortwards of $\sim$200 $\mu$m \citep{spilker05}. \citet{duffard14} infer the presence of
water ice in Chariklo's rings. This might be taken as an argument that a similar roll-off 
in brightness temperature with increasing wavelengths occurs for Chariklo's rings as well. 
We note however, that the albedo inferred above for Chariklo rings (I/F = 0.010-0.035)
is much weaker than Saturn's rings (I/F $\sim$ 0.3 -- 0.7) and actually more similar to Uranus' rings (I/F 
$\sim$ 0.05), which do not show evidence for water ice \citep{dekleer13}. Thus, water ice might be present in Chariklo's rings
only as a trace component. We note in passing that \citet{duffard14} base their argument for water ice on the detection
of the 1.5 and 2.0 $\mu$m H$_2$O features in four spectra taken in 1997-2002 and again in 2013, but not in 2007-2008 when rings were edge-on. The identification of H$_2$O ice in 1997-2002 seems secure \citep[see][]{guilbert11}, but we
feel that the single 2013 spectrum -- which does not show directly the H$_2$O bands -- is less convincing \citep[1.3 $\sigma$
detection according to Table~2 of][]{duffard14} and would deserve confirmation. If not present in Chariklo's
rings, an alternative explanation to the data could be that H$_2$O is located in restricted areas at high southern 
latitudes of Chariklo.

\begin{figure}[ht]
\centering
\includegraphics[width=8cm,angle=0]{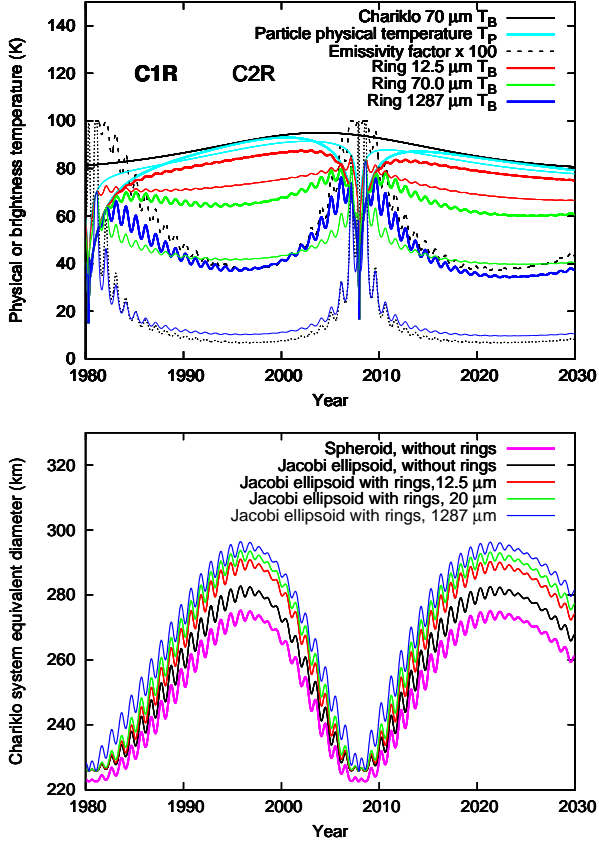}
\caption{Thermal contribution of Chariklo rings. {\em Top: Ring temperature model}: Particle temperature (light blue) and ring brightness temperature at three wavelengths (red: 12.5 $\mu$m,
green: 70 $\mu$m, dark blue: 1287 $\mu$m) as a function of time over 1980-2030 for Chariklo's C1R $\tau$=0.4 (thick lines)
and C2R $\tau$=0.06 (thin lines) rings. They are compared to the 70 $\mu$m brightness temperature of Chariklo itself (black).
The dotted line represents the emissivity factor (1 - e$^{-\tau/B}$), multiplied by 100. {\em Bottom}: Effect of rings
on the apparent diameter. See text for details.}

\label{fig:ringstb}
\end{figure}

Model results in terms of ring particle physical temperature and ring brightness temperature at three
wavelengths (12.5, 70 and 1287 $\mu$m) are
shown in Fig.~\ref{fig:ringstb} for Chariklo's C1R (thick lines) and C2R (thin lines) rings, and compared to the 
mean brightness temperature $T_{B,Ch}$ of Chariklo itself. The latter, which was calculated for the spherical 
NEATM solution ($\eta$=1.20, p$_v$ = 0.036), is slightly wavelength-dependent and is here shown
at 70 $\mu$m. Ring particle temperatures are colder than T$_{B,Ch}$ due to ring low opacity and mutual
shadowing. As expressed by Eq.~\ref{eqn:tbrings}, ring emission as seen from Earth is further reduced by a geometrical emissivity factor (1 - e$^{-\tau/B}$), which is shown by the dotted lines in Fig.~\ref{fig:ringstb} (after multiplication by 100).
Although this factor is wavelength-independent, it leads to a wavelength-dependent reduction of the ring
$T_B$, as shown by the various curves in the top panel of Fig.~\ref{fig:ringstb}.
 
The ring model was then used to calculate the flux emitted by the ring as a function of epoch and 
wavelength. To get a feeling about its significance, the ring flux was compared to the Chariklo flux
calculated for the spherical NEATM solution, and then expressed in terms of a (wavelength-dependent) effective area,
(i.e. the area that, with the same thermal properties as Chariklo itself, would produce a thermal flux equal to
that of the rings). This effective area was then combined with the time-dependent projected 
area of Chariklo in the Jacobi ellipsoid model, and the result finally expressed in terms of the (epoch and wavelength-dependent) equivalent diameter for the Chariklo+rings system. Results
are shown in the bottom panel of Fig.~\ref{fig:ringstb}. We find that ring emission can bias the diameter measurement
by up to $\sim$15 km (5 \%) for wide open ring geometries. Although the ring $T_B$ decreases with increasing 
wavelength, the bias turns out to be more important at the longest wavelengths due to a Planck-function effect.
This conclusion is however sensitive to our assumption that the long wavelength emission of Chariklo's rings is not
affected by scattering due to water ice.

With the above model, the contribution of Chariklo's rings at the relevant epochs and wavelengths 
was subtracted from the thermal data, which were then refit with the spherical and elliptical NEATM models. 
The flux correction amounted to 1-10~\% depending on each particular observation and generally increasing with wavelength. 
As detailed in Table~\ref{othermodelfitresultsChariklo}, compared to the ``no-ring" case, the ring correction leads to slightly 
smaller values for $\eta$ ($\sim$0.05), the equivalent diameter (by $\sim$5 \%), and the relative emissivities (by
$\sim$ 0.04). For the elliptical case, the best fit of the ring-corrected Chariklo thermal data is obtained for a scaling factor of 0.98$\pm$0.04 to the Jacobi ellipsoid model of \citet{leiva17}, still fully consistent with that model.

Combining the effect of shape with the ring correction, we infer emissivities of 0.89$\pm$0.07 from our ALMA
data and of 0.71$\pm$0.12 from the IRAM data. We did not explicitly run elliptic thermophysical models, but 
based on our analysis in section \ref{sec:TPM},  the temperature effect associated to the more pole-on orientation 
in 2016 (ALMA) and 1999-2000 (IRAM) than in 2006-2010 causes an apparent $\sim$2 \% increase in the emissivity.
Correcting for this, we obtain our best estimate for the relative emissivity ($\epsilon_\lambda$ / $\epsilon_b$) as 0.87$\pm$0.07 from the ALMA data and 0.69$\pm$0.10 from IRAM. In summary, the above study shows that the apparent anomalous emissivity
for Chariklo is the combined result of (i) change in the apparent cross section for an elliptical object (ii)
change in the sub-solar latitude (iii) possible ring contribution, the first effect being by far the dominant one. 

We estimate Chariklo's ring thermal emission to be $\sim$0.13 mJy at 1287 $\mu$m (for 2016). While such a flux is not out of reach
of ALMA in terms of sensitivity, the difficulty for detection is that the emission is azimuthally distributed, leading to very low surface
brightnesses. Given also the proximity of the body at $\sim$30 milli-arcseconds, a direct imaging Chariklo's rings with ALMA 
seems rather difficult.

In what follows, we apply the previous models for two other Centaurs, Chiron and Bienor, for which pole orientation parameters
have been proposed \citep{ortiz15,fernandez17}.

\begin{table*}
\caption{Additional thermal modelling fit results: Chiron and Bienor} 
\label{othermodelfitresultsChironBienor}      
\centering                          
\begin{center}
\begin{tabular}{llcccccc}
Object  & Model  & Diameter or  & Geometric           &  Beaming & Thermal & Ref. for & Relative submm\\
 & & a,b,c (km) & albedo & factor & inertia (MKS) & submm/mm data & emissivity\\
\hline 
\\

 Chiron &  TPM$^a$, small rough. & 214$_{-7}^{+9}$  & 0.167$_{-0.034}^{+0.024}$ & N/A  & 0.7$_{-0.7}^{+2.6}$ & This work &  0.62$_{-0.06}^{+0.06}$ (1.29 mm) \\
\\

 Chiron &  TPM$^a$, interm. rough. & 211$_{-9}^{+11}$  & 0.171$_{-0.032}^{+0.032}$ & N/A  & 2.1$_{-2.1}^{+3.6}$ & This work &  0.64$_{-0.06}^{+0.06}$ (1.29 mm) \\
\\

Chiron &  TPM$^a$, large rough. & 210$_{-12}^{+10}$  & 0.180$_{-0.039}^{+0.037}$ & N/A  & 4.9$_{-3.3}^{+4.9}$ & This work &  0.65$_{-0.07}^{+0.07}$ (1.29 mm)\\
\\

Chiron & NEATM, elliptical & 114 x 98 x 62 & 0.10 & 0.93$_{-0.12}^{+0.12}$ & N/A & This work &  0.70$^{+0.09}_{-0.08}$ (1.29 mm) \\

\\

 &  &  &  &  &  & A95$^b$ &  0.56$_{-0.13}^{+0.14}$ (1.20 mm)\\

\\

Chiron, ring-corr. & NEATM  & 186$_{-14}^{+13}$  & 0.100$_{-0.021}^{+0.026}$ & 0.87$_{-0.11}^{+0.15}$ & N/A & This work &  0.55$_{-0.08}^{+0.09}$ (1.29 mm) \\

\\
 &  &  &  &  &  & A95$^b$ &  0.41$_{-0.14}^{+0.18}$ (1.20 mm)\\

\\
Chiron, ring-corr. & NEATM, elliptical  & 100 x 86 x 54  & 0.13 & 0.84$_{-0.10}^{+0.12}$ & N/A & This work &  0.62$_{-0.09}^{+0.10}$ (1.29 mm)\\

\\

 &  &  &  &  &  & A95$^b$ &  0.46$_{-0.18}^{+0.17}$ (1.20 mm)\\

\\

Chiron, ring-corr.$^c$ & NEATM, elliptical  & 100 x 86 x 54  & 0.13 & 0.84$_{-0.10}^{+0.12}$ & N/A & This work &  0.87$_{-0.13}^{+0.12}$ (1.29 mm) \\

\\
Bienor & TPM$^a$, no rough. & 184$_{-6}^{+6}$ & 0.050$_{-0.016}^{+0.019}$ & N/A & 8$_{-2}^{+3}$ & This work &  0.66$_{-0.06}^{+0.07}$ (1.29 mm)\\

\\
Bienor & TPM$^a$, small rough. & 182$_{-7}^{+5}$ & 0.049$_{-0.015}^{+0.016}$ & N/A & 10$_{-3}^{+4}$ & This work &  0.67$_{-0.07}^{+0.06}$ (1.29 mm)\\

\\
Bienor & TPM$^a$, interm. rough. & 180$_{-6}^{+6}$ & 0.050$_{-0.013}^{+0.016}$ & N/A & 12$_{-3}^{+4}$ & This work &  0.68$_{-0.06}^{+0.06}$ (1.29 mm)  \\

\\
Bienor & TPM$^a$, big rough. & 179$_{-7}^{+5}$ & 0.053$_{-0.013}^{+0.018}$ & N/A & 15$_{-4}^{+6}$ & This work &  0.68$_{-0.06}^{+0.07}$ (1.29 mm)\\

\\

Bienor & NEATM, elliptical & 144 x 65 x 51 & 0.041 & 1.24$\pm$0.10 & N/A & This work &  0.64$_{-0.07}^{+0.07}$ (1.29 mm) \\

\\

\hline
\multicolumn{8}{l}{$^a$ TPM: thermophysical model}\\
\multicolumn{8}{l}{$^b$ 
A95: \cite{altenhoff95}} \\ 
\multicolumn{8}{l}{$^c$ In this case, the 1.29 mm flux is not corrected for a ring contribution, representing
a case with extreme roll-off in brightness temperature }\\
\multicolumn{8}{l}{\hspace*{1mm}   towards long wavelengths.} 

\end{tabular}
\end{center}
\end{table*}

\subsection{Chiron}
Following the discovery of Chariklo's rings, \citet{ortiz15} proposed that Chiron harbours a ring system as well.
Their argumentation was based on (i) the evidence for absorption features in occultations observed 
in 1993, 1994 and 2011 \citep{bus96,elliot95,ruprecht15} (ii) the
time variability of Chiron's H$_v$ magnitude, optical lightcurve amplitude, and water ice
spectral signature. Although the occultation features have been attributed before to a combination of narrow jets 
and a broader dust component, and the changing optical properties to photometric contamination by 
cometary-like activity, the general similarity of the widths, depths, and distance to the body of the 
occultation features with those of Chariklo led \citet{ortiz15} to favor the ring explanation, for which they determined
a preferred pole orientation ($\lambda$~=~144$^{\circ}$,  $\beta$=24$^{\circ}$, i.e. J2000 $\alpha_P$ = 156$^{\circ}$, $\delta_P$ = 36$^{\circ}$). 

Although the radio-emissivity derived for Chiron from simple NEATM modelling does not show any ``anomaly" 
(see Table~\ref{fitresults}), it is worth investigating the effects of pole orientation, shape and a possible ring system on the 
thermal photometry. 

As for Chariklo, we assumed that the object's pole orientation coincides with that of the proposed
rings, and first ran a spherical thermophysical model, using a 5.92 hr rotation period. This led to
best fit thermal inertias of 0.7 / 2.4 / 5.6 MKS for small / moderate / large roughness (no good fit was found without surface
roughness), and to a relative radio
emissivity of  (0.62--0.65)$\pm$0.08, again insignificantly different from that obtained from spherical NEATM
(0.63$\pm$0.07).

Shape information for Chiron is more uncertain than for Chariklo, given that none of the three occultations 
observed so far included multiple chords across the object \footnote{The 1993 occultation included a $\sim$158 km chord
and a grazing event, the 1994 one missed the body itself, and the 2011 one has a single 158$\pm$14 km chord.}.
Assuming a Jacobi ellipsoid, a shape model can still be adopted based on other arguments. Following \citet{luu90}, \citet{groussin04} showed convincingly
the anti-correlation of Chiron's apparent lightcurve amplitude $\Delta m$ and apparent visible brightness and derived the 
true (i.e. body-only) lightcurve amplitude to be $\Delta m_0$ = 0.16$\pm$0.03. This is valid whether the
``diluting factor" for the lightcurve amplitude is a coma or a ring system. In this latter case, and assuming identical
pole orientation for the body and the ring system, $\Delta m_0$ occurs when the system is equator-on, implying $a$/$b$ = 1.16.
Assuming the object to be a Jacobi figure in hydrostatic equilibrium, this implies also $c$/$a$ = 0.54. The spherical
NEATM fit indicates an equivalent diameter D = 210 km (Table~\ref{fitresults}). Identifying the latter to 2~$a^{1/4}b^{1/4}c^{1/2}$ would lead 
to $a$~=~148 km, $b$~=~127 km, $c$~=~80 km. However, 
it can be anticipated that this initial guess is too large because, if the pole orientation proposed by \cite{ortiz15} is adopted, 
the geometry of the {\em Spitzer}/{\em Herschel} observations is considerably pole-on (sub-earth latitude 
$\beta$ $\sim$51-60\deg). Indeed, fitting these data with an elliptical NEATM model leads to a scaling factor 
$f_{scale}$ = 0.77$\pm$0.05 from the initial guess (i.e. best fit dimensions of 114 x 98 x 62 km)  and a beaming factor $\eta$ = 0.93$^{+0.10}_{-0.09}$.
Running the model to the geometry of the mm observations then implies a relative emissivity
of 0.56$^{+0.14}_{-0.13}$ for the 1994 IRAM observations
of \citet{altenhoff95} and 0.70$^{+0.09}_{-0.08}$ for our 2016 ALMA observations. These values are only
marginally different from those inferred previously from the spherical NEATM model. This is because, with
the assumed pole orientation, the projected areas at the epochs of the IRAM (sub-solar/sub-earth latitude $\beta$$\sim$-57$^{\circ}$)
and ALMA ($\beta$$\sim$+51$^{\circ}$) data were similar to that at the {\em Spitzer} and {\em Herschel} epochs ($\beta$$\sim$51$^{\circ}$
and $\beta$ $\sim$60$^{\circ}$, respectively), a situation markedly different from the Chariklo case.

\begin{figure}[ht]
\centering
\includegraphics[width=6.5cm,angle=-90]{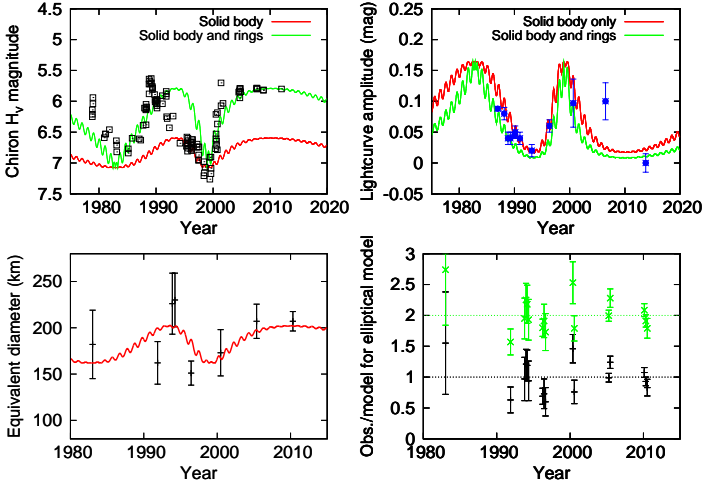}
\caption{Models of: (top left) H$_v$ magnitude; (top right) optical lightcurve amplitude and; (bottom left) apparent diameter
of Chiron for a 114 x 98 x 62 km Jacobi ellipsoid with p$_v$ = 0.10 and the orientation proposed by \citet{ortiz15}. 
The modelled H$_v$ and apparent diameter are calculated at a phase corresponding to mean lightcurve brightness. The first two panels
include the effect of rings with a 20 \% I/F reflectivity. In the top right panel, the blue symbols are the
measurements as collected by \citet{groussin04}, augmented by more recent measurements by \cite{ortiz15} and
\cite{galiazzo16}. In the bottom left panel, the model is compared to
values of the diameter derived from our modelling of past thermal infrared observations \citep[see][]{groussin04}. 
In the bottom right panel, the black points show ratios of individual measured thermal fluxes (some are shifted by $\pm$0.1 year for better visibility) by the elliptical NEATM model. In this panel, the green points (shifted by +1 in y) show the ratio
of ring-corrected thermal fluxes by fluxes calculated from the elliptical model (see text for details). 
}
\label{fig:chiron}
\end{figure}

Fig.~\ref{fig:chiron} shows models of Chiron's H$_v$ magnitude, lightcurve amplitude, and equivalent diameter
based on this Jacobi ellipsoid model. The large variations (up to 1.5 magnitude) of H$_v$ cannot be fit with
a realistic shape model, and require adjunction of a varying coma, a ring system, or more likely a combination
of two \citep{ortiz15}, as the ring model fails to reproduce the evolution of H$_v$ over 1987-1997. 
As done for Chariklo above, we fit the H$_v$ data by the combination of the body and a ring system. 
We find a best-fit I/F of 0.20 for the rings, consistent with 0.17 from \citet{ortiz15}, but stress
that this is rather different from the I/F inferred above for Chariklo (I/F = 0.01-0.035),
putting into question the similarity of Chiron's putative ring system with Chariklo's. We also note
that with the rather elongated shape model we have adopted ($a$/$c$ = 1.85), the adjunction of a ring system has 
a relatively minor effect on the modelled lightcurve amplitude (top right panel of Fig.~\ref{fig:chiron}). \citet{ortiz15} further claimed that the presence of a ring system can explain the variability in the near-IR spectrum of Chiron, with the water ice spectral features being detected by \cite{foster99,luu00} in April 1998 and April 1999, but not by \citet{romon03} in June 2001. However, this argument is weak because with their preferred ring pole orientation, the sub-earth/solar latitude was approximately -9$^{\circ}$ and +3$^{\circ}$ for the first two spectra, vs +26$^{\circ}$ for the last one, i.e. the rings were more edge-on when H$_2$O was detected than when it was not.

Numerous measurements of Chiron's thermal flux and associated diameter have been obtained in the past. In the most detailed study before {\em Spitzer}/{\em Herschel}, \citet{groussin04} analyzed  multiple-band ISOPHOT observations from 1996 with a spherical thermophysical model coupled with a standard beaming factor $\eta$ = 0.756. They inferred a diameter D = 142$\pm$10 km and a thermal inertia $\Gamma$ = 3$^{+5}_{-3}$ MKS. On the basis of this model, they reanalyzed all anterior thermal flux measurements from Chiron,
and noted a significant dispersion between the individually inferred diameters. Doing the same using our spherical NEATM results (i.e. fixing $\eta$ to 0.93$^{+0.13}_{-0.10}$, see Table~\ref{fitresults}) we obtain very similar diameters as 
\citet{groussin04}. Part of the dispersion can be explained by an elliptical body, as can be seen in the bottom
left panel of Fig.~\ref{fig:chiron}. However, as shown in the bottom right panel of Fig.~\ref{fig:chiron}, calculating individual fluxes with the elliptical NEATM solution applied to the condition of each observation still shows significant residuals from the observations, suggesting that additional factors come into play
(presumably, measurement errors).

\begin{figure}[ht]
\centering
\includegraphics[width=8cm,angle=0]{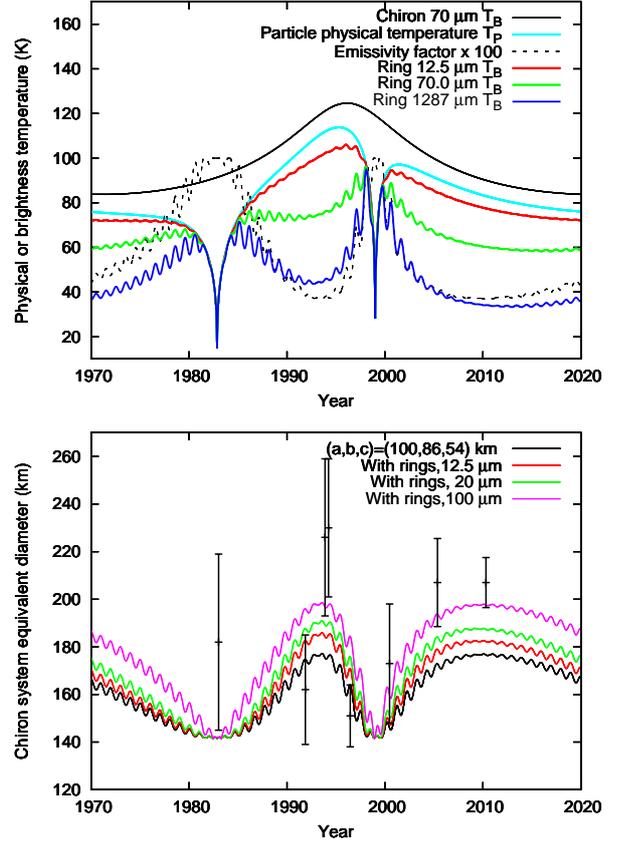}
\caption{Thermal contribution of Chiron putative ring. {\em Top: Ring temperature model}: Particle temperature (light blue) and ring brightness temperature at three wavelengths (red: 12.5 $\mu$m,
green: 70 $\mu$m, dark blue: 1287 $\mu$m) as a function of time over 1970-2020 for a Chiron ring with $\tau$=0.4. They are compared to the 70 $\mu$m brightness temperature of Chiron itself (black).
The dotted line represents the emissivity factor (1 - e$^{-\tau/B}$), multiplied by 100. {\em Bottom}: Wavelength-dependent effect of rings on the apparent diameter. See text for details.}

\label{fig:chironrings}
\end{figure}

We also applied the ring thermal model to the Chiron case. For this we used the ring parameters considered
by \citet{ortiz15}, i.e. a 324 km diameter, a 10 km width and a 0.4 opacity\footnote{\citet{ortiz15} quoted a 0.7-1.0 opacity
based on the depth of the occultation events, but as for Chariklo, this value must be divided by
two to account for diffraction effects.}.   Similar to Fig.~\ref{fig:ringstb}, Fig.~\ref{fig:chironrings}
shows (i) the ring particle physical temperature and ring brightness temperature vs time (top panel) and (ii) the putative effect
of ring emission to the Chiron equivalent diameter. In this case, Chiron's absolute dimensions were tuned so as to 
fit the smallest diameter determination  (i.e., from \citet{groussin04} in 1996), multiplying
the ``no-ring" Jacobi ellipsoid solution by 0.875 (i.e., giving $a$ = 100 km, $b$ = 86 km, $c$ = 54 km) and increasing its geometric albedo to 0.13 to continue to match the H$_v$ magnitude. These calculations qualitatively show that Chiron's rings, if existent, would affect diameter
determinations. However, Fig.~\ref{fig:chironrings} suggests that including the effect of Chiron's rings does not permit to 
fully reconcile the discrepant diameter determinations; in particular, the ring contribution appears maximal over 1991--1995, yet
diameter measurements over this period indicate large dispersions. We nonetheless explored this further by fitting thermal fluxes  corrected by our estimates of the ring contribution.
A spherical NEATM fit of the {\em Herschel}/{\em Spitzer} ring-corrected fluxes indicates a best-fit diameter of 186$^{+13}_{-14}$ km and beaming factor $\eta$ = 0.87$^{+0.15}_{-0.11}$. 
Using instead the elliptical model and starting from the 148 x 127 x 80 km initial guess model, we found a best fit scaling factor $f_{scale}$ = 0.69$\pm$ 0.05, giving nominally 102 x 87 x 55 km, and a beaming 
factor $\eta$ = 0.84$^{+0.12}_{-0.10}$. For both cases, the associated radio emissivities, based either on our own data or on the \citet{altenhoff95} data, are summarized in Table~\ref{othermodelfitresultsChironBienor}     
\footnote{In Table~\ref{othermodelfitresultsChironBienor}, we also considered a case where all the {\em Herschel} / {\em Spitzer} fluxes are corrected
for ring contribution, but not the mm flux. This case is meant to represent an extreme case of a ``Saturn-like" ring brightness
temperature roll-off with increasing wavelength, as could be relevant given the high I/F reflectivity 
invoked for Chiron's rings in Fig.~\ref{fig:chiron}.}.
The above figures illustrate quantitatively the extent to which a ring system would affect the diameter determinations. However, applying this ring-corrected elliptical model to past (ring-corrected) thermal observations leads to observed/modelled residuals comparable to the ``no-ring" case (compare green and black points in the bottom right panel of Fig.~\ref{fig:chiron}). 

In summary, we feel that the case for a ring system around Chiron with properties similar to Chariklo's remains doubtful in several
aspects, and is not particularly supported by the overall analysis of the thermal data. In contrast, the characteristics and variability of Chiron's optical lightcurve are consistent with a triaxial body in hydrostatic equilibrium. 
We thus do not adopt the ``ring-corrected" emissivities, favoring instead values from the uncorrected elliptical NEATM,
and finally obtaining a relative emissivity of 0.70$\pm$0.09 at 1.29 mm based on the ALMA flux measurement.

\subsection{Bienor}
Recently, \cite{fernandez17} presented time series photometry of (54598) Bienor over 2013-2016, and by comparison with
earlier literature found a strong variation of both the lightcurve amplitude (from $\Delta m$$\sim$0.6 mag in 2000 to $\sim$0.1 mag
currently) and the absolute magnitude (changing abruptly from H$_v$$\sim$8.1 in 2000 to H$_v$$\sim$7.4 in 2008 and beyond).
Interpreting the lightcurve amplitude evolution as due to a change of the sub-observer latitude, they proposed a solution
for the rotation axis ($\lambda$~=~35$\pm$8$^{\circ}$,  $\beta$=50$\pm$3$^{\circ}$, i.e. J2000 $\alpha_P$ $\sim$ 4$^{\circ}$, $\delta_P$ $\sim$ 58$^{\circ}$) along with a shape model. Assuming hydrostatic equilibrium, the axis ratios are
$b$/$a$ = 0.45$\pm$0.05 and $c$/$b$ = 0.79$\pm$0.02. \citet{fernandez17} however noted that this model does not simultaneously 
reproduce the H$_v$ evolution and considered a variety of more complex models, such as non-hydrostatic equilibrium, a time-variable albedo, model including a ring contribution, etc. None of the models -- except apparently the albedo model, but the latter is not
really described -- achieves a good fit of the H$_v$ evolution \citep[see Fig. 3 of][]{fernandez17}. We do not
attempt here to interpret in detail the photometric behavior of Bienor, but note that it is reminiscent of
that of (139775) 2001 QG$_{298}$, convincingly demonstrated to be a contact binary with changing viewing geometry \citep{lacerda11}\footnote {For example, two equal spheres in contact viewed from the mutual orbit plane produce a lightcurve amplitude of 0.75, and the system mean brightness at this epoch
is 0.375 magnitude fainter than when the mutual eclipses do not occur; this effect is somewhat weaker but comparable
to the observed behavior of Bienor, and can be enhanced because contact binary components are tidally elongated along the line joining their centers. Note also that the shape of Bienor's lightcurve, as observed in 2001 when it had a large
amplitude, is also reminiscent of that of a contact binary, with a peaked lightcurve minimum \citep[see][]{ortiz03}.}.

\begin{table*}
\caption{Makemake model fits} 
\label{fitmakemake}      
\centering                          
\begin{center}
\begin{tabular}{l|ccc|ccc|ccc|c}
 Model  & \multicolumn{3}{c}{Makemake} & \multicolumn{3}{c}{Moon} & \multicolumn{3}{c}{Dust} & Relative submm\\
        & D (km) & p$_v$ & $\eta$      &  D (km) & p$_v$ & $\eta$ & T$_{dust}$ (K) & Area (km$^2$) & Mass (kg)   & emissivity\\
\hline 
& & & & & & & & & & \\
Makemake alone & 1430 & 0.795 & 2.0$^{+0.75}_{-0.65}$ & N/A & N/A & N/A & N/A & N/A & N/A & 1.03$^{+0.10}_{-0.09}$ \\

& & & & & & & & & & \\


Makemake + moon &  1430 & 0.795 & 2.64 & 265$^{+166}_{-193}$ & 0.017$^{+0.21}_{-0.010}$ & 0.34$^{+0.18}_{-0.27}$ & N/A & N/A & N/A & 1.05$^{+0.10}_{-0.10}$ \\

& & & & & & & & & & \\

Makemake + dust &  1430 & 0.795 & 2.23$^{+0.76}_{-0.80}$  & N/A & N/A & N/A & 54.6 & 4.1$\times$10$^5$ & 7.2$\times$10$^6$ & 1.07$^{+0.10}_{-0.09}$\\
& & & & & & & & & & \\

\hline 
\end{tabular}
\end{center}
\end{table*}

\begin{figure*}[ht]
\centering
\includegraphics[width=4.2cm,angle=-90]{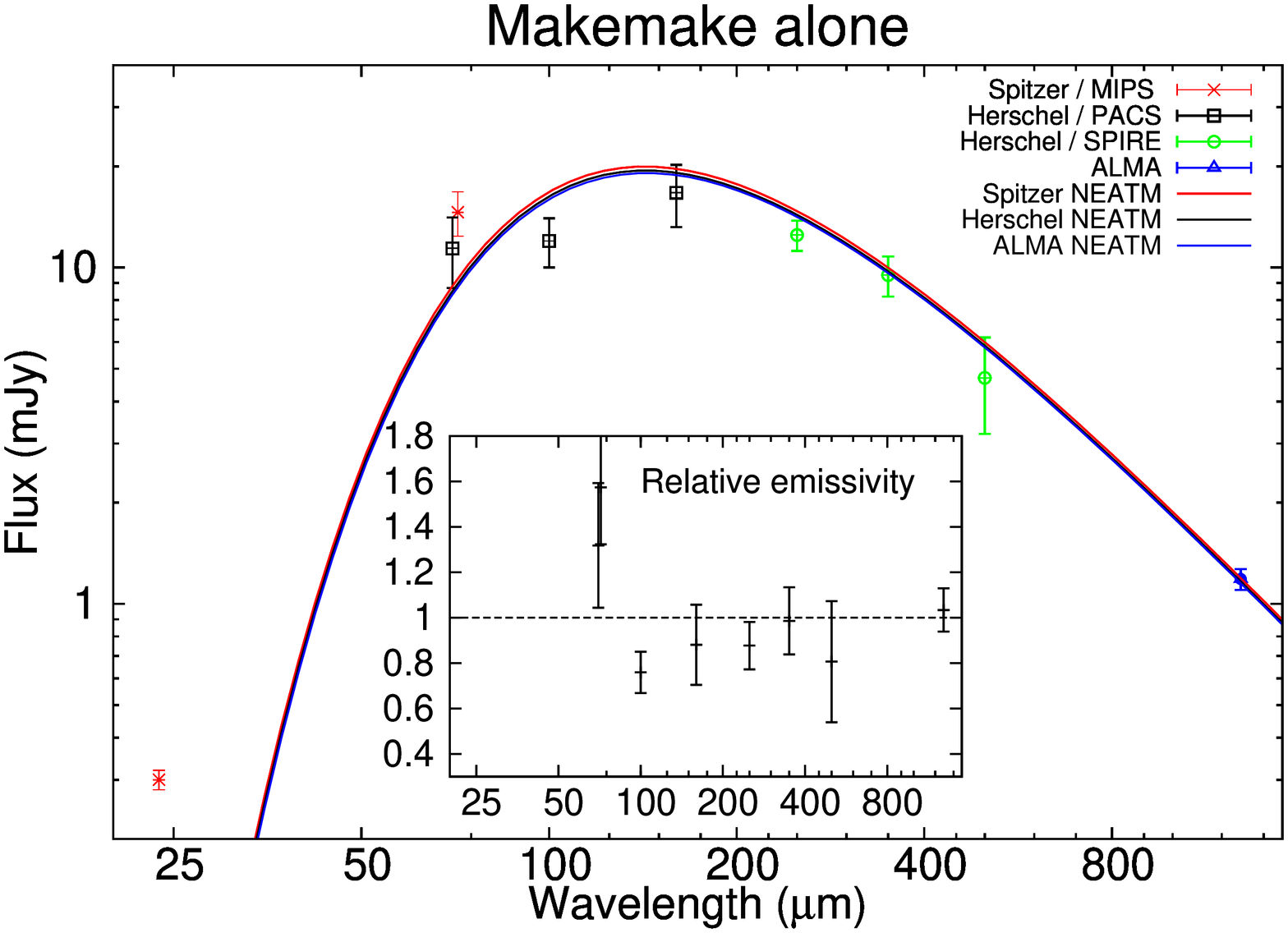}
\includegraphics[width=4.2cm,angle=-90]{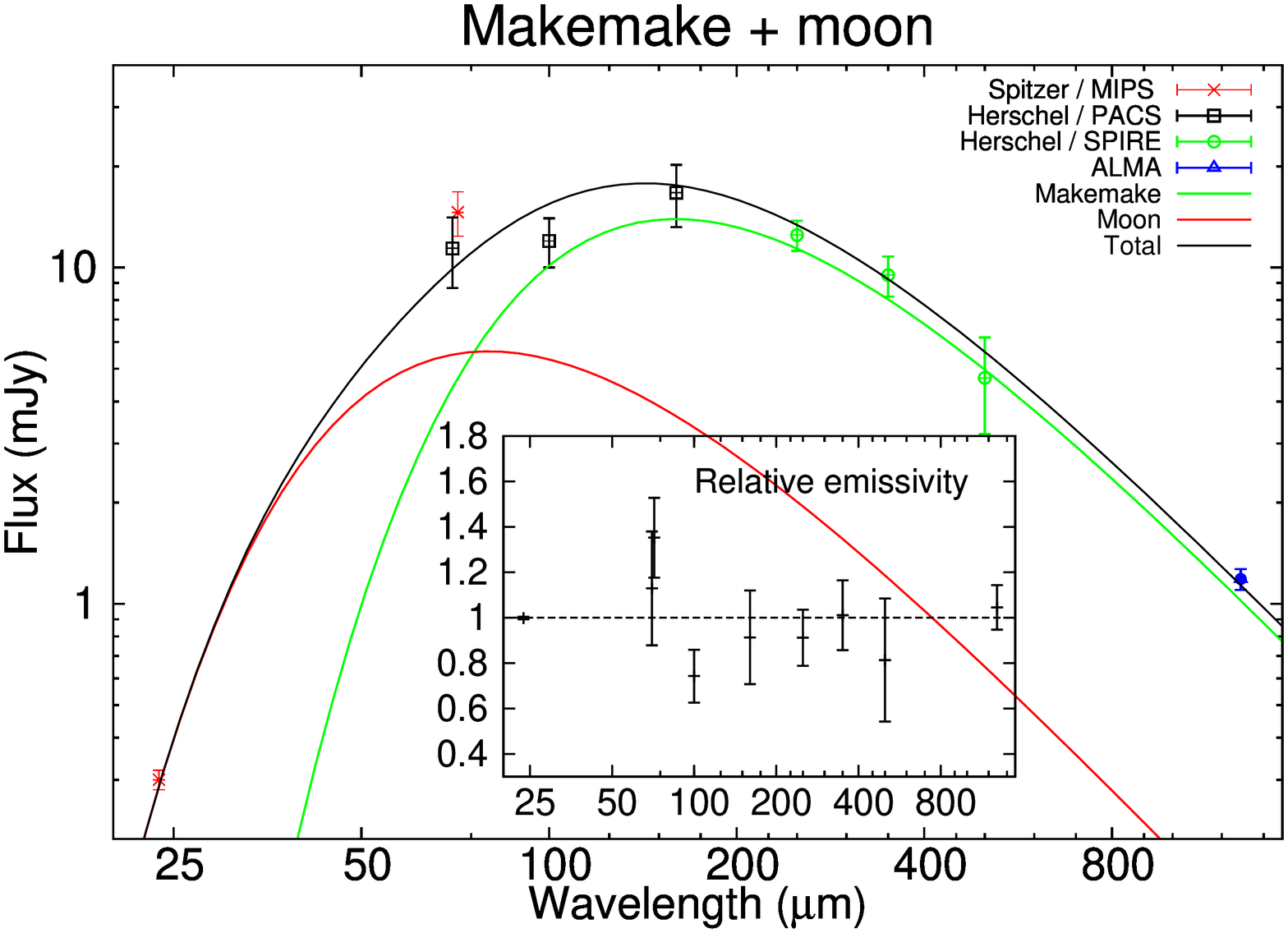}
\includegraphics[width=4.2cm,angle=-90]{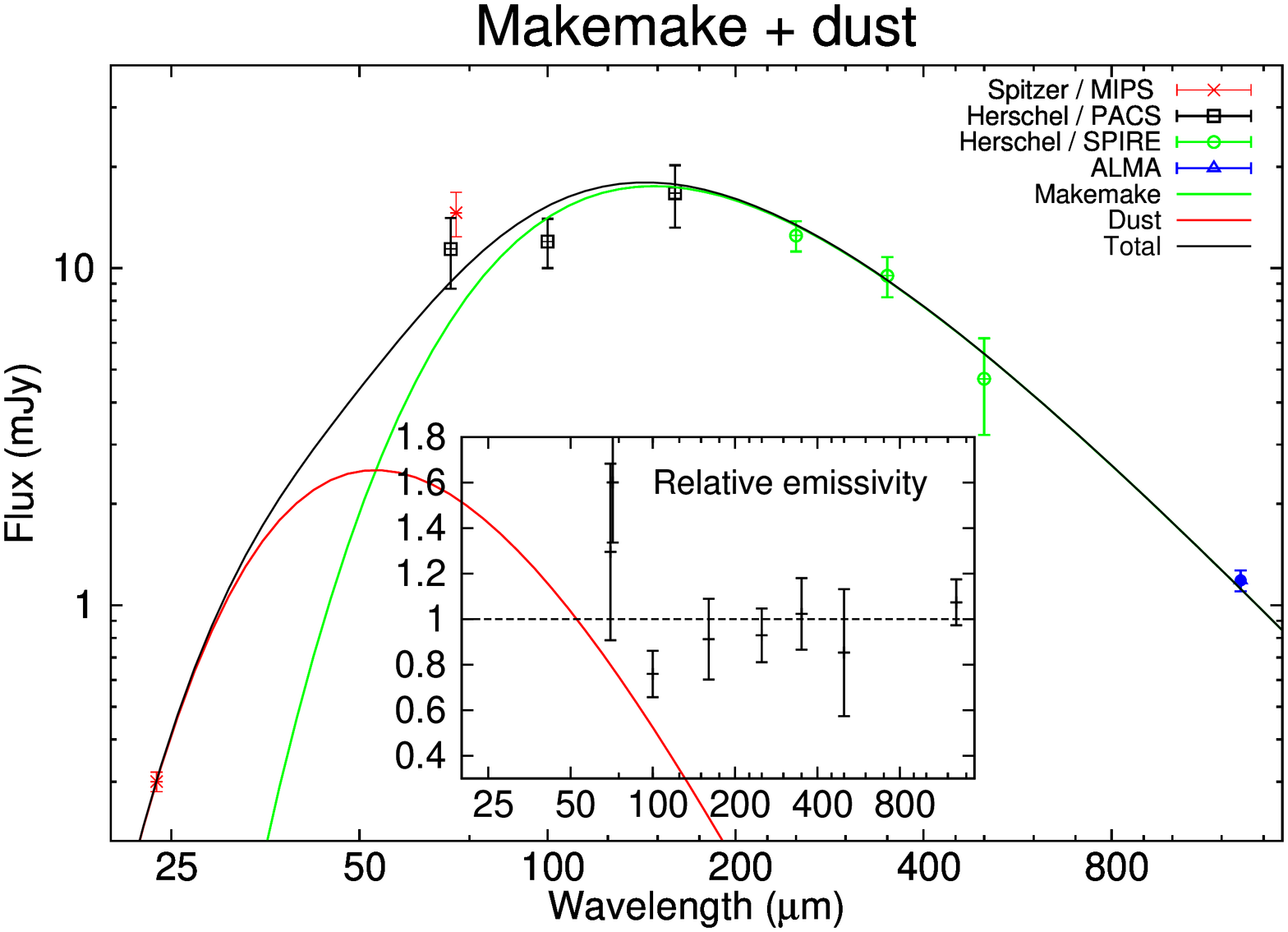}
\caption{Flux measurements, spectral energy distribution and relative emissivity spectra (in inset) of Makemake, for the
three considered models. 
Model fit parameters are as follows. (Left): ``Makemake alone" model: D = 1430 km, p$_v$ = 0.795, $\eta$ = 1.69.  
(Center): ``Makemake + moon" model: For Makemake, D = 1430 km, p$_v$ = 0.795, $\eta$ = 2.64; for the moon, D = 323 km, p$_v$ = 0.012,  $\eta$ = 0.406. (Right): ``Makemake + dust" model: For Makemake, D = 1430 km, p$_v$ = 0.795, $\eta$ = 1.95; for the dust, T = 54.6 K, mass = 7.2$\times$10$^9$ kg. In the left panel, different colors show NEATM fits applied to the conditions of the different
observations ({\em Spitzer}, {\em Herschel}, and ALMA). In the other two panels, the models are shown for the {\em Herschel} conditions. 
}
\label{fig:makemake}
\end{figure*}

Still,  all models presented in  \citet{fernandez17} call for generally similar pole orientations and 
shape models. We thus adopt the above solution parameters, and refit the {\em Spitzer}/{\em Herschel} thermal data with
elliptical NEATM. Free parameters are a scaling factor to the shape model and the beaming factor  
(the albedo is kept at 0.041 from Table~\ref{fitresults}, which may not be accurate but is unimportant). The best fit
solution is found for a = 144 km, b = 65 km and c = 51 km, with 4 \% uncertainties, and $\eta$ = 1.24$\pm$0.10, noticeably
different from the value obtained before from the spherical model (1.58$^{+0.15}_{-0.11}$). With these values, 
the equivalent diameter of Bienor at the {\em Spitzer} (July 2014, sub-solar latitude $\beta$ = -29$^{\circ}$) and
{\em Herschel} (January 2011,  $\beta$ = -49.5$^{\circ}$) epochs is 158 and 177 km, respectively, considerably smaller than derived from the spherical model (199 km) and illustrating the limited value of a direct scaling of a shape model to a diameter determined from a spherical model. Nonetheless, applying the model to the geometry of the ALMA observations ($\beta$ = -58$^{\circ}$) leads to a relative 1.29 mm
emissivity of 0.64$\pm$0.07, virtually identical to the value from  spherical NEATM (see Table~\ref{fitresults}).

Application of a spherical thermophysical model with the above orientation parameters, a rotational period of 9.17 hr
\citep{fernandez17}, and various levels of roughness, leads to best fit thermal inertia of 8-15 MKS, a diameter of (179-184)$\pm$6 km, and a relative 1.29 mm emissivity of 0.68$\pm$0.07 (Table~\ref{othermodelfitresultsChironBienor}). Although the pole direction
of Bienor remains to be confirmed, we adopt this as the best guess estimate of the object's mm/submm emissivity.

We finally note that a stellar occultation by 2002 GZ$_{32}$ was successfully observed from several locations in Europe
on May 20, 2017. When a shape model becomes available, this will provide the opportunity to reinterpret the thermal measurements, including our ALMA data.

\subsection{Makemake}
Before our measurement with ALMA, Makemake had been measured in the thermal range by {\em Spitzer} \citep{stansberry08}
and {\em Herschel} \citep{lim10}. As demonstrated at that time, standard NEATM fits fail to reproduce 
Makemake's SED, and a multi-component model is required. Within NEATM, \citet{lim10} found solutions combining
a large, high-albedo unit making up most of Makemake's surface, and a localized, very dark unit, which is responsible
for Makemake's elevated 24 $\mu$m flux; in these models, the two units have very different values
of the beaming factor ($\eta$= 1.3-2.2 and 0.4-0.5 respectively), suggesting sharply different thermal properties.
\citet{lim10} further suggested that the dark terrain could represent a satellite, an hypothesis
that gained credit with the discovery of S/2015 (136472)1 by \citet{parker16}. These latter authors revisited the 
thermal models by including new constraints from (i) the H$_V$ magnitude of the satellite, 7.80 mag fainter than Makemake
and (ii) the occultation-derived diameter of Makemake \citep{ortiz12,brown13b}, confirming the essential conclusion
that S/2015 (136472)1 may contribute a large fraction of the flux associated to the dark terrain. As noted
by \citet{parker16}, their models still do not fully reproduce the system 24 $\mu$m flux; yet they invoke
$\eta$ values of 0.25-0.40. Such values are hard to understand physically, as even a surface saturated with hemispherical 
craters does not produce beaming factors lower than 0.6 \citep{spencer90,lellouch13}.

To determine the mm emissivity of Makemake in spite of these complications, we considered three end-member models,
termed ``Makemake-alone",  ``Makemake + moon", and ``Makemake + dust".
In all cases, Makemake diameter and geometric albedo were held fixed at values determined from
the stellar occultation. As a compromise between the (anyway similar) values
given in \citet{ortiz12} and \cite{brown13b}, we adopted D = 1430 km, which for H$_v$ = 0.091$\pm$0.015 \citep{parker16}, 
gives p$_v$ = 0.795. As for the other objects, the models were applied to the set of  {\em Spitzer}, {\em Herschel},
and ALMA fluxes. We used the {\em Spitzer} and {\em Herschel}/PACS flux values as published in 
\citet{lim10}, but updated the {\em Herschel}/SPIRE fluxes, accounting for (i) a new reduction pipeline (HIPE version 8.2) 
and destriping methods \citep{fornasier13} (ii) the availability of new Makemake observations from June 2010, in addition to the
Nov./Dec. 2009 data presented in \cite{lim10}. Averaging the two epochs, the new SPIRE fluxes are 
12.5$\pm$1.3 mJy, 9.5$\pm$1.3 mJy, and 4.7$\pm$1.5 mJy  at 250, 350 and 500 $\mu$m respectively.

In the first model (``Makemake-alone"), a standard NEATM fit was performed on the 3-band {\em Herschel}/PACS
and the {\em Spitzer} 71.42 $\mu$m fluxes, ignoring the {\em Spitzer} 23.68 $\mu$m flux, with $\eta_{Make}$ as the only free parameter. As before, the model was then applied to other wavelengths to determine the corresponding spectral emissivities.
The best fit solution, shown in the left panel of Fig.\ref{fig:makemake}, has $\eta_{Make}$ = 1.69, but even though the diameter and geometric albedos are fixed, the error bar on $\eta_{Make}$ is large ($\eta_{Make}$ = 2.0$^{+0.75}_{-0.65}$; see
Table~\ref{fitmakemake}), a consequence of the strong effect of the phase integral uncertainty for a bright object. Our ALMA
flux measurements then implies a 1.03$^{+0.10}_{-0.09}$ relative mm emissivity.

In a second model (``Makemake + moon"), the two {\em Spitzer} and the three {\em Herschel}/PACS measurements were fit with the sum
of emission from Makemake and its moon. This model had initially four parameters,  $\eta_{Make}$ and (D, p$_v$ and $\eta$) for its moon, for which we adopted H$_v$ = 7.89$\pm$0.3 (to allow for possible lightcurve variation).  However, the system was somehow degenerate and best fit solutions tended to call for very large values of 
$\eta_{Make}$ ($>$3-4), while values larger than $\sim$2.7 are unphysical \citep[see e.g. Fig.~4 of][]{lellouch13}.
$\eta_{Make}$ was then fixed at a maximum value of 2.64. Even then, the problem remained somewhat under constrained, 
leading to broad ranges for (D, p$_v$ and $\eta$) of the moon (Table~\ref{fitmakemake}). More worryingly 
but consistent with the previous studies \citep{lim10,parker16}, the solution range of $\eta_{moon}$ remained unphysical
(0.34$^{+0.18}_{-0.27}$) suggesting that Makemake's moon alone might not entirely explain the 24-$\mu$m flux. Since the 
temperatures scale as ($\eta$$\epsilon_b$)$^{-1/4}$, a possible way
to alleviate the problem would be to invoke a {\em bolometric} emissivity much lower than 1 for Makemake's moon,
but we note that restoring physically-reasonable values of $\eta$ (say $\eta$ $>$ 0.65) would then require $\epsilon_b$$<$0.5.
In any case, with this ``Makemake + moon" model, we infer a 1.05$^{+0.10}_{-0.10}$ relative mm emissivity for the system.

\begin{figure}[ht]
\centering
\includegraphics[width=8cm, angle=0]{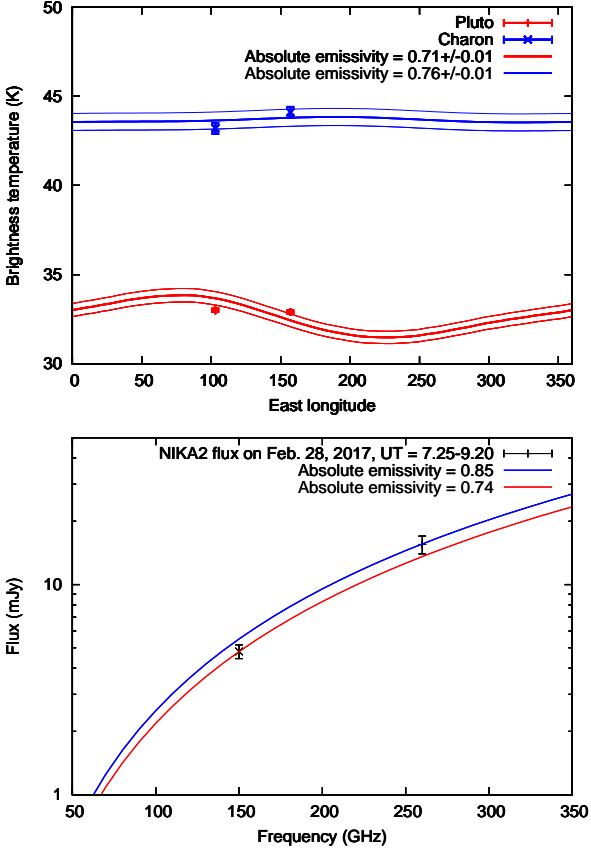}
\caption{{\em Top}: Separate 0.86 mm brightness temperatures of Pluto and Charon measured by ALMA on June 12 and 13, 2015 \citep{butler17}, compared to calculations based on the \citet{lellouch16} model, and adjusting the absolute 0.86 mm emissivity to
0.71$\pm$0.01 for Pluto and 0.76$\pm$0.06 for Charon. {\em Bottom}: Flux density of the Pluto+Charon system for the conditions of the IRAM/NIKA2 observations of Feb. 28, 2017, and two values (0.72 and 0.85) of the system absolute emissivity. The statistical and calibration error bars have been combined quadratically.}
\label{fig:Pluto_ALMA_NIKA_detailedmodel}
\end{figure}

In a third model, the 24 $\mu$m flux excess from Makemake's system was attributed to dust emission.
For this we adopted the dust absorption coefficients of \citet{lidraine01}, giving e.g. $\kappa$ = 56 m$^2$ kg$^{-1}$
at 23.68 $\mu$m, and specified a dust grain temperature of 54.6 K, which corresponds  to the 
instantaneous equilibrium of a sun-facing surface with zero albedo at Makemake's 52 AU distance. 
This choice is somewhat arbitrary as small grains could be poor radiators (hence warmer than the above value), 
and/or radiating over 4$\pi$ (hence cooler). With these values, we infer an effective emitting area of 
4.1x10$^5$ km$^2$ for the grains, equivalent to a 720 km diameter, and a total dust mass of 7.2$\times$10$^6$ kg. These
figures do not seem unreasonable, although there is no evidence for dust activity at Makemake. Once corrected
for the dust contribution at all wavelengths, the {\em Spitzer} and {\em Herschel} fluxes were refit with a Makemake model, again
with $\eta_{Make}$ as the only free parameter, leading to a best fit value of $\eta_{Make}$ = 1.94 and a range of
2.23$^{+0.76}_{-0.80}$. We note that significantly cooler grain temperatures than the above value would not permit
to match the data. For example, assuming slowly rotating large dust grains (radiating over 2$\pi$), i.e. a grain temperature 2$^{1/4}$ lower (45.9 K) would require a total dust mass $\sim$8 times larger, but then the corrected fluxes would
lead to an unphysical $\eta_{Make}$ = 3.7$^{+1.2}_{-1.3}$.

While each of the three models had its own issues, they all lead to remarkably similar values of the relative
mm emissivity, consistent with unity, as summarized in Table~\ref{fitmakemake} and Fig.~\ref{fig:makemake}. Complications associated with non-spherical shape and/or with changing apparent orientations can be dismissed as (i) stellar occultation
indicates that Makemake has a low oblateness \citep[projected elongation $<$1.06;][]{ortiz12,brown13b}; and (ii) Makemake
has travelled only 10$^{\circ}$ along its orbit in the 2006-2016 timeframe covered by the thermal measurements, limiting
changes in the sub-solar latitude to less than this value. We conclude that, unlike all other objects in our sample
and that of \citet{brown17}, Makemake does have a relative mm emissivity close to 1. Interestingly, in spite of considerably larger
error bars, the SPIRE fluxes are also consistent with a relative emissivity of 1 over 250-500 $\mu$m (see insets
of Fig.~\ref{fig:makemake}).

\begin{table*}
\caption{Pluto/Charon NEATM fits} 
\label{fitpluto}      
\centering                          
\begin{center}
\begin{tabular}{l|cc|cc|cc|c}
 Model  & \multicolumn{2}{|c|}{Diameter (km)} & \multicolumn{2}{|c|}{$p_v$} & \multicolumn{2}{|c|}{$\eta$}  & Relative submm\\
        & Pluto & Charon & Pluto      &  Charon & Pluto & Charon     & emissivity\\
\hline 
& &  & & & & & \\
1-component & \multicolumn{2}{|c|}{2279$^{+112}_{-108}$} & \multicolumn{2}{|c|}{0.66$^{+0.12}_{-0.12}$} & \multicolumn{2}{|c|}{1.28$^{+0.52}_{-0.70}$} &    0.99$^{+0.05}_{-0.04}$ (0.86 mm) \\

& & & & & & & \\
2-component & 2238$^{+248}_{-190}$ & 1207 & 0.61$^{+0.11}_{-0.13}$  & 0.362 & 2.62$^{+1.51}_{-1.47}$ & 1.23$^{+0.23}_{-0.17}$ & 0.88$^{+0.06}_{-0.07}$ (0.86 mm) \\

& &  & & & & & \\

\hline 
\end{tabular}
\end{center}
\end{table*}

\begin{figure*}[ht]
\centering
\includegraphics[width=6.3cm,angle=-90]{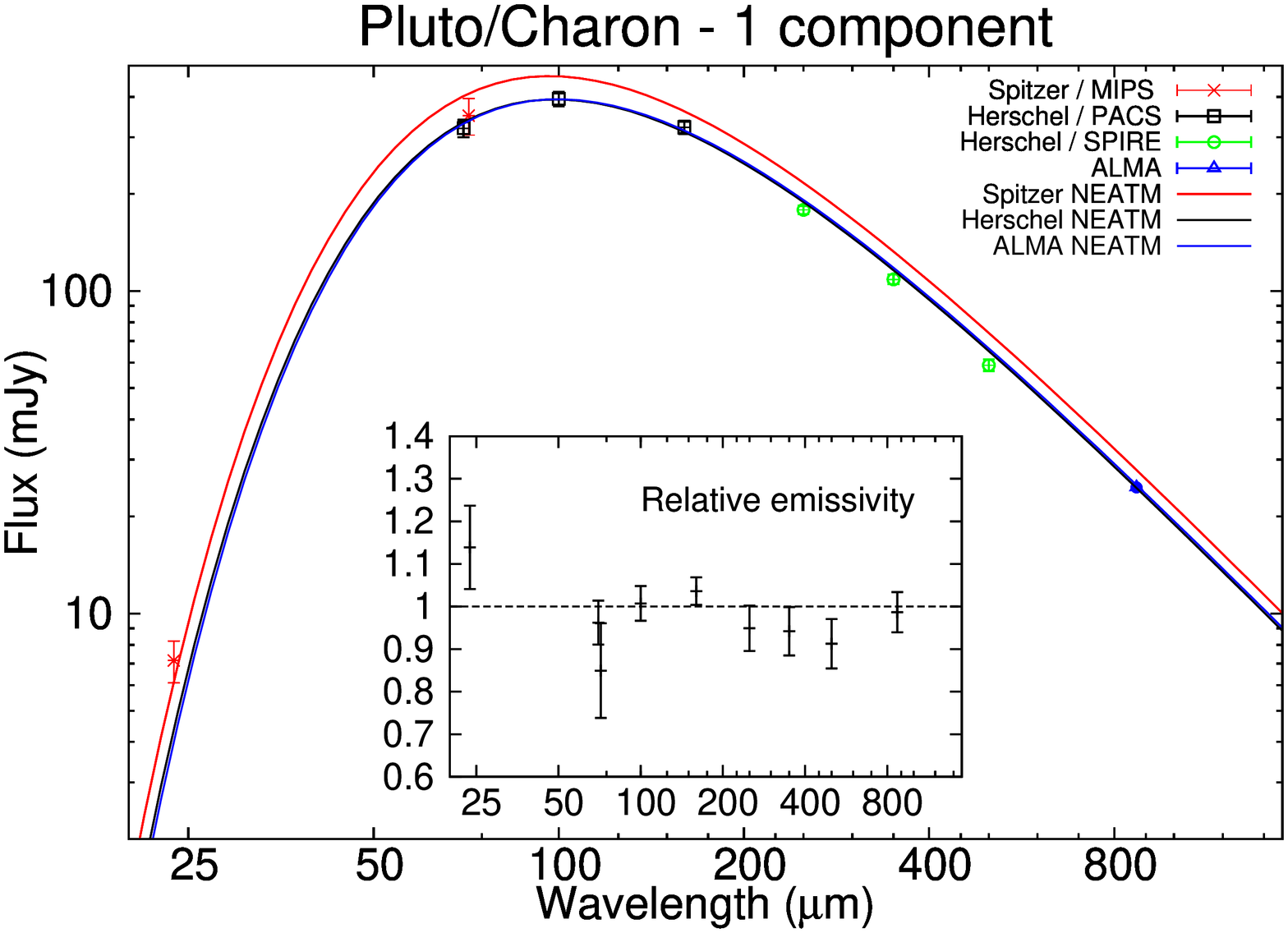}
\includegraphics[width=6.3cm,angle=-90]{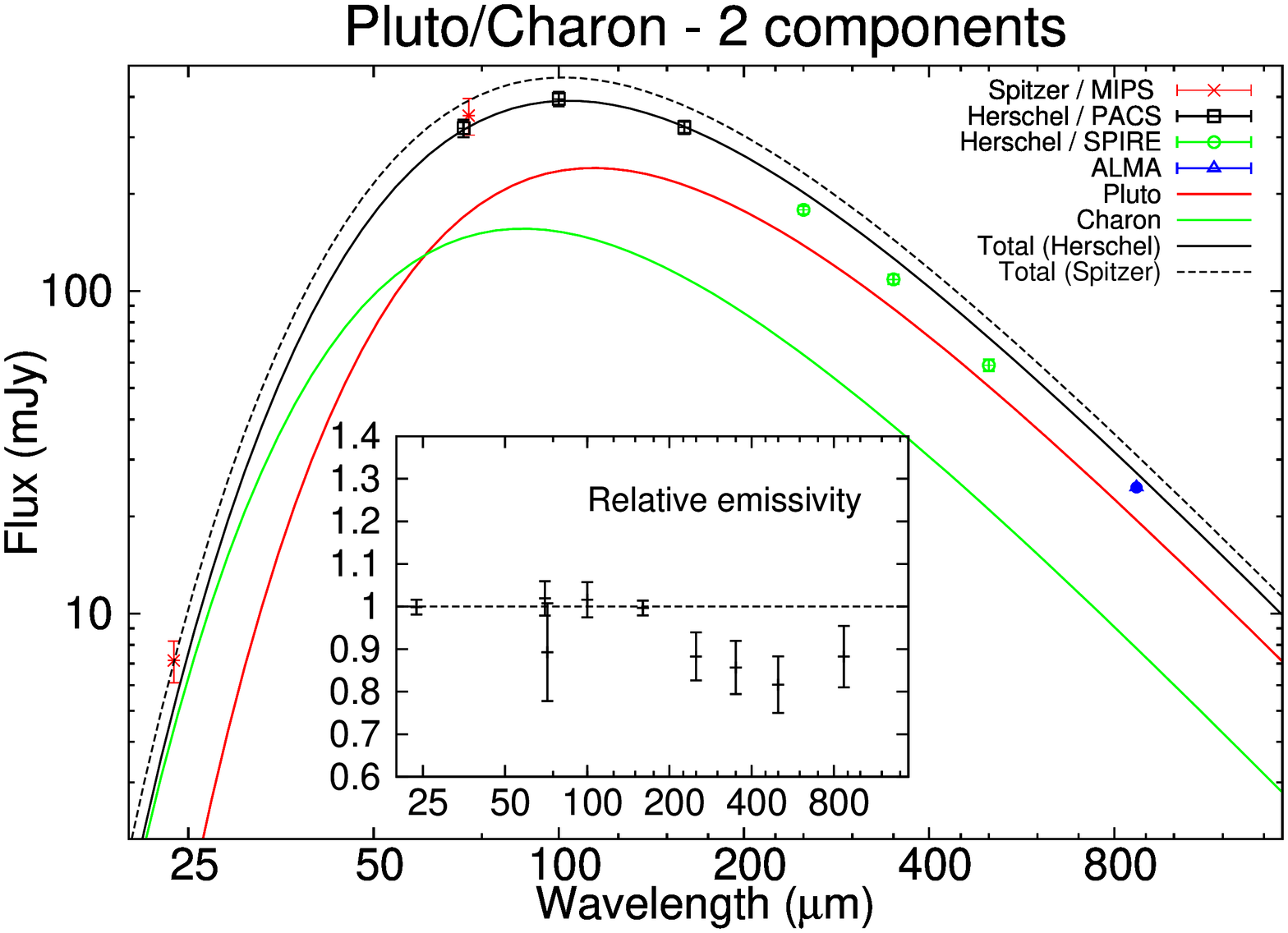}
\caption{Flux measurements, spectral energy distribution and relative emissivity spectra (in inset) of the Pluto/Charon system, for the 1- and 2- component NEATM fits.  (Left): 1-component fit, with following parameters: D = 2294 km, p$_v$ = 0.64, $\eta$ = 1.38.  
(Right): 2-component fit, with fixed diameter/albedo for Charon. Model parameters are: Charon, D = 1207 km, p$_v$ = 0.362, $\eta$ = 1.20; Pluto, D = 2225 km, p$_v$ = 0.61,  $\eta$ = 2.56.  In the left panel, different colors show NEATM fits applied to the conditions of the different observations ({\em Spitzer}, {\em Herschel}, and ALMA). In the ring panel, the individual model Pluto (red) and Charon (green) components are shown for the {\em Herschel} conditions, and the total model flux (black) is shown both for {\em Herschel} (solid line) and {\em Spitzer} (dashed line).
}
\label{fig:pluto}
\end{figure*}

\subsection{Pluto and Charon}

\subsubsection{ALMA observations}
The thermal emission from the Pluto/Charon system has been investigated in considerable details in the last 20 years. This
includes spatially unresolved but rotationally repeated far-IR measurements (i.e. thermal lightcurves) using ISO, {\em Spitzer} and {\em Herschel} over 20-500 $\mu$m \citep{lellouch00a,lellouch11,lellouch16} and 
numerous but more widespread ground-based mm/submm data. Initially, the latter did not resolve the
Pluto/Charon system \citep[see][and references therein]{lellouch00b}, but 
the advent of interferometers (SMA, VLA, ALMA) permitted the thermal emission from Pluto and Charon to be measured
separately \citep{gurwell11,butler15,butler17}.  In particular, based on ALMA observations acquired on June 
12 and 13, 2015 (i.e. at two separate longitudes) which also led to the detection of CO and HCN in Pluto's atmosphere \citep{lellouch17},
\citet{butler17} reported 860 $\mu$m brightness temperatures 
of $\sim$33 K for Pluto and $\sim$43.5 K for Charon, providing a definitive demonstration that Charon is warmer
than Pluto.  Making use of the multi-terrain thermophysical models presented by \citet{lellouch11,lellouch16}, these measurements were interpreted  in terms of separate {\em absolute} 860 $\mu$m emissivities of the two objects, found to be 0.69-0.72 for Pluto and 0.75-0.77 for Charon \citep{butler17}. Fits of the ALMA fluxes for the two observing dates are shown in the top panel 
of Fig.\ref{fig:Pluto_ALMA_NIKA_detailedmodel}. For a bolometric emissivity of 0.9, as is characteristic of the models of \citet{lellouch16}, this implies relative emissivities of 0.78$\pm$0.015 and 0.84$\pm$0.01, respectively, i.e. a surface 
weighted-average value of 0.80 for the Pluto/Charon system. 

As these values were obtained from a model that makes use of considerably more information (thermal lightcurves, precise object sizes, distribution of ices, etc.) than is available for other TNOs, it is enlightening to explore what could be derived on the mm emissivity of the Pluto system by applying a simple NEATM analysis as for other objects. 
For this purpose, we first created a Pluto/Charon dataset comparable to what we have for the other TNOs by averaging the {\em Spitzer}/MIPS fluxes (8 visits in 2004, sampling the Pluto lightcurve in 2004) and the  {\em Herschel}/PACS and /SPIRE data (9 visits each in 2012). In doing so, the error bar on each flux value was defined by the (real) dispersion of the measurements at a given wavelength due to thermal lightcurve, rather than by the noise of each individual measurement. We then performed a simple NEATM analysis of the {\em Spitzer},  {\em Herschel}  and ALMA fluxes, considering, as for Makemake, 1- and 2- components fits. Results and fits are summarized in Table~\ref{fitpluto} and Fig.~\ref{fig:pluto}. In the 1-component fit, the relative 860 $\mu$m emissivity of the Pluto/Charon system is close to 1, but the solution
diameter (2280$\pm$110 km) is as much as $\sim$15 \% smaller than the equivalent diameter of the Pluto-Charon system (2670 km), a problem already encountered by \citet{mommert12}. The inability of the 1-component NEATM model to fit
the SED of Pluto/Charon is related to the fact that the two bodies have comparable sizes, yet different
albedos. In the 2-component model, Charon's diameter and geometric albedo were 
held fixed at their observed values (D = 1207 km, p$_v$ = 0.36), and the SED of the system was fit in terms of four
parameters, $\eta_{Charon}$ and (D, p$_v$ and $\eta$) for Pluto. In this case, the Pluto best fit diameter was found
to be 2225 km. This is still 6 \% smaller than the known Pluto diameter (2380 km)  -- one likely reason being that part of Pluto's surface is covered by isothermal N$_2$-ice at $\sim$37.4 K, which contributes little to the dayside
thermal emission at the shorter wavelengths -- but error bars (Table~\ref{fitpluto}) 
encompass the true value. Thus the introduction of Charon essentially reconciles the radiometric Pluto diameter with its established value. In this 2-component fit, the ALMA-derived
relative emissivity for the Pluto/Charon system is 0.88$^{+0.06}_{-0.07}$ at 860 $\mu$m. 
It is gratifying that even with this simple model, the emissivity is only 1-$\sigma$
different from that obtained using the more detailed thermophysical models.

\subsubsection{IRAM/NIKA2 observations}
Most recently, the Pluto system was observed during the commissioning of the new NIKA2 wide-field camera (6.5 arcminutes covered by 2900 detectors) dual-band (1.15 mm/2 mm) camera installed at the IRAM-30 m radio-telescope on Pico Veleta, Spain \citep{calvo16}\footnote{We had ourselves acquired Pluto dual-band 1.15/2 mm data on Feb. 19-20, 2014 with NIKA (the prototype of NIKA2), in the framework of IRAM proposal 118-13, but data reduction for those observations, taken in Lissajous mode, is fraught with complexity and did not so far provide a reliable photometry.}. Observations were taken on Feb. 29, 2017, UT 7.25-9.20, corresponding
to 1.44 hr on-source time an observing longitude L = 87$\pm$2. Although, unlike the SMA, VLA and ALMA observations of \citet{butler17}, these do not resolve Pluto from Charon, they provide the first detection of the system at 2 mm. The 1.15 and 2 mm bands cover approximately $\sim$230-285 GHz and $\sim$125-175 GHz at mid-power, with effective frequencies of 150 and 260 GHz. The Pluto+Charon fluxes were reported to be 4.8$\pm$0.2$\pm$0.3 mJy at 2 mm and 15.1$\pm$1.0$\pm$1.1 mJy at 1.15 mm \citep{adam17},  where the first (second) error bar indicates the statistical (calibration) uncertainty. We combine quadratically these error bars.

Applying the thermophysical model of \citet{lellouch16} to the geometry of these observations, we find that 
the ALMA-derived model, which has a system-weighted absolute emissivity of 0.72, is fully
consistent with the 2 mm flux but falls below the measured 1.15 mm flux by 1.5~$\sigma$. Fine tuning the model
to the observations, optimal fits of the 2 mm and 1.15 mm fluxes are achieved for absolute emissivities of 
0.74$\pm$0.06 and 0.85$\pm$0.07, respectively. In the submm range, \citet{lellouch16} found that
the emissivity of the Pluto system decreases from $\sim$1 at 20-25 $\mu$m to $\sim$0.7 at 500 $\mu$m.  
This, and the fact that a non-monotonic variation of the emissivity over 0.86-1.15-2 mm does not
seem plausible, makes us suspect that the 1.15 mm flux reported by \citet{adam17} is slightly overestimated. In this respect, an additional source of uncertainty is the yet unknown telescope/instrument gain response with elevation, that would affect the 260 GHz flux at low elevation, but not
the 150 GHz flux (J.-F. Lestrade, priv. comm.), an issue that will be revisited when more observations of calibrators with NIKA2 are
available. For the time being, the ALMA-derived separate emissivities and Pluto and Charon are retained.

\subsection{Other objects}
We finally briefly reconsidered a few additional mm/submm observations of TNOs/Centaurs from literature.  Few of these
measurements led to actual detections. Exceptions are Varuna \citep{jewitt01,lellouch02}, 1999~TC$_{36}$
\citep[][who also obtained upper limits on 6 other objects]{altenhoff04}, and Eris \citep{bertoldi06}.

{\em Varuna:} We first focus on the case of Varuna for which the above two measurements at radio wavelengths provided best fit diameters of
900-1060 km, in sharp contrast with the {\em Spitzer} \citep[500$\pm$100 km,][]{stansberry08} and {\em Spitzer/Herschel} \citep 
[668$^{+154}_{-86}$ km,][]{lellouch13} values. Performing the same NEATM analysis as for all the TNOs observed by ALMA, we find an updated diameter of 654$^{+154}_{-102}$ km, but unphysical mm/submm emissivities of 2.49$^{+1.24}_{-1.00}$  at 0.850 mm and 1.91$^{+1.00}_{-0.82}$ at 1.20 mm, using the fluxes reported by \citet{jewitt01} and \citet{lellouch02}. This indicates a gross inconsistency between the far-IR and mm/submm measurements.

\citet{sicardy10} reported on a stellar occultation by Varuna on February 19, 2010. This occultation, which was acquired

near maximum lightcurve of the object, provided one 1003 km long chord across the object -- in itself sharply inconsistent with the spherical solution from {\em Spitzer/Herschel} --  along with a non-detection just
225 km south of that chord. Although this is insufficient to determine a complete shape and orientation model, probability considerations indicate that the most likely figure of the object is strongly elongated, with $a$ $\sim$ 860 km and 
$c$ $\sim$375 km  \citep{sicardy10}.  Varuna is also characterized by a marked optical lightcurve ($\Delta$m = 0.42 mag, period = 6.344 hr). Assuming hydrostatic equilibrium and considering different surface optical properties and pole orientation, \citet{lacerda07} constrained axis
ratios to be in the range $b/a$ = 0.63-0.80 and $c/a$ = 0.45-0.52. We here adopt for definitiveness $a$ = 860 km, $b$ = 550 km, $c$ = 390 km 
 \citep[the lunar model and aspect angle = 75$\deg$ from][]{lacerda07}), giving p$_v$ = 0.048. As was done above for Chariklo and Chiron, we refitted the {\em Spitzer/Herschel} data allowing for an overall scaling factor $f_{scale}$ on this shape model. Assuming that all observations referred to
an intermediate phase between lightcurve maximum and minimum, we inferred $f_{scale}$ = 0.60$^{+0.14}_{-0.08}$ and $\eta$ = 2.0$^{+1.0}_{-0.5}$. The {\em Spitzer} data for Varuna include two epochs 
while the {\em Herschel} data correspond to yet another epoch.
Even if all three epochs corresponded to lightcurve minimum, the required scaling factor would
still be only 0.68$^{+0.14}_{-0.08}$, grossly
inconsistent with  $f_{scale}$~=~1.

In contrast, the mm/submm measurements of \citet{jewitt01} and \citet{lellouch02}, appear consistent with the above shape model. We fitted them {\em separately} (i.e. not including the {\em Spitzer/Herschel} data),
assuming that they sample rotationally-averaged conditions\footnote{Each of the two observations included several long ($\sim$0.75-hr to $\sim$2-hr) integrations on 2 to 5 separate dates.}, and adopting a beaming factor $\eta$ = 1.175$\pm$0.42. This number comes
from fitting a gaussian distribution to the 85 beaming factor values reported by \cite{lellouch13} -- and is similar to the ``canonical"
1.20$\pm$0.35 value from \citet{stansberry08}. A bolometric emissivity of 0.90$\pm$0.10 was still specified. These assumptions
lead to relative emissivities of 0.81$\pm$0.23 and 0.62$\pm$0.23 at 0.85 and 1.20 mm, in full agreement with the values derived for other objects. It thus appears that the $\sim$3$\sigma$  detections reported by 
\citet{jewitt01} and \citet{lellouch02} were likely to be real, although we regard these derived emissivity values as tentative. The inconsistency between the occultation-derived size and the far-IR thermal measurements remains to be elucidated; for this a more detailed shape model from future
stellar occultations will be welcome.

{\em 1999~TC$_{36}$:} From stacking of IRAM observations gathered over Dec. 2000 -- Mar. 2003 ($\sim$4 h integration total), \cite{altenhoff04} reported the detection of  1999TC$_{36}$ with a 1.11$\pm$0.26 mJy flux at 250 GHz. Their inferred diameter was 
609$^{+93}_{-47}$ km, largely inconsistent with the {\em Spitzer/Herschel} preferred solution \citep[D~=~393$^{+25}_{-27}$ km,][]{mommert12}. Our joint NEATM analysis of the data would indeed lead to a similar D = 395$^{+22}_{-31}$ km and an unphysical relative emissivity of 2.08$^{+0.64}_{-0.49}$ at 250 GHz. 1999~TC$_{36}$ is a triple system \citep{benecchi10} with rather similar relative sizes, estimated to  $\sim$272, 251 km, and 132 km assuming equal albedos \citep{mommert12}. The largest two components (A1, A2) orbit each other with distance $a$ $\sim$ 867 km and period P $\sim$1.9 days. However the orbit orientation is such that they are not mutually eclipsing, and the overall photometric variability of the both the central pair (A1+A2) and of the third component (B) is small and shows no evidence for a lightcurve. Note also that 1999~TC$_{36}$ has traveled $\sim$22\deg\ on its orbit over the entire time interval (Dec. 2000-July 2004-July 2010) covered by the thermal observations, restricted any changes of the sub-solar latitude to less than this number. All of this effectively rules out geometrical or orientation considerations as the cause of the discrepant far-IR and mm data. As the {\em Spitzer} and {\em Herschel} detections have high S/N \citep[see][]{mommert12}, this
suggests that the \citet{altenhoff04} detection might not be real.  ALMA observations (resolving the system) should confirm the mm fluxes. 

Eris was detected from IRAM in 2006 \citep{bertoldi06} and later by {\em Spitzer} \citep{stansberry08} and {\em Herschel} 
\citep{santos12,lellouch13}. These data, along with the occultation results from \citet{sicardy11} are analyzed jointly by Kiss et al. (in prep.). Note also that new ALMA measurements of the Eris-Dysnomia system
have become recently available (program 2015.1.00810.S, PI. M.E. Brown).

We note finally that \citet{margot02} reported a successful detection of 2002 AW$_{197}$ at 1.2 mm from IRAM and estimated a $\sim$890$\pm$120 km diameter, but did not quote the measured fluxes, which prevents us for performing a reanalysis.

\section{Discussion and conclusions}

\begin{table}
\caption{Emissivity correlation searches and statistics} 
\label{emissivity}      
\centering                          
\begin{center}
\begin{tabular}{lccc}
Variables & N$_{values}$ & $\rho$ $^1$  & p $^2$ \\
\hline 
& & &  \\
$\epsilon_r$ {\it vs} D & 18 & 0.293 &  0.238 (1.2 $\sigma$)  \\
& & &  \\
$\epsilon_r$ {\it vs} p$_v$ & 18 & 0.245 &  0.326 (1.0 $\sigma$)  \\
& & &  \\
$\epsilon_r$ {\it vs} $\eta$ & 18 & -0.066 &  0.794 (0.3 $\sigma$)  \\
& & &  \\
$\epsilon_r$ {\it vs} T$_{SS}$ & 18 & 0.009 &  0.971 (0.0 $\sigma$)  \\
& & &  \\
$\epsilon_r$ {\it vs} color & 18 & -0.287 &  0.249 (1.2 $\sigma$)  \\
& & &  \\
$\epsilon_r$ {\it vs} $f_{water}^3$ & 10 & 0.459 &  0.182 (1.4 $\sigma$)  \\
& & &  \\
\hline

\multicolumn{4}{l}{\footnotesize $^1$ $\rho$: Spearman rank correlation coefficient} \\
\multicolumn{4}{l}{\footnotesize $^2$ p: significance p-value of correlation}\\
\multicolumn{4}{l}{\footnotesize $^3$ H$_2$O ice content, as defined by \citet{brown12}}
\end{tabular}

\vspace*{1cm}

\begin{tabular}{lccc}
Wavelength & N$_{values}$ & Median $^1$  & Mean$\pm$stdev \\
\hline 
& & &  \\
0.87 mm & 6 & 0.71$^{+0.07}_{-0.01}$ &  0.72$\pm$0.08  \\
& & &  \\
1.3 mm &  12 & 0.70$^{+0.17}_{-0.16}$ &  0.73$\pm$0.15   \\
& & &  \\
Both &  18 &  0.70$^{+0.13}_{-0.13}$ &  0.73$\pm$0.13 \\
& & &  \\
\hline

\multicolumn{4}{l}{\footnotesize $^1$ Error bars encompass central 68.2 \% of values}
\end{tabular}

\end{center}
\end{table}

\begin{figure}[ht]
\centering
\includegraphics[width=6cm,angle=-90]{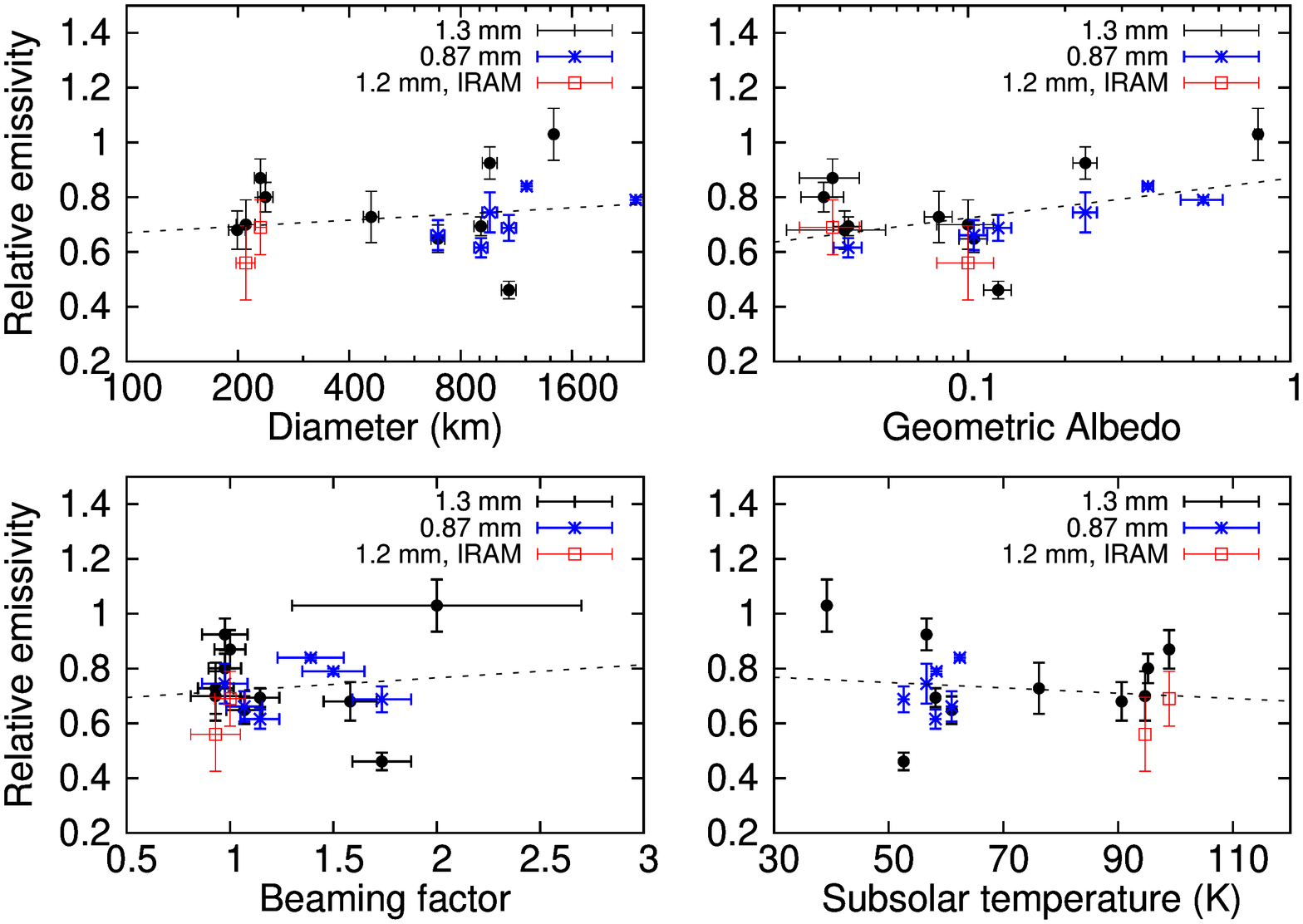}
\includegraphics[width=6cm,angle=-90]{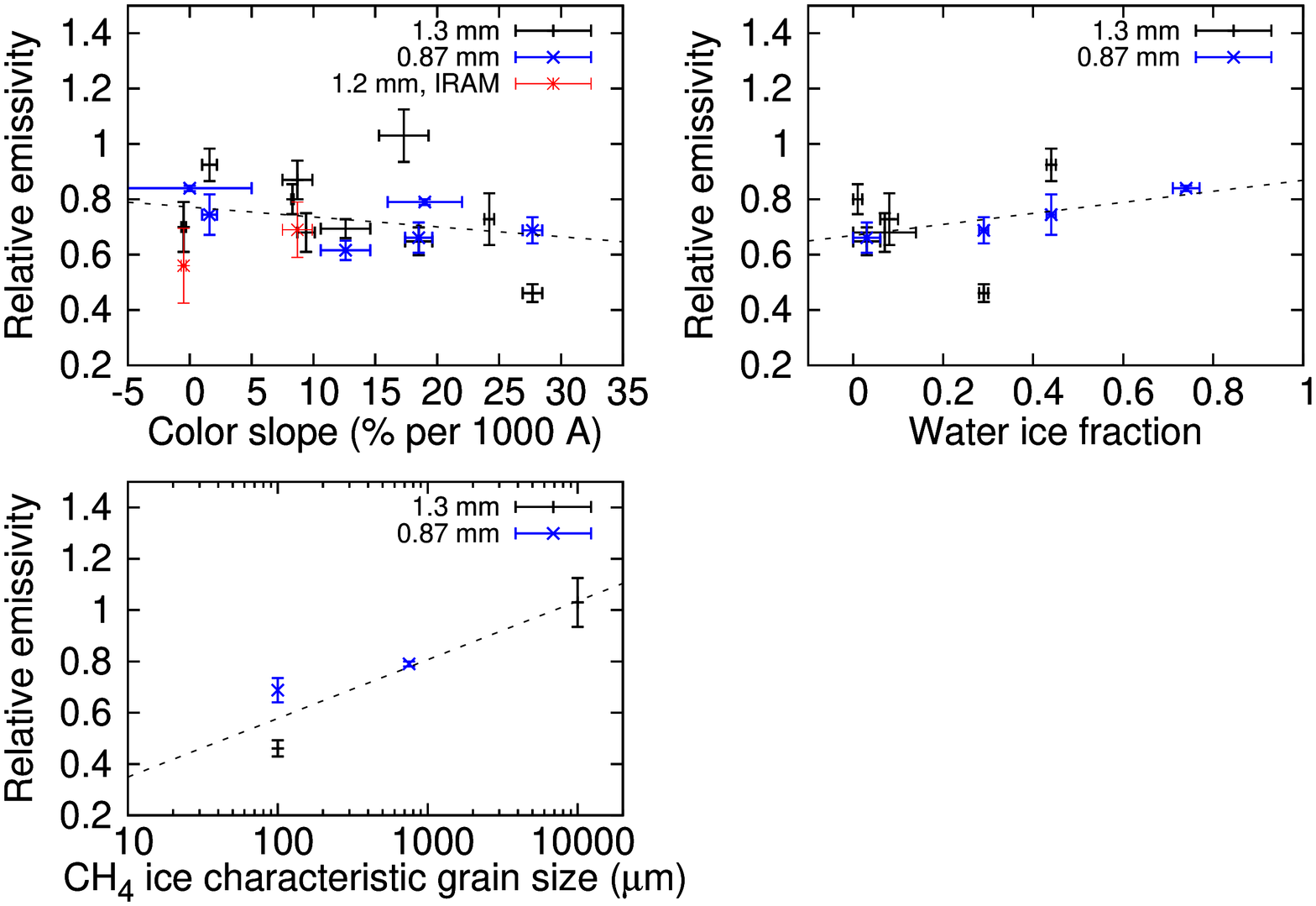}
\caption{Derived radio emissivities plotted as a function of diameter, geometric albedo, beaming factor, sub-solar temperature, color slope, H$_2$O ice content, and CH$_4$ ice characteristic grain size (see text). Dashed lines indicate simple linear fits.}
\label{fig:emissivity_trends}
\end{figure}
Including the two-wavelength measurements from \citet{brown17} (4 objects, 8 measurements), and the values for Pluto and Charon based on the \citet{butler17} ALMA data -- but not our more tentative inferences for Varuna -- we now have radio (mm/submm) emissivities for 12 Centaurs/TNOs (18 independent values). Fig.~\ref{fig:emissivity_trends}
displays these emissivities as a function of conceivably relevant parameters, i.e. diameter, geometric albedo, beaming factor\footnote{For Pluto and Charon, the equivalent beaming factor was recalculated by using the range of thermal inertia proposed by \citep{lellouch16}},
sub-solar temperature as per the NEATM model, 
and color (i.e. the slope of the visible spectrum, taking values from \citep{lacerda14}).
These plots, along with a simple Spearman rank correlation analysis (Table~\ref{emissivity}), reveal no significant correlation of the emissivity with any of these five parameters, yet the dispersion exceeds individual error bars, suggesting real variability. Table~\ref{emissivity} also summarizes statistics for the emissivities, considering either separately or together the 0.87 and 1.3 mm values. Noting that the dual-band emissivity measurements of \citet{brown17} do not suggest any consistent emissivity trend with wavelength, we recommend using a $\epsilon_{r}$ = 0.70$\pm{0.13}$ median value for interpreting further measurements at mm/submm wavelengths.
\\ 

Subdued emissivities at radio wavelengths are commonly measured in Solar System bodies, yet interpreted in various ways. For asteroids, absolute emissivities of 0.6-0.7 were reported \citep{redman92,muller98}. They were
initially attributed to grain size dependent sub-surface scattering processes, but later shown to most likely result from
a combination of surface roughness (differentially enhancing the short-wavelength fluxes), sub-surface sounding to deeper, colder
temperatures relative to the dayside surface at the longest wavelengths, and specular reflection at surface with a moderate ($\sim$2.3) dielectric constant \citep{keihm13}. Radio-observations of cometary nuclei also indicate emissivities lower than 1,
e.g. $\sim$0.5 for C1995 OI Hale-Bopp to reconcile sizes derived from IR and mm data \citep{fernandez02}, and $<$0.8 (with 0.67 as the best guess value) for 8P/Tuttle \citep{boissier11}. 

The NEATM (and thermophysical) model approaches we have used here implicitly define the emissivity in reference to the fluxes emitted from the surface itself. However, the analysis performed for Pluto/Charon by \citet[][ see their Table 2]{lellouch16} indicates that for this system at least, allowing for the decrease of the dayside temperature in the sub-surface layers would lead to an increase of the emissivity by 1-4~\% only at 500 $\mu$m (and beyond), even if the sub-diurnal layer is reached and its temperature affected by seasonal thermal inertia effects. A similar analysis cannot be done for the other objects under study here because we do not have a good handle on their individual thermal inertias. Still, the magnitude of the thermal gradient in the subsurface depends on (decreases with) the thermal inertia $\Gamma$ through the thermal parameter $\Theta=\frac{\Gamma\sqrt{\omega}}{\epsilon_b\sigma T_{ss}^3} $. TNOs as a population have thermal inertias of $\Gamma$ = (2.5$\pm$0.5) J m$^{-2}$s$^{-1/2}$K$^{-1}$ \citep{lellouch13}, i.e. 5-10 times less than Charon and Pluto, but this effect is largely compensated in the expression of $\Theta$ by a $\sim$20 times larger rotation rate $\omega$. Thus, although detailed calculations cannot be done, we suspect that sub-surface sounding is not the prime cause of the low radio emissivities. We also note that if it were the case, we might except a negative
correlation of the radio emissivity with T$_{ss}$ (for given values of $\Gamma$ and $\omega$, a higher T$_{ss}$
corresponds to a lower $\Theta$, i.e. a larger thermal gradient in the subsurface, and thus a lower apparent emissivity if sub-surface sounding was the cause), and a positive correlation with the beaming factor $\eta$, which is a proxy for $\Theta$; those are not apparent in  Fig.~\ref{fig:emissivity_trends}.

Dielectric reflection of the upward thermal radiation at the surface interface may contribute to the
some of the depressed emissivities, but this effect is unlikely to be the dominant explanation. For example,
a disk-averaged absolute radio emissivity of 0.54 ($\epsilon_{r}$ = 0.60) for a smooth surface 
implies a dielectric constant of $\sim$24, much higher than e.g. the value for solid organics  \citep[(1.2,][]{paillou08}
or H$_2$O ice ($\sim$3.1, {\it ibid.}). Although some materials have much higher dielectric constants, 
e.g. $\sim$25 for some iron oxides, this does not seem to be a realistic explanation. Applying the \citet{ostro09}
relationship relating the dielectric constant and the near surface density $\rho$ would also lead to an unrealistic $\rho$ = 5.1~g~cm$^{-3}$ density. Furthermore, the dielectric effect should be spectrally independent over the mm/submm range, unlike what is seen 
for Orcus and Quaoar in the \citet{brown17} data. 

The low emissivities observed here are reminiscent of similar findings on Giant Planet icy satellites, including
(i) 3-40 mm brightness temperatures of Europa and especially Ganymede well below the subdiurnal temperatures \citep{muhleman91}
(ii) apparent emissivities of 0.4-0.9 in most of Saturn's satellites at 2.2 cm \citep[e.g.][]{ostro06,janssen09,legall14,ries15,legall17} 
as well as of composition and texture-dependent emissivities  of ice and snow on Earth \citep{hewison99}.
Scattering effects are usually invoked, either at the surface in presence of microscopic surface roughness, or,
for transparent media, in volume, due to inhomogeneities or voids on scales comparable to~$\lambda$. 

Scattering effects are particle-size dependent, and produce emissivity minima for sizes $a$ $\sim$ $\lambda$/4$\pi$, i.e.
$\sim$100 $\mu$m for $\lambda$ $\sim$1 mm. As was already mentioned in \citet{lellouch00b}, this behavior was verified in models of terrestrial snow emissivity at 5-90 GHz as a function of particle size, absorption coefficients and medium density \citep{sherjal95}. These calculations are relevant for comparison with our measured emissivities because the absorption coefficient of ice at 260 K and 6 mm (50 GHz) is similar to that at 50 K and 1.2 mm \citep{matzler98,lellouch16}. They indicate that for a snow density of 0.15-0.20, a minimum nadir emissivity of 0.65-0.70 is reached at $a$ $\sim$ $\lambda$/4$\pi$. Emissivity increases with
increasing density, but low densities in  upper surface layers are likely for TNOs, given that their low thermal inertia suggests considerable near-surface porosity \citep{lellouch13,ferrari16}, and that the bulk density
of TNOs smaller than $\sim$1000km is already significantly less than 1 \citep{brown13a,vilenius14}.

We speculate that part of the diversity we are seeing in TNO emissivity may be due to a variability in the mean particle size in the surface/upper sub-surface layers probed by the thermal radiation. Many KBOs show the spectral signatures of H$_2$O ice \citep{barucci11,brown12}. While spectra bear in principle information on particle size, the latter is generally hard to disentangle with effects due to H$_2$O abundance, geographical coverage and state of mixing (geographical vs intimate). For this reason, we classify our objects of interest
according to the $f_{water}$ proxy defined by \citet{brown12} for the overall ``water ice amount" in the spectrum.
Plotting the radio emissivities vs $f_{water}$\footnote{In doing this we exclude Chiron and Chariklo due to the variability of the ice features in their spectrum.}  suggests (Fig.~\ref{fig:emissivity_trends}) a possible increase of $\epsilon_{r}$ with $f_{water}$, especially visible at 0.87 $\mu$m. The Spearman rank correlation coefficient and statistical significance are low (0.46 and 1.5$\sigma$, respectively) but still the strongest in Table~\ref{emissivity}. Thus, objects with the lowest radio emissivities
may have effective particle sizes of $\sim$100 $\mu$m, and the positive correlation between
$\epsilon_{r}$ with $f_{water}$, if real, could be explained if an increased $f_{water}$ traces an increase in particle size.
As suggested by the last panel in Fig.~\ref{fig:emissivity_trends}, a similar correlation between grain size and radio emissivity -- but restricted to three objects -- may occur for the methane-rich bodies in our sample (Makemake, Pluto, Quaoar). 
Here we use CH$_4$ grain sizes of 1 cm for Makemake \citep{brown07,tegler07}, and 100 $\mu$m for Quaoar \citep{schaller07}; for
Pluto, based on {\em New Horizons} data, \citet{protopapa17} report a variety of grain sizes (from 70 to 1140 $\mu$m) for the 
CH$_4$-dominated $\overline{\mathrm{CH}_4}$:N$_2$ unit and we adopt a weighted mean of 750 $\mu$m. We finally
note that based on a study of the 3.2 mm lightcurve of (4)~Vesta, \cite{muller97} similarly concluded that the grain size
distribution in the 100 $\mu$m range seems to control the emissivity at this wavelength, with spatial variations in grain size 
causing large perturbations on the rotational thermal curve.

Probably the most surprising aspect of the emissivity results is the spectral variation of emissivity for Quaoar and Orcus seen
in the data of \citet{brown17},  with both objects showing marked but opposite emissivity trends over the short 0.87-1.3 mm wavelength interval. This is unexpected because in the framework
of scattering effects, the emissivity minimum at $\lambda$  $\sim$ 4$\pi$ $a$ is expected to be broad in conjunction with a particle size distribution \citep[see e.g. Fig. 4 of][]{redman92}. A similar and equally unexplained behavior was observed for
asteroid (2867) Steins by  \citet{gulkis10}, who reported emissivities of   0.6--0.7 and 0.85-- 0.9 at 0.53 
and 1.6 mm respectively. Whether this behavior is frequent in transneptunian objects and deserves particular 
attention will require a much more extensive survey than presented here. 

As a practical conclusion, until these aspects are clarified and the possible above trends consolidated, we recommend, 
for the analysis of isolated mm/submm fluxes of TNOs, to use a bolometric emissivity of 0.90$\pm$0.06, a relative mm/submm emissivity of $\epsilon_r$ = 0.70$\pm$0.13, and a beaming factor of 1.175$\pm$0.42. Applying
this to the ALMA 233 GHz measurement of 2014 UZ$_{224}$ \citep{gerdes17}, we infer D = 633$^{+98}_{-80}$ km. Here 
the model uncertainty contributes to $^{+75}_{-60}$ km, i.e. $\sim$10 \%. This is about twice larger than the systematic uncertainty quoted by the authors ($^{+32}_{-39}$ km), who used $\epsilon_r$ = 0.68 without error bars. While a $\sim$10 \% model uncertainty
on the diameter induces a 30 \% error on the volume (i.e. density), errors on the volume associated to unknown shape / orientation, may be even bigger, as exemplified by the diverse values reported in literature for 
Chariklo's diameter. For precise determination of KBO densities, a combination of rotational lightcurves, stellar occultations, and thermal data is required.

\begin{acknowledgements}
This paper makes use of the following ALMA data: ADS/JAO.ALMA\#2015.1.01084.S (PI: E. Lellouch). ALMA is a partnership of ESO (representing its member states), NSF (USA) and NINS (Japan), together with NRC (Canada), MOST and ASIAA (Taiwan), and KASI (Republic of Korea), in cooperation with the Republic of Chile. The Joint ALMA Observatory is operated by ESO, AUI/NRAO and NAOJ.
 B.S. and R.L. received funding from the
European Research Council under the European Community's H2020
2014-2020 ERC grant Agreement n°669416 "Lucky Star".
E.L., S.F. and R.M. were supported by the
French Programme National de Plan\'etologie. 
T.M. and P. S.-S. acknowledge financial support by the
European Union's Horizon 2020 Research and Innovation Programme, under
Grant Agreement no 687378. P. S.-S. has further received funding from the Spanish grant AYA-2014-56637-C2-1-P and
the Proyecto de Excelencia de la Junta de Andaluc\'ia J.A. 2012-FQM1776.
We thank Juan Mac\'{\i}as-P\'erez and Jean-Fran\c{c}ois Lestrade for communicating us with the Pluto/Charon
NIKA2 fluxes prior to publication and for important discussions.

\end{acknowledgements}



\appendix

\section{Absolute flux calibration}
 \label{sec:calibration}

\begin{figure*}[ht]
 \label{fig:calibration}
\centering
\includegraphics[width=4.2cm,angle=-90]{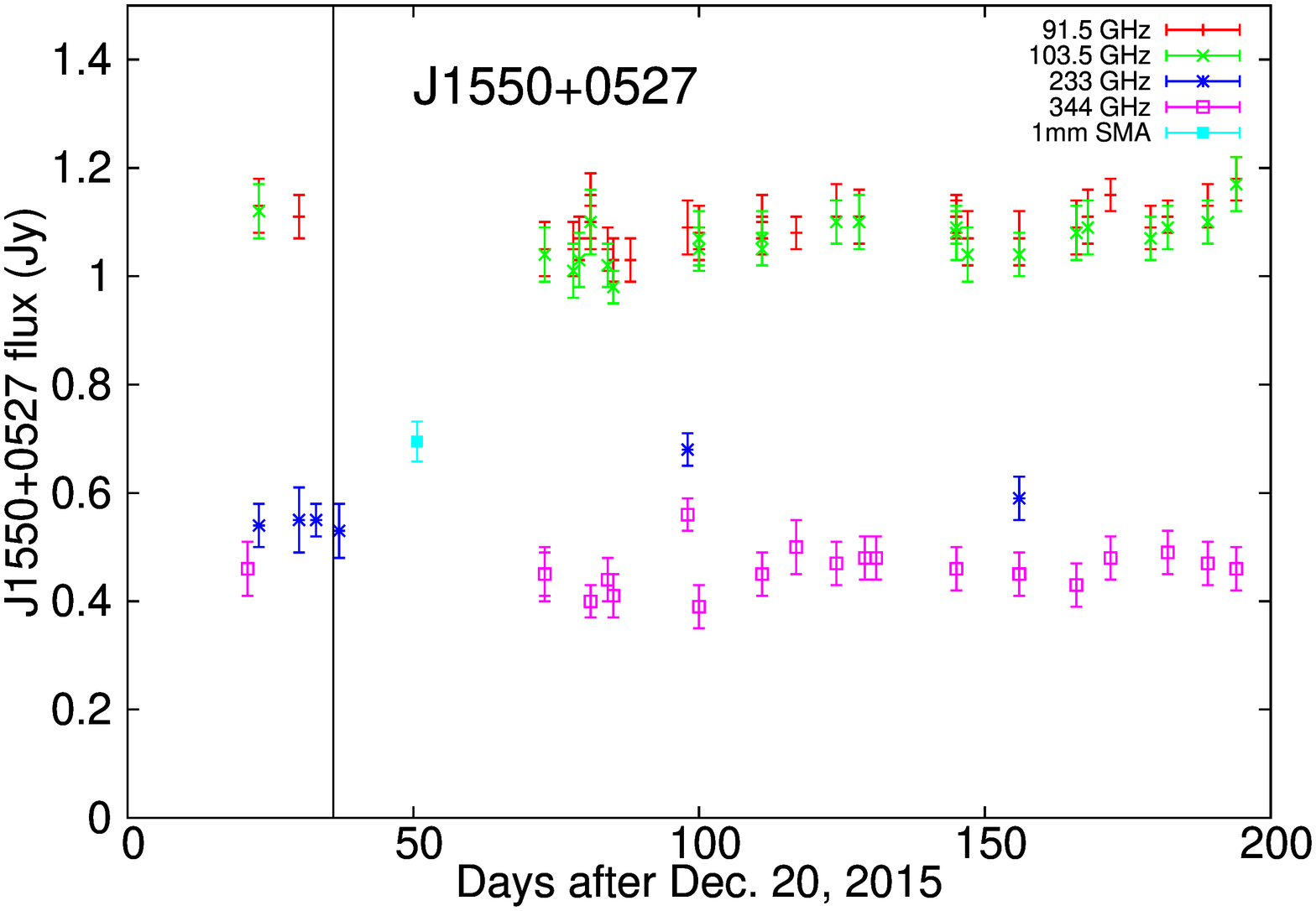}
\includegraphics[width=4.2cm,angle=-90]{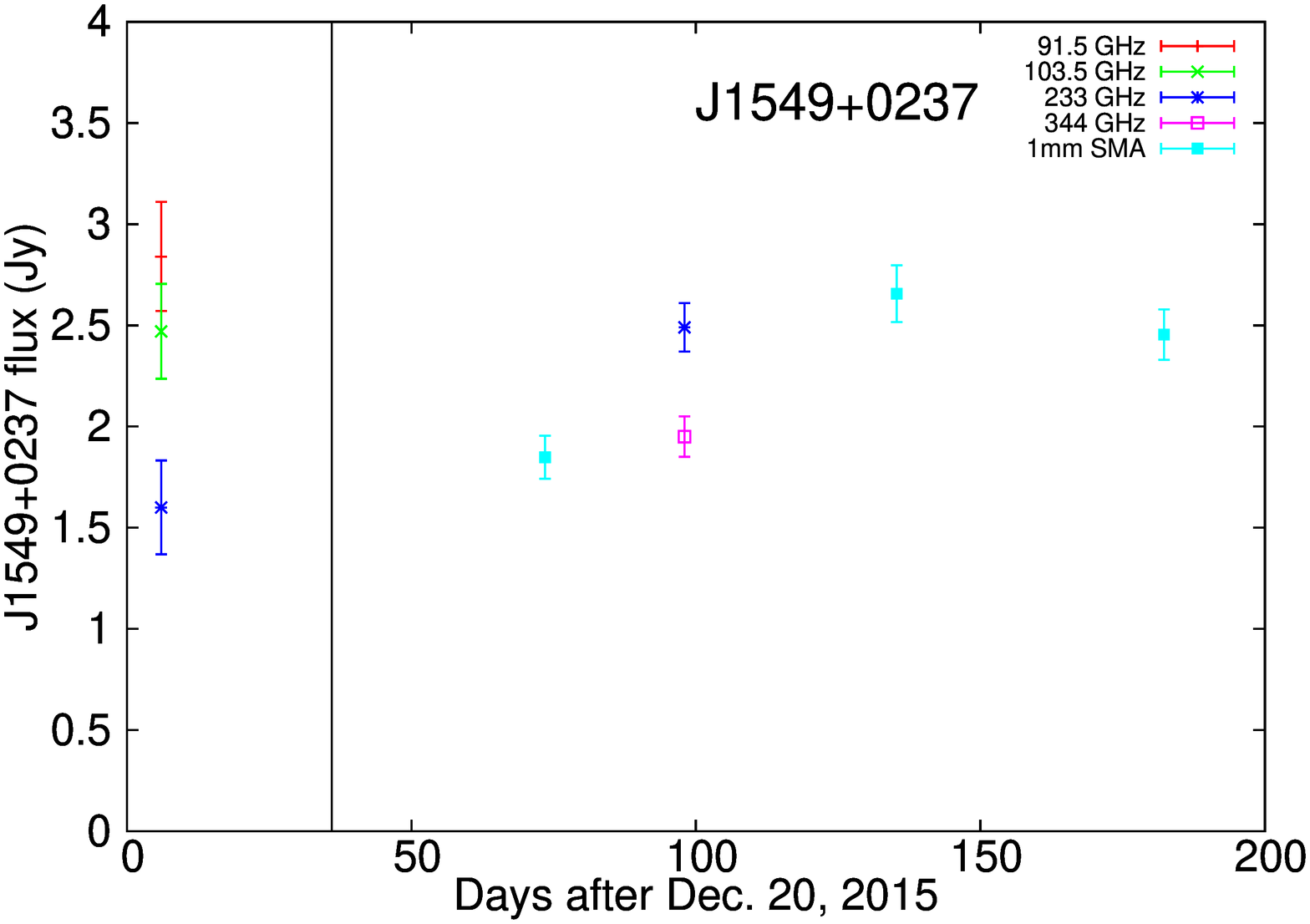}
\includegraphics[width=4.2cm,angle=-90]{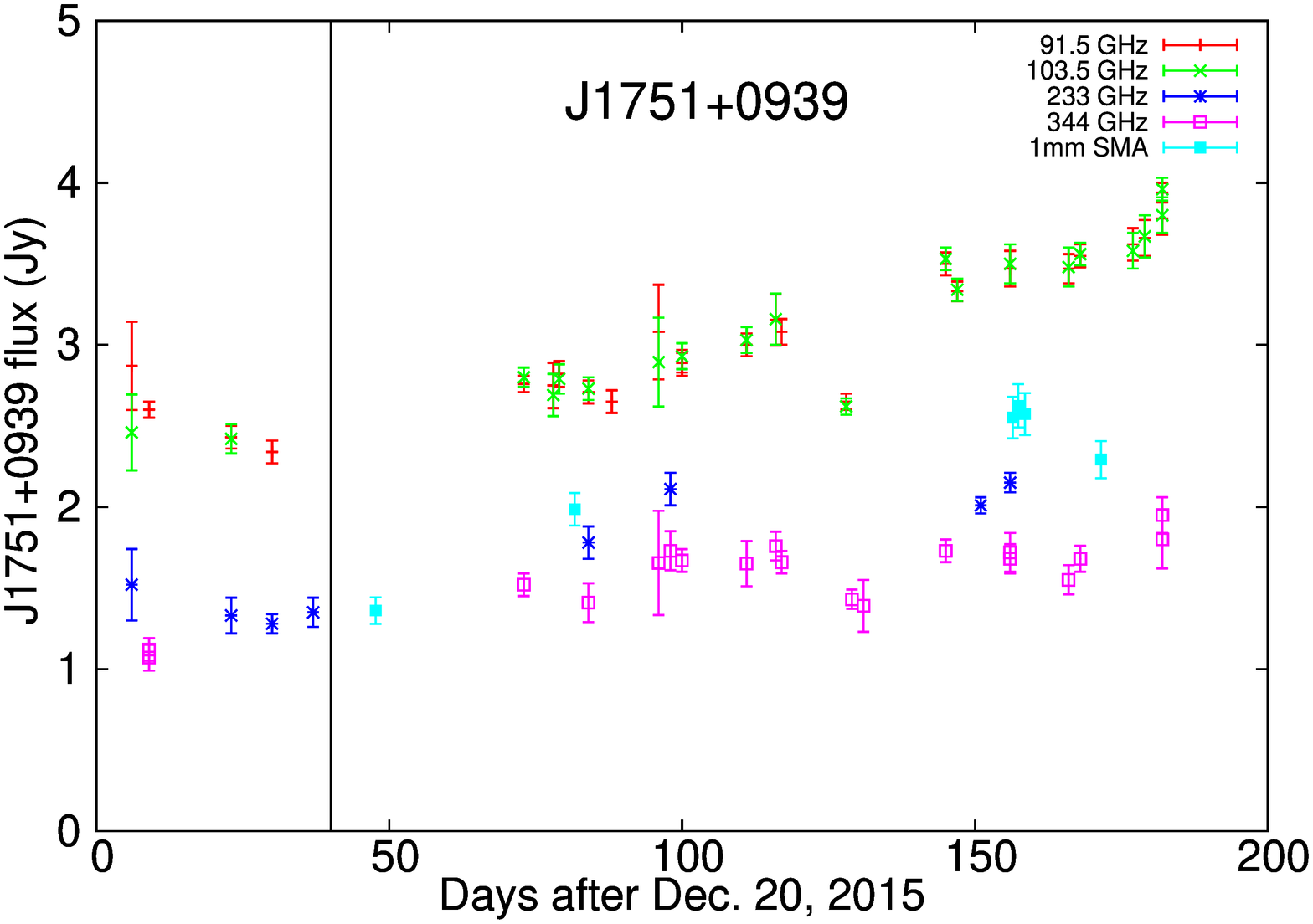}
\includegraphics[width=4.2cm,angle=-90]{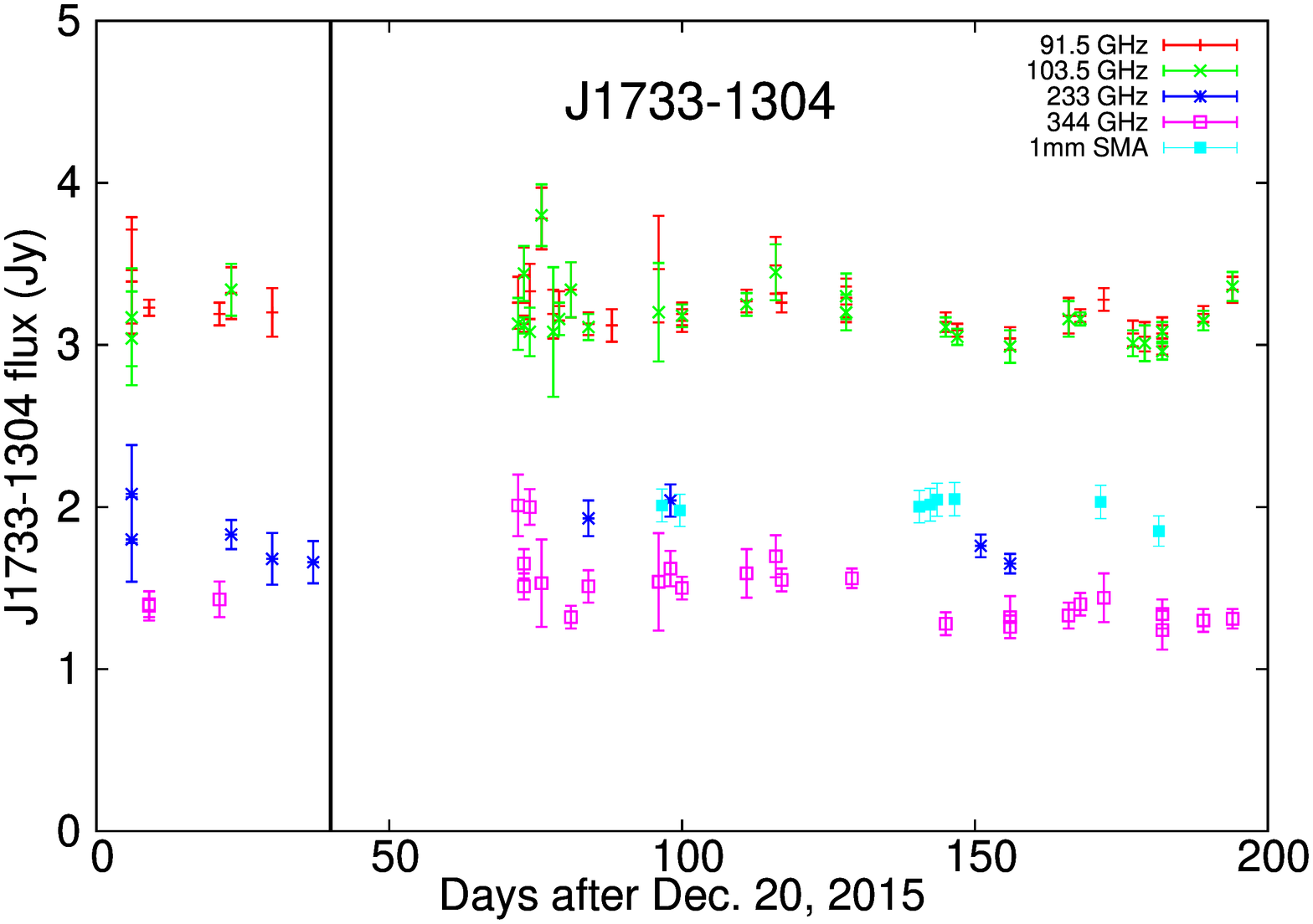}
\includegraphics[width=4.2cm,angle=-90]{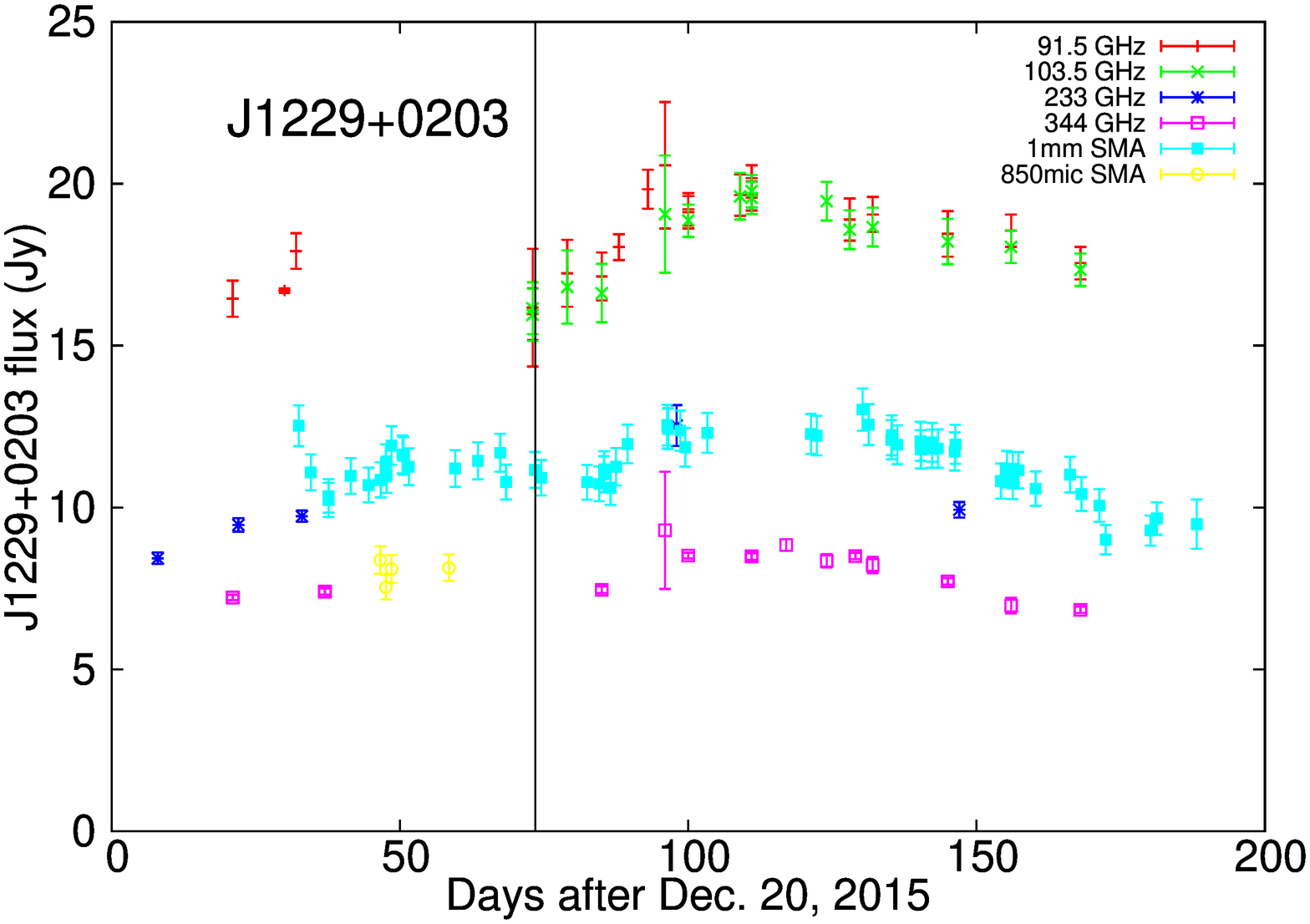}
\includegraphics[width=4.2cm,angle=-90]{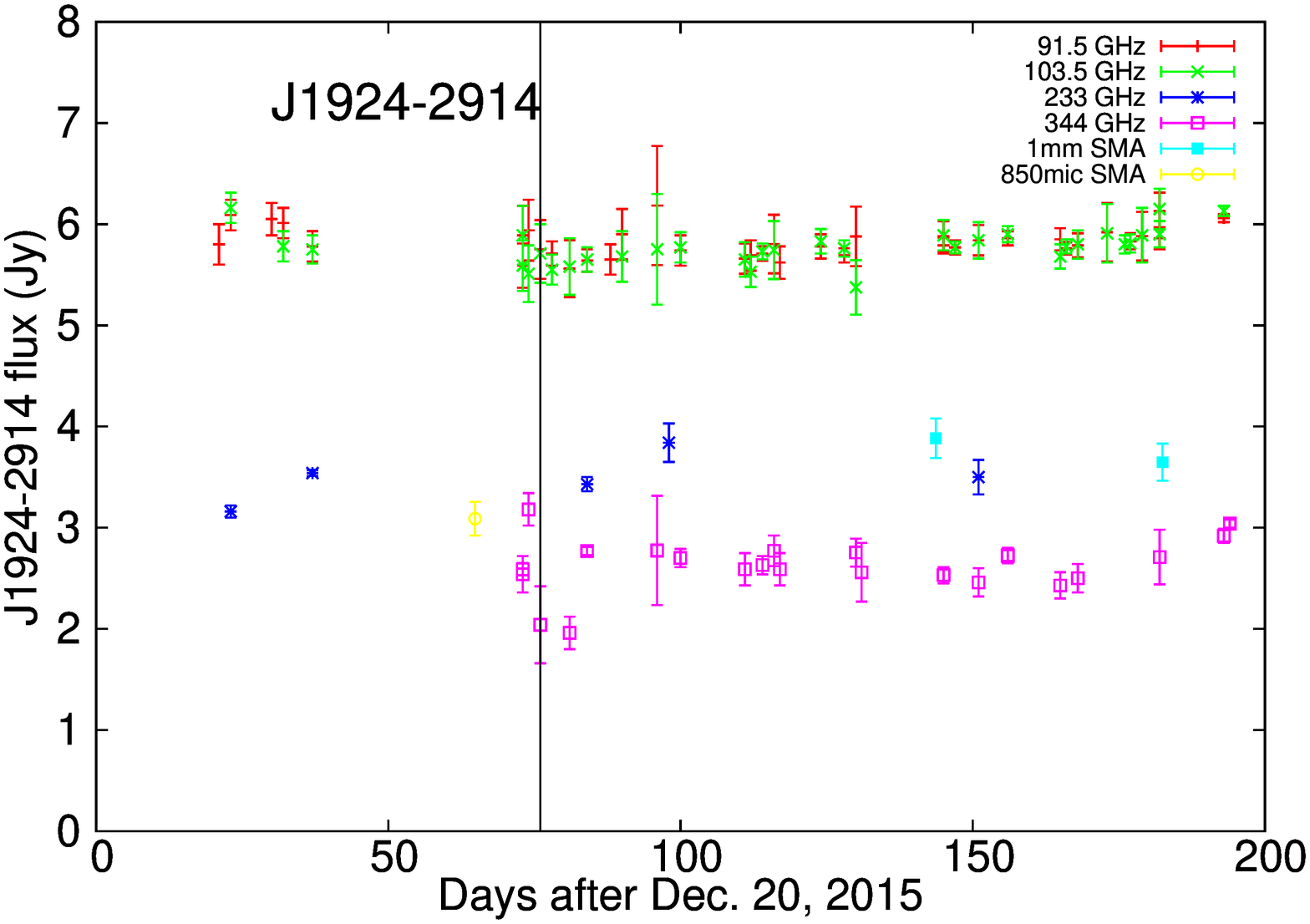}
\includegraphics[width=4.2cm,angle=-90]{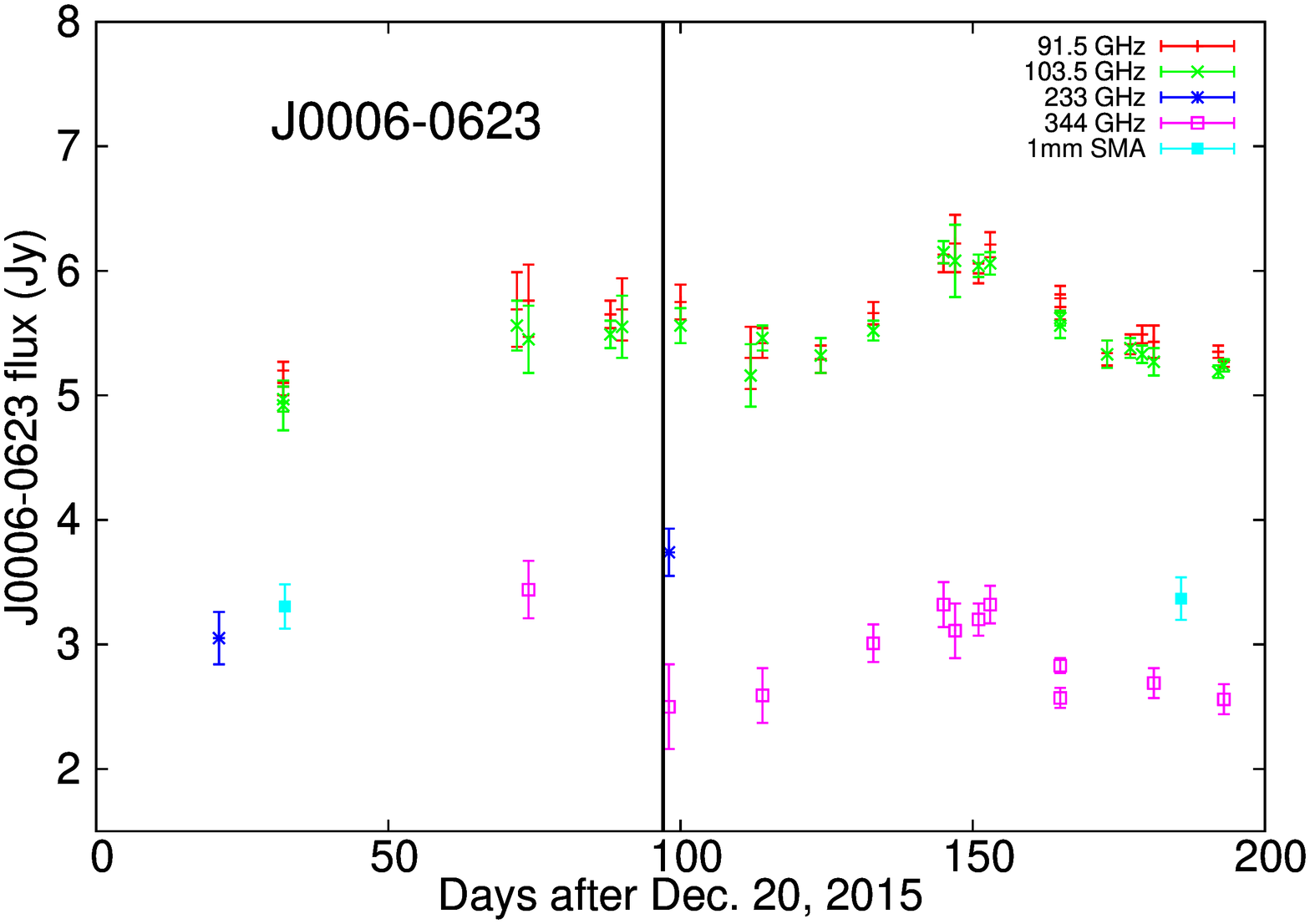}
\includegraphics[width=4.2cm,angle=-90]{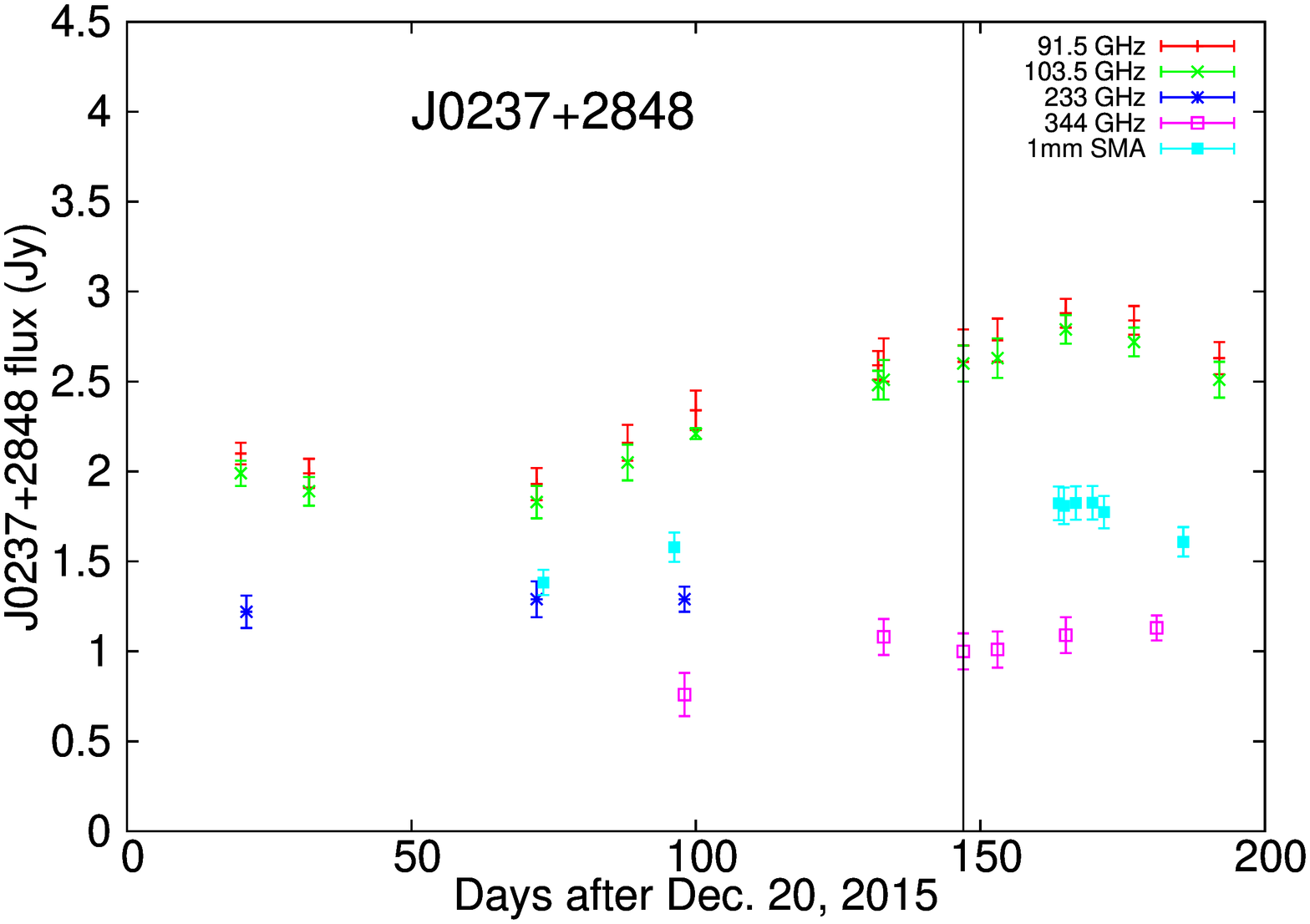}
\includegraphics[width=4.2cm,angle=-90]{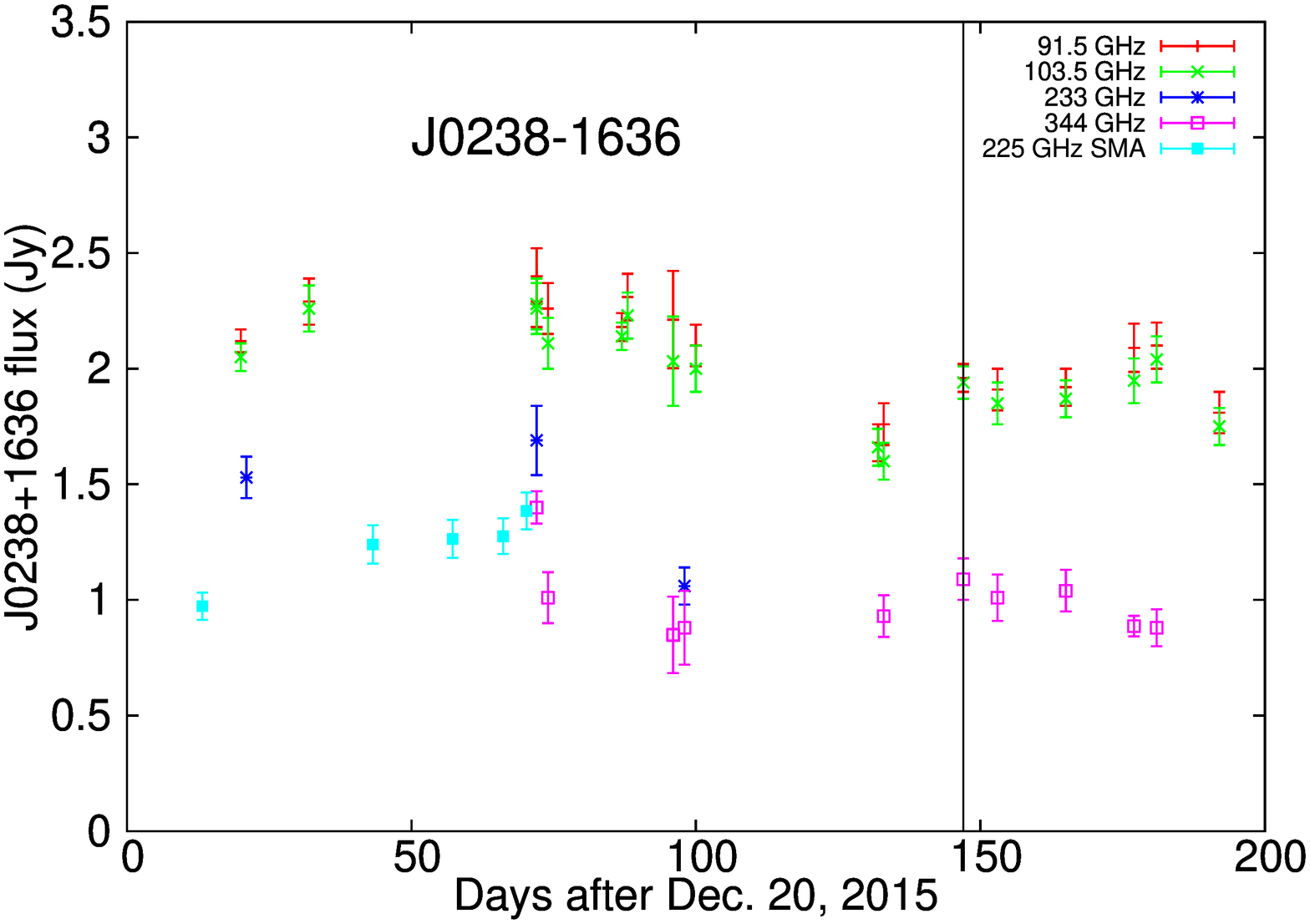}
\caption{Flux measurements of nine quasars used as flux or secondary calibrators, over a 200 day period (Dec 20, 2015 - July 7, 2016) spanning our observations. These measurements are taken from the
ALMA Calibrator Source Catalogue ({\it https://almascience.eso.org/sc/}) at 91.5, 103.5, 233 and 344 GHz or
from the SMA Calibrator list ({\it http://sma1.sma.hawaii.edu/callist/callist.html}) at $\sim$225 GHz. 
The vertical bar corresponds to the date of our TNO observations.
}
\end{figure*}

A good knowledge of the 233 GHz flux of the calibrators, which may be time-varying, is critical for the reliability of the TNO flux scale. Out of the 12 quasars observed in our program, 9 are monitored from ALMA (normally at 91.5, 103.5, 233 and 344 GHz) and from the SMA at 1 mm ($\sim$225 GHz) and more occasionally at 0.85 mm ($\sim$345 GHz). These measurements are displayed in Fig.~\ref{fig:calibration}.1 over a 200 day period (Dec. 20, 2015 - July 7, 2016) spanning our observing period. ALMA 233 GHz measurements
are the most direct source of information for our purpose, but are less abundant than at other bands. To estimate
the 233 GHz fluxes at the relevant dates (marked by the vertical bars in Fig.~\ref{fig:calibration}.1), we combined this
information with (i) $\sim$225 GHz SMA measurements and (ii) flux values in ALMA Band 3 and Band 7  and the associated
spectral indices. Details follow:

\begin{itemize}
\item {\bf J1924-2914} (Chariklo). This object is very stable in time. Direct 233 GHz measurements from ALMA indicate a mean flux of
3.47$\pm$0.024$\pm$0.24 Jy, where the first error bar is the formal uncertainty calculated from the individual errors, and the second one is the standard deviation between the individual measurements. Spectral index (spix) fitting of the mean 91.5, 103 and 343 GHz 
fluxes gives spix = -0.56 and a 233 GHz flux of 3.55 Jy, fully consistent with above. Because we do not have any other quasar available to calibrate the Chariklo data, we conservatively adopt a 3.51$\pm$0.24 Jy flux, i.e. a 6.8~\% uncertainty.\\

\item {\bf J1733-1304} and {\bf J1751+0939} (2002 GZ$_{32}$). J1733-1304 is also very stable in time, with a mean flux of 
1.89$\pm$0.026$\pm$0.19 Jy. Spectral index fitting of the 91.5, 103 and 343 GHz fluxes gives spix = -0.61 and a 233 GHz flux of 1.86 Jy. We initially adopt a 1.87 Jy flux. J1751+0939 is more variable, but its 233 GHz flux was measured on Jan. 26, 2016 (3 days before 2002 GZ$_{32}$) to be 1.35$\pm$0.09 Jy. Specifying a 1.87 Jy for the prime calibrator J1733-1304 leads to 1.28 Jy for 
J1751+0939 in our data. A better overall compromise is achieved by specifying 1.93 Jy for J1733-1304 (+3.2 \% higher than
the initial estimate, in which case the measured J1751+0939 flux is 1.31 Jy (3.3 \% too low). We thus adopt 1.93$\pm$0.06 Jy 
(3 \% uncertainty).  \\

\item{\bf J1550+0527} (Huya). This object shows variability and ALMA and SMA data do not appear fully consistent.
For February 8, 2016, the SMA provides a 0.695$\pm$0.037 Jy flux at 225.52 GHz, i.e. 0.680$\pm$0.037 Jy at 233 GHz given a spix of -0.69$\pm$0.05, derived from 12 sets of 91, 103.5 and 343.5 GHz ALMA measurements.
In contrast, ALMA provides 0.54$\pm$0.01 Jy over January 19-26, 2016. The Huya data were taken on Jan. 25. Adopting 0.68 (resp. 0.54 Jy) for J1550+0527 would yield 1.08 Jy (resp. 0.85) Jy for Titan, which was observed on the same date. Our Titan models 
\citep[e.g.][]{moreno12} predict 0.96 Jy for that date. We finally adopt 0.61$\pm$0.07 Jy (ie 11.4 \% uncertainty) for J1550+0527.
Note that J1549+0233 was observed as another phase calibrator but this object is 
not well enough characterized (see 
Fig.~\ref{fig:calibration}) to serve as flux calibrator. \\

\item{\bf J1229+0203} (Makemake).  This variable quasar, a.k.a. 3C273, is particularly monitored at the SMA, and
measurements on March 2 (i.e. the same day as our Makemake observations), indicate a 225.47 GHz flux of 11.160$\pm$0.559 Jy. ALMA 91, 103.5, and 343 GHz measurements indicate a spix of -0.67$\pm$0.05, from which we derive 10.92$\pm$0.55 Jy at 233 GHz for that date. For conservativeness, and because we do not have catalog values for the second calibrator (J1303+2433), we finally adopt 10.9$\pm$0.75 Jy for J1229+0203 (6.9 \% uncertainty). \\ 

\item{\bf J0006-0623} and {\bf Pallas} (Chiron). Asteroid (4) Pallas was observed as the prime calibrator for the Chiron observations. Detailed shape and thermophysical models \citep{muller14} for this object (as well as  for (1) Ceres, (2) Vesta and (21) Lutetia)  have been demonstrated to match the ensemble of the available (and independently calibrated) {\em Herschel} PACS and SPIRE
measurements over 70-500 $\mu$m  within 5 \%. As these models are less constrained at longer wavelengths,
we prefer to use the secondary calibrator
(J0006-0623) as flux calibrator, as it was remeasured at ALMA on March 27, 2016, just 1 day after the Chiron observations, 
with a 233-GHz flux of 3.74$\pm$0.19 Jy. On another hand, the spectral index, derived from 10 sets of ALMA 91.5, 103.5 and 343 GHz dat, is spix = -0.50$\pm$0.05. Applying this to the 343 GHz flux measured on the same date (2.5$\pm$0.34 Jy) would provide 3.03$\pm$0.41 Jy, only 1.2$\sigma$ consistent with the above. The weighted-mean average of the two values is 3.61$\pm$0.17 Jy. We still adopt the directly measured value of 3.74 Jy, but inflate somewhat the uncertainty to 0.25 Jy (6.7 \%). With this, we derive a Pallas flux of 159$\pm$11 mJy for March 26, 2016, UT =14h40. The application of the above models from \citet{muller14}
to the UT date of the Pallas measurement gives a flux of 153.7 mJy (with significant thermal lightcurve variability: 152$\pm$8 mJy), fully within the measurement error, and giving further confidence in both the flux scale and the Pallas model. \\

\item{\bf J0238+1636} and {\bf J0237+2848} (Bienor). These two variable quasars have few 233 GHz measurements, but numerous
91.5, 103.5 and 343 GHz data from ALMA, including for the precise date of the Bienor data (May 15, 2016). For this date,
spectral index fitting results in 233 GHz fluxes of 1.31$\pm$0.08 and 1.37$\pm$0.11 Jy respectively. We adopt
a 1.29 Jy for J0238+1636 (i.e. 1.7 \% less than the above flux), which using our own measurements, returns 1.40 Jy for
J0237+2848 (i.e. 2.4 \% more than above). Given the excellent consistency between the two quasars, we adopt 
a 2.5 \% uncertainty on their fluxes. \\

\end{itemize}

\end{document}